\documentclass[reprint,
preprintnumbers,
 amsmath,amssymb,
 aps,
pra,
]{revtex4-2}
\usepackage{graphicx}
\usepackage{dcolumn}
\usepackage{multirow}
\usepackage{dsfont}
\usepackage{braket}
\usepackage{verbatim}
\usepackage{xcolor}
\usepackage{float}
\makeatletter
\let\newfloat\newfloat@ltx
\makeatother
\usepackage{algorithm}
\usepackage{algpseudocode}
\usepackage{array}
\usepackage{hyperref}
\usepackage{bm}
\hypersetup{
    colorlinks=true,
    linkcolor=blue,
    citecolor=blue,
    urlcolor=blue
}

\newcommand{\im}{\textrm{Im}}

\begin{document}

\preprint{MIT-CTP/5939}
\title{Robust multiparameter estimation using quantum scrambling}

\author{Wenjie Gong}

 \author{Bingtian Ye}%
\author{Daniel K. Mark}%
\author{Soonwon Choi}%
\affiliation{%
Center for Theoretical Physics---a Leinweber Institute, Massachusetts Institute of Technology, Cambridge, MA 02139, USA
}%

\date{\today}

\begin{abstract}
We propose and analyze a versatile and efficient multiparameter quantum sensing protocol, which simultaneously estimates many non-commuting and time-dependent signals that are coherently or incoherently coupled to sensing particles. Even in the presence of control imperfections and readout errors, our approach can detect exponentially many parameters in the system size while maintaining the optimal scaling of sensitivity. To accomplish this, scrambling dynamics are leveraged to map distinct signals to unique patterns of bitstring measurements, which distinguishes a large number of signals without significant sensitivity loss. Based on this principle, we develop a computationally efficient protocol utilizing random global Clifford unitaries and evaluate its performance both analytically and numerically. Our protocol naturally extends to scrambling dynamics generated by random local Clifford circuits, local random unitary circuits (RUCs), and ergodic Hamiltonian evolution---commonly realized in near-term quantum hardware---and opens the door to applications ranging from precise noise benchmarking of quantum dynamics to learning time-dependent Hamiltonians.

\end{abstract}

\maketitle

Quantum sensing is one of the most promising frontiers of quantum science~\cite{Degen2017, Pirandola2018, Aslam2023}.
It underpins technologies such as atomic clocks~\cite{Ludlow2015} and magnetometers~\cite{Taylor2008, Kominis2003} and has the potential to transform areas ranging from medical imaging~\cite{Aslam2023} to gravitational wave and dark matter detection~\cite{Barsotti2019, Tobar2024, Ebadi2022, Schnabel2010, Shi2023}. Many of these applications --- including  vector electromagnetic field imaging~\cite{Wang2021, Niethammer2016, Yang2020} and astronomical detection involving multiple sources~\cite{Dailey2020} --- inherently require the estimation of multiple parameters.

Quantum multiparameter metrology---or the simultaneous estimation of more than one parameter with a quantum system---has been extensively studied, with particular focus on fundamental precision limits~\cite{Li2022, Gessner2018, Goldberg2021, Sidhu2020, ZhouChen2025, Demkowicz-Dobrzaski2020, Tsang2011, Chen2024Inc}. Many practical protocols have primarily targeted improving sensitivity for estimating only a few parameters, such as the three components of a vector field~\cite{Isogawa2023, Vasilyev2024, Omanakuttan2024, Kaubruegger2023}. While highly valuable, these studies do not consider settings in which a large number of parameters---possibly coherent, incoherent, or time-dependent---must be simultaneously determined. A representative example is waveform estimation~\cite{Tsang2011, Tritt2025, Cooper2014, Xu2016}, where the full spatiotemporal profile of a signal must be reconstructed. Only very recently have variational strategies been proposed for robustly sensing arbitrary parameters~\cite{Meyer2021, Kaubruegger2023, Le2023}, yet  such methods require re-optimization for each new signal set. These considerations calls for a single, broadly applicable, and robust protocol for sensing many parameters simultaneously.

In this work, we introduce a simple and robust multiparameter sensing protocol that can detect exponentially many signals with respect to the number of sensors. The key idea is to subject sensor qubits to known entangling dynamics---for example, generated by random Clifford unitaries---interspersed with unknown time-dependent signals. As we show, such scrambling dynamics encodes the signals in a form that, although complex, is recoverable from simple bitstring measurements in the computational basis via efficient classical post-processing.
\begin{figure}
    \centering
    \includegraphics[width=1\columnwidth]{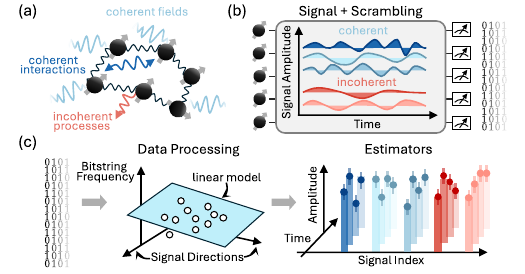}
    \caption{(a) Our multiparameter sensing protocol subjects sensor qubits to coherent signals (blue), including external fields or multi-body interactions, as well as incoherent dissipation (red). (b) The system evolves under known random dynamics while accumulating signal, followed by possibly noisy measurement in the computational basis. (c) The resulting empirical bitstring distributions are processed classically using a multivariate least-squares regression procedure (described in the main text) to simultaneously estimate many signal parameters. Estimated amplitudes and their uncertainties are indicated schematically by markers, in close agreement with their true values (bars). }
    \label{fig:intro}
\end{figure}

Our approach enables sensing both coherent and incoherent signals. 
Specifically, coherent signals are described by the unitary evolution $\exp{(-i \sum_\alpha\theta_{\alpha} P_{\alpha})}$, where $P_\alpha$ is a Pauli operator (possibly acting on many qubits), and $\theta_\alpha$ is the strength of the signal. Incoherent signals arise from dissipative processes $\rho \rightarrow (1-\gamma_\alpha) \rho + \gamma_\alpha P_\alpha \rho P_\alpha$ on the state of the system $\rho$, where $\gamma_\alpha$ is the rate to be estimated. The sensors are exposed to both coherent and incoherent signals at discrete times $t$, and $\theta_\alpha(t)$ and $\gamma_{\alpha}(t)$ are determined independently and simultaneously for all time steps.
In physical settings, signals associated with single-body Pauli operators may represent external fields, while multi-body terms can arise from interactions between the sensing qubits (Fig. \ref{fig:intro}).
\begin{figure*}
    \centering
    \includegraphics[width=2\columnwidth]{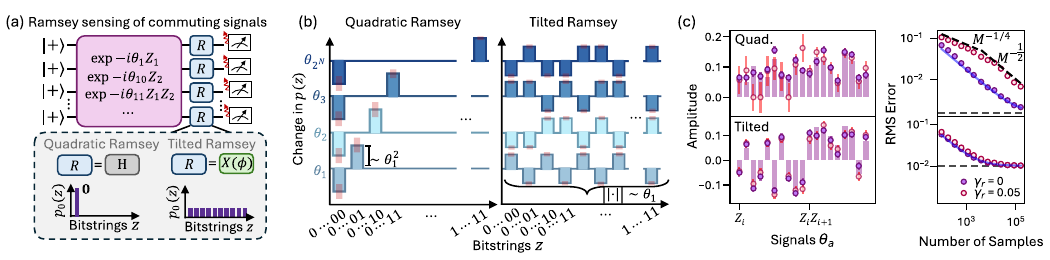}
    \caption{(a) Quadratic Ramsey protocol and tilted Ramsey protocol for the estimation of commuting signals generated by Pauli $Z$ operators. (b) The change of the output distribution $p(z)$ after signal accumulation with each Ramsey protocol. The effect of readout error is indicated by red bars. In the tilted Ramsey protocol, unlike the quadratic Ramsey protocol, the change in $p(z)$ scales \emph{linearly} with the strength of the signal, providing weak robustness against readout error. (c) Numerical demonstrations of the quadratic and tilted Ramsey sensing protocols. Left: Reconstructed strengths for all single-body $Z_i$ and nearest-neighbor $Z_iZ_{i+1}$ signals. Bars show the ground truth signal, while purple and red markers indicate estimates at readout error $\gamma_r = 0$ and $\gamma_r = 0.05$, respectively. The simulation is on a system $N=10$ with $M=2000$ samples. Right: Scaling behavior of the root-mean-square (RMS) error for the Ramsey estimators. 
    In quadratic Ramsey sensing (top), $\gamma_r = 0.05$ is shown for only single-body signals. The blue solid line indicates the theoretical prediction for the typical sample complexity (derived in the SM~\cite{supp}), while the horizontal dashed line indicates the bias of the estimator, where RMS error saturates for large enough sample size.
    In the tilted Ramsey protocol (bottom), the error scaling remains robust even when $\gamma_r=0.05$.}
    \label{fig:ramsey}
\end{figure*}

Our approach can detect up to exponentially many signals $\{\theta_\alpha(t), \gamma_\alpha(t)\}_{\alpha, t}$ in the system size $N$. In general, estimating all of these parameters to high multiplicative precision necessitates exponentially many measurements, since each measurement provides at most $N$ bits of information. 
Instead, we estimate the signals to additive precision $\epsilon$ in the regime where the total strength of the signals is weak and perturbative: $\sum_{\alpha, t} \theta_{\alpha}(t)^2 + \sum_{\alpha, t} \gamma_{\alpha}(t)\ll 1$.
Practically, this regime is most meaningful when only a small subset of signals have appreciable magnitude, and our goal is to identify and estimate the dominant signal sources among many candidates. Interestingly, the number of measurements required to achieve a target error $\epsilon$ depends only logarithmically on the number of signals and is largely independent of $N$, as we show in this work.



In our setting, standard quantum limit (SQL) scaling is information-theoretically optimal: all parameters to be estimated are independently encoded, with no temporal or spatial correlations available to amplify sensitivity. We therefore target the \textit{robust} SQL: even in the presence of control imperfections---which can be treated as additional signals---and extensive readout error---which can be mitigated through classical error correction techniques---our protocol maintains SQL sensitivity $\epsilon\sim 1/\sqrt{M}$. Here, $M$ is the total number of measurement samples. Thus, our protocol constitutes a foundational building block to accomplish more structured sensing or learning tasks such as benchmarking or time-dependent Hamiltonian learning~\cite{manole2025, Boixo2018, Mark2023, Choi2023, Shaw2024, Han2021}. 

\textit{Multiparameter sensing via Ramsey protocols ---} 
To illustrate the key conceptual ideas behind our approach, we start by considering the simplest possible toy example: the standard Ramsey protocol~\cite{Ramsey1950,Degen2017}. This simple method, which uses unentangled sensors, can determine the magnitudes of exponentially many commuting coherent signals, albeit with several limitations: it assumes no time-dependence, lacks robustness to readout errors, and cannot be extended to incoherent or non-commuting signals.
We will progressively resolve these shortcomings by first presenting a modified Ramsey protocol and then introducing our Clifford-circuit protocol.

For simplicity, we present our analysis at leading order in the signal strength. Higher-order corrections can be removed by post-processing and introduce only a small bias to our estimators; as we operate in the regime where statistical error from finite $M$ dominates, bias is neglected in our sample complexity analyses. A detailed analysis of the bias and how it can be corrected is provided in the Supplemental Material (SM)~\cite{supp}.


The standard (linear) Ramsey procedure~\cite{Degen2017}
senses a field along the $Z$ direction by preparing the sensor in the $|+\rangle \equiv (|0\rangle+|1\rangle)/\sqrt{2}$ state, allowing it to accumulate a small phase $\exp{(-i \theta Z)}$ under the field, and then measuring in the $y$ basis to infer the magnitude of $\theta$.
The \emph{quadratic} Ramsey protocol~\cite{Degen2017} proceeds similarly, except with measurement in the $x$ basis (Fig.~\ref{fig:ramsey}(a)). 
Both protocols yield the same signal-to-noise ratio in the absence of any errors, while the linear protocol achieves better sensitivity scaling under readout error~\cite{Degen2017}.
As we explain below, the $N$-qubit generalization of the quadratic Ramsey protocol, unlike the linear one, can detect exponentially many signals. In this generalization, the initial state and signal unitary are replaced by $\ket{+}^{\otimes N}$ and $\exp{(-i \sum_a \theta_a Z_a)}$ respectively, where signals are generated by commuting $N$-qubit Pauli strings $Z_a = \otimes_{k=1}^{N} Z^{a_k}$. Here, $a = a_1 \cdots a_N \in \{0, 1\}^N$ is a length $N$ bitstring, and each bit $a_k$ indicates whether the operator on qubit $k$ is the identity or Pauli $Z$. 

The key feature that enables multiparameter sensing is the one-to-one mapping of  
each signal $\theta_a$ onto a unique measurement outcome. 
In the absence of signal, the quadratic Ramsey protocol results in a deterministic outcome: $\ket{0}^{\otimes N}$, which we associate with the all-zeros bitstring $z=\pmb{0}$. In the presence of $\theta_a$, some measurement outcomes may flip. These different bitflip patterns are enumerated via bitstrings $z$.
The probability of measuring the bitstring $z=a$ is given by $p(z=a|\theta_a) \sim \theta ^2_a + O(\theta^3_a)$.
Thus, the magnitude of $\theta_a$ can be deduced from the empirical probability $\hat p(z=a) =  \frac{\hat N_a}{M}$, where $\hat N_a$ is the number of times the bitstring $a$ is measured in $M$ total shots. This simple approach  achieves worst-case error $\max_a |\hat \theta_a -|\theta_a|| \leq \epsilon$ with high probability $1-\delta$ using $M= O(\log(K/\delta)/\epsilon^2)$, where $K$ is the number of signals (see SM~\cite{supp} for proof). 





While the quadratic Ramsey protocol is nominally capable of multiparameter sensing, its usefulness is limited in practice by its fragility to readout errors.
When the qubit readout error rate $\gamma_r$ is nonzero, estimating a single-body signal of magnitude $\theta$ proceeds slowly over a large non-asymptotic regime in $M$, with the additive error scaling as $\epsilon \sim M^{-1/4}$. The standard quantum limit, $\epsilon \sim 1/\sqrt{M}$, is recovered only asymptotically for $M \gg 1/\theta^{4}$ (see SM~\cite{supp} and Fig.~\ref{fig:ramsey}(c)).

We introduce a \textit{tilted Ramsey} protocol that overcomes this fragility to readout error. Instead of the $x$-basis measurement used in quadratic Ramsey sensing, this protocol applies a layer of rotations about the $x$-axis $X(\phi)$, followed a $z$-basis measurement (Fig.~\ref{fig:ramsey}(a)). The angle $\phi$ can be chosen arbitrarily, provided that $\phi/\pi$ is irrational~\footnote{ The choice of $\phi$ satisfying $s\phi = m\pi$ for integers $s, m$ may lose sensitivity to certain signals.}. While this choice of measurement basis can result in a larger sample-complexity prefactor, the overall sensitivity scaling remains robust to readout error.

In the absence of signal, the tilted Ramsey protocol leads to an exactly uniform distribution of measurement outcomes $p_0(z)$.
The presence of signal $\theta_a$ linearly perturbs $p_0(z)$, leading to the distribution $p(z|\theta_a)$. Crucially, the \emph{probability differences} $p(z|\theta_a)- p_0(z)$ form a unique pattern over bitstrings $z$ (Fig.~\ref{fig:ramsey}(b)).
The magnitude and sign of each probability difference can be efficiently precomputed. Estimators $\hat \theta_a$ are then obtained by fitting the observed $\hat p(z) = \frac{\hat N_z}{M}$ to these patterns using least-squares regression~\cite{Hastie2001}. With probability at least $1-\delta$, these estimators similarly achieve $\max_a |\hat \theta_a -\theta_a|\leq \epsilon$ for small $\epsilon$ when $ M = O(\log(K/\delta)/\epsilon^2)$ (see SM~\cite{supp} for the proof). As the estimator $\hat \theta_a$ combines statistics from many bitstring measurements, a small amount of readout error will increase the sample complexity, but does not alter its SQL scaling---we call this property \textit{weak robustness} against readout error (see Fig.~\ref{fig:ramsey}(c) and SM~\cite{supp}).

\textit{Robust multiparameter sensing via scrambling dynamics ---} While the previous protocol used unentangled resources to detect commuting signals, achieving optimal precision for non-commuting signals must involve entanglement~\cite{Proctor2018, Chen2022}. Here, we generate the necessary entanglement using scrambling dynamics implemented by random Clifford unitaries. 
Intuitively, by applying a scrambling unitary before and after signal accumulation, the information carried by each signal is delocalized across the exponentially large Hilbert space. As we describe below, this makes all signals---including non-commuting signals---distinguishable by analyzing measurement outcomes.
This leads to our general protocol, capable of estimating many signals---$K_c$ coherent and $K_{ic}$ incoherent---with a worst-case additive error $\epsilon$ using $M = O( \log(K_{c}K_{ic})/\epsilon^2)$ samples and efficient classical post-processing.
A formal theorem and its proof is presented in the SM~\cite{supp}.
Our approach stands in contrast to shadow-based methods for multiparameter estimation, which rely on randomized measurements rather than fixed scrambling dynamics~\cite{ZhouChen2025,Huang2020}.

To formalize this intuition, we consider signals $\{\theta_\alpha(t), \gamma_\alpha(t)\}_{\alpha, t}$ corresponding to potentially non-commuting Pauli strings $P_\alpha$. Similar to the Ramsey case, the presence of $\theta_\alpha(t)$ 
or $\gamma_\alpha(t)$ 
will introduce linear changes in the output measurement distribution $p(z)$, 
\begin{align}
    \delta p_{\alpha, t}(z) &\equiv \partial_{\theta_\alpha(t)} p(z|\vec{\theta}, \vec{\gamma}) \vert_{\vec{\theta}=0, \vec{\gamma}=0},  \label{eq:dp} \\
    k_{\alpha, t}(z) &\equiv \partial_{\gamma_\alpha(t)} p(z|\vec{\theta}, \vec{\gamma}) \vert_{\vec{\theta} = 0, \vec{\gamma}=0} \label{eq:k},
\end{align}
where $p(z|\vec{\theta}, \vec{\gamma})$ is the distribution of measurement outcomes in the presence of signals $\vec{\theta} = \{\theta_\alpha(t)\}_{\alpha, t}$ and $\vec{\gamma}= \{\gamma_\alpha(t)\}_{\alpha, t}$. Expanding to first order yields
\begin{align}\label{eq:linmodel}
p(z|\vec{\theta}, \vec{\gamma}) 
&\approx p_0(z)
 + \sum_{\alpha,t}\theta_\alpha(t)\,\delta p_{\alpha,t}(z)  \notag\\
&\quad + \sum_{\alpha,t}\gamma_\alpha(t)\,k_{\alpha,t}(z) + O(\gamma_\alpha^2, \theta_\alpha^2).
\end{align} Provided that the scrambling dynamics can be simulated,  the effects of each signal can be computed in advance. Estimators $\hat \theta_\alpha (t)$, $\hat \gamma_\alpha (t)$ are then constructed by fitting $\hat p(z|\vec{\theta}, \vec{\gamma}) = \frac{\hat N_{z}}{M}$ to the linear model Eq.~\eqref{eq:linmodel} via least-squares regression \cite{Hastie2001}. In this procedure, the unknown parameters can be estimated up to the desired precision without having to estimate $\hat{p}(z)$ for all $z$. Notably, our approach  only works if the perturbations $\delta p_\alpha(z)$, $k_\alpha(z)$ are linearly independent across signals: for instance, if two signals produce linearly dependent effects, only their sum can be inferred. Our use of random dynamics ensures that this condition is satisfied with high probability (see SM~\cite{supp}). 

\begin{figure*}
    \centering
    \includegraphics[width=2\columnwidth]{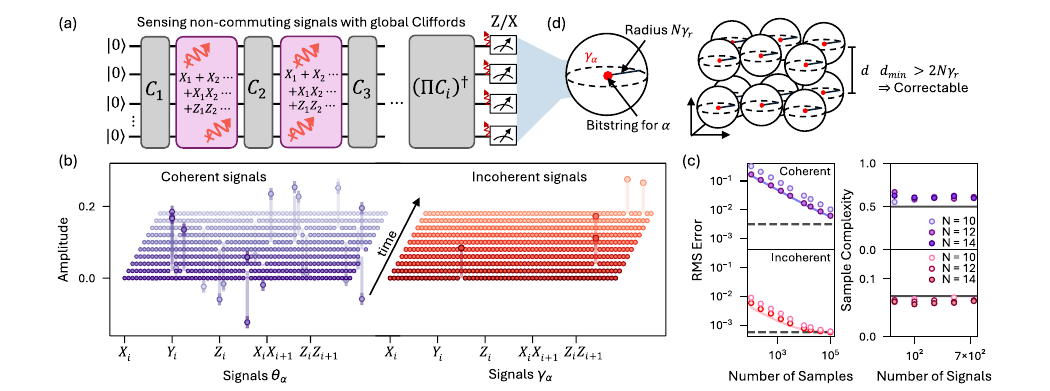}
    \caption{(a) Schematic of the multiparameter sensing protocol based on random Clifford circuits.
    (b) Numerical demonstration of our protocol in estimating a large number of  time-dependent, coherent and incoherent signals ($K_{c} = K_{ic}=580 $ and $T=10$ time steps) using only $N=12$ qubits and $M = 10^4$ measurements. Circuit repetitions $n_c = 10$ and $3$ are used for coherent and incoherent signals, respectively. We choose the signals to be on-site local fields ($X_i$, $Y_i$, and $Z_i$) and nearest-neighbor interaction terms ($X_iX_{i+1}$ and $Z_iZ_{i+1}$). Near-zero signals below a certain threshold are set to zero using standard statistical regularization~\cite{Wainwright2019} (see SM~\cite{supp}), and error bars on such signals are omitted for clarity. Error bars are smaller than the marker size for incoherent signals. 
    (c) Our protocol robustly achieves multiparameter SQL scaling, even in the presence of noise, and has a typical sample complexity that does not depend on $N$ or the number of signals. Left: RMS error versus $M$ for coherent (top) and incoherent (bottom) signals from (b). Light and dark markers indicate the protocol with ($\gamma_r = 0.05$) and without readout error, respectively. 
    Blue and pink solid lines indicate theory curves for the typical sample complexity (see SM~\cite{supp}), and the horizontal dashed line marks the estimator bias. Right: Scaling of typical sample complexity---defined as the slope of RMS error versus $M$---with the number of signals for different system sizes. The typical sample complexity of our protocol is independent of the number of signals; imposing worst-case performance guarantees leads to the $\log(K_cK_{ic})$ dependence discussed in the main text. We consider a single layer $T = 1$, and in order to simulate a large number of signals, we consider one-, two- and three-body signals which need not be geometrically local. Circuit repetitions $n_c$ = 15 and 3 are used for coherent and incoherent signals. Black solid lines indicate the theory approximation for typical signals. (d) Our protocol maps incoherent signals into random bitstrings, which can be treated as the codewords of a classical error correcting code. This enables classical error correction for readout error per qubit up to $d_{min}/2N$, where $d_{min}$ is the closest Hamming distance between two codewords.}
    \label{fig:cliff}
\end{figure*}


While the scrambling dynamics can be generated from generic processes, from random unitary circuits (RUCs) to local Hamiltonian evolution (see SM \cite{supp}), we utilize randomly chosen $N$-qubit  Clifford unitaries to ensure that computation is efficient and thus scalable to large system sizes. Concretely, the system is  initialized in $\ket{0}^{\otimes N}$ and undergoes alternating random Clifford $C_t$ and signal accumulation steps indexed by time $t$. Similar to mirror benchmarking~\cite{Mayer2021, Proctor2022, Colombo2022, Brady2024, Macr2016, Yin2024}, the final Clifford layer applies the inverse of all previous Clifford operators~\footnote{One could similarly adopt a structure where signals are also interspersed between layers of inverse Clifford operators, e.g. $C_1$, $C_2$, $C_2^\dagger$, $C_1^\dagger$; our choice is a matter of convention.}, thus returning the system to $\ket{0}^{\otimes N}$. To estimate incoherent signals, measurements are performed in the $z$-basis, while coherent signals are estimated using measurements in the $x$-basis. The procedure is then carried out for $n_c$ different sets of Clifford circuits, denoted as $\{C^{(n)}_t\}_{t=1}^T$ for $n = 1, \cdots, n_c$. $M/n_c$ measurements are taken for each circuit, for a total of $M$ samples. We now analyze the steps of our protocol for estimating incoherent and coherent signals.

\textit{Incoherent signals---} 
The output of the incoherent procedure ($z$-basis measurements) is deterministic in the absence of signal: $p_{0}(\pmb{0}) = 1$. Moreover, the presence of an incoherent signal $\gamma_\alpha(t)$ produces an output distribution supported on a single bitstring $z'$, $p(z | \gamma_\alpha(t)) \sim \gamma_\alpha(t) \delta_{z, z
'}$, similar to the quadratic Ramsey protocol (cf. Fig. \ref{fig:ramsey}(b)). This arises because, for any set of random $\{C^{(n)}_t\}_{t=1}^T$ and any incoherent Pauli signal $P_\alpha$, the support of $k^{(n)}_{\alpha, t}(z)$ is a single bitstring. This can be seen from the fact that for a Clifford circuit, $k_{\alpha, t}(z) = |\braket{z|P'_\alpha|0}|^2$, where $P'_\alpha$ is the conjugation of $P_\alpha$ by a Clifford operator and therefore remains a Pauli operator~\cite{Nielsen_Chuang_2010}.

Choosing random sets of Clifford circuits $\{C^{(n)}_t\}_{t=1}^T$ ensures that the mapping from signal to bitstring is uniformly random. Thus, the probability that two specific signals get associated with the same bitstring---which would cause our procedure to fail---is low. More precisely, upon repeating the procedure for $n_c$ sets of random Clifford circuits, all $K_{ic}$ incoherent signals are mapped to unique combinations of bitstrings across circuits with probability more than $1-\delta$ as long as $n_c \geq \log_2(K^2_{ic}/2\delta)/N$ (see SM~\cite{supp}).

An estimator $\hat \gamma_\alpha(t)$ is computed from the observed probabilities of the bitstrings corresponding to the support of each $k^{(n)}_{\alpha, t}(z)$. This estimator achieves worst-case error $\max_{\alpha ,t}|\hat \gamma_\alpha(t) -\gamma_\alpha(t)| \leq \epsilon$ with probability at least $1-\delta$ using $M=O(\log(K_{ic}/\delta)/\epsilon^2)$ samples (see SM \cite{supp}). 
Notably, coherent signals have no measureable effect in the $z$ measurement basis to first order. 

\textit{Coherent signals---} 
To detect coherent signals, we require that a non-zero $\theta_\alpha(t)$ modify the output measurement distribution $p(z)$---this occurs with probability $\frac{1}{2}$ over a random choice of Clifford circuits. More specifically, $\delta p_{\alpha, t}(z) = 2\im\big[\bra{z}\mathrm{H}^{\otimes N}P'_\alpha\ket{0} \braket{0|\mathrm{H}^{\otimes N}|z}\big]$, where $P'_\alpha$ is the conjugation of $P_\alpha$ by a random Clifford circuit, and $\mathrm{H}$ is the Hadamard operator. This perturbation $\delta p_{\alpha, t}(z)$ is non-zero precisely when $P'_\alpha$ contains an odd number of Pauli $Y$s, which happens with probability $\frac{1}{2}$ for large $N$. Repeating the procedure over $n_c \geq \log_2(K^2_c/\delta)$ random Clifford circuits, with $\delta$ the failure probability, ensures that all signals are mapped to a non-zero $\delta p^{(n)}_{\alpha, t} (z)$ for at least one circuit (see SM~\cite{supp}). 

When $\delta p^{(n)}_{\alpha, t} (z) \neq 0$, a nonzero $\theta_\alpha(t)$ will perturb $p(z)$ linearly across all bitstrings, analogous to the tilted Ramsey procedure (c.f. Fig. \ref{fig:ramsey}(b)). Amplitudes $\theta_\alpha(t)$ can therefore be estimated by least-squares regression of the empirical distribution onto the precomputed perturbation patterns $\delta p^{(n)}_{\alpha, t} (z)$~\cite{Hastie2001}. Operating under the same principle as cross-entropy benchmarking~\cite{Arute2019,Boixo2018} and its applications to parameter estimation~\cite{Choi2023,manole2025}, only a constant number of bitstrings sampled from the circuit’s output distribution is required to achieve the target estimation error, so sample complexity does not scale with system size. Specifically, the estimator $\hat \theta_\alpha(t)$  has worst-case error $\max_\alpha |\hat \theta_\alpha(t) -\theta_\alpha(t)|\leq \epsilon$ with probability at least $1-\delta$ when $M= O(\log (K_{c}/\delta)/\epsilon^2)$ (see SM~\cite{supp}). A numerical demonstration of sensing coherent and incoherent signals is shown in Fig. \ref{fig:cliff}(b, c).

\textit{Strong robustness to readout error}--- For the protocol that senses incoherent signals, readout errors can be mitigated by classical error-correction. Recall that each incoherent signal is identified with a specific bitstring. As these bitstrings are randomly distributed in $\{0, 1\}^N$, they can be viewed as codewords of a random code~\cite{Shannon1948,Cover2005} in $\{0, 1\}^N$ Hamming space. A readout error rate of $\gamma_r$ per qubit expands each codeword into a ``sphere" of bitstrings with Hamming distance $\gamma_r N$ from the original codeword. As long as the spheres for each signal do not overlap, i.e. the minimum distance between pairs of codewords $d_{min} > 2\gamma_r N$, readout error can be corrected by replacing measured bitstrings with the closest codeword (Fig. \ref{fig:cliff} (d)). Our protocol can achieve $d_{min}/N > \alpha$ for a target relative distance $\alpha$ with high probability $1-\delta$ when the number of qubits is sufficiently large $N=O(\log(K_{ic}^2/\delta)/(1-H(\alpha)))$, implying correction of readout errors up to rate $\gamma_r < \alpha/2$ per sensing qubit (see SM~\cite{supp}). Here, $H(\cdot)$ is the binary entropy. We call the ability to use classical error-correcting techniques \textit{strong robustness}. Unlike the incoherent protocol ($z$-basis measurement), the coherent protocol ($x$-basis measurement) is not strongly robust. However, measurements in the $z$ basis are sensitive to coherent signals at second order (see SM~\cite{supp}). This property can be exploited to first robustly identify nonzero signals from a large candidate set, after which they can be further characterized to determine their sign.

\textit{Discussion---} 
Our robust multiparameter estimation protocol is versatile and feasible to implement in near-term experiments.
Although we focus on random global Clifford unitaries, the protocol generalizes to local Clifford unitaries, random unitary circuits, and Hamiltonian evolution (see SM~\cite{supp}).
Specifically, the procedure based on ergodic Hamiltonian dynamics~\cite{Mark2023, Choi2023, Zhang2023} remains accessible on platforms such as solid-state spin ensembles~\cite{Rovny2025, Zhou2025, Lei2024, Biswas2025, Lee2025} with limited control, albeit with inefficient classical simulation.
Importantly, such protocols are \emph{self-calibrating}: even if the circuits have errors or the precise Hamiltonian parameters are unknown, the sensing procedure remains effective by simply treating control errors as additional signals to be determined.

Our protocol can also naturally be applied to the task of Hamiltonian learning from dynamics~\cite{Yu2023, Huang2023, Bakshi2024, Ma2024, Hu2025, Chen2023, Mirani2024, Li2024}, and we discuss in the SM how it can serve as a subroutine for Heisenberg-limited time-independent Hamiltonian learning, which infers Hamiltonian amplitudes with precision scaling as $1/T$, where $T$ is the total evolution time~\cite{Huang2023, Bakshi2024, Hu2025, Mirani2024, Li2024, Rouz2023, brahmachari2026}. Moreover, our protocol also learns a specific class of time-dependent Hamiltonians in a time-optimal manner (see SM~\cite{supp}); extending these ideas to a general theory of time-dependent Hamiltonian learning remains an interesting open direction.


Finally, our approach can also be used to benchmark quantum devices. 
By treating device parameters as signals, the rates of various imperfections ranging from cross-talk to decay can be estimated simultaneously. 
Indeed, our work Ref.~\cite{manole2025} studies the related task of learning incoherent errors from bitstring measurements of RUCs, deriving provably optimal estimators in this setting.


\textit{Acknowledgments---} We thank Milan Kornjaca, Pedro Lopes, Susanne Yelin, Manuel Endres, Tudor Manole, and Yury Polyanskiy for insightful discussions. WG is supported by the Hertz Foundation Fellowship. We acknowledge support by the NSF QLCI Award OMA-2016245, the NSF QuSeC-TAQS Award 2326787,
the Center for Ultracold Atoms, an NSF Physics Frontiers Center (NSF Grant PHY-2317134), and the
NSF CAREER award 2237244. 

\bibliography{refs}

\end{document}



\title{Supplemental Materials for `Robust multiparameter estimation using quantum scrambling'}
\author{Wenjie Gong}

\author{Bingtian Ye}
\author{Daniel K.~Mark}
\author{Soonwon Choi}%
\affiliation{%
Center for Theoretical Physics---a Leinweber Institute, Massachusetts Institute of Technology, Cambridge, MA 02139, USA
}%

\date{\today}
\maketitle
\tableofcontents
\setcounter{figure}{0}
\renewcommand{\thefigure}{S\arabic{figure}}
\section{Multiparameter sensing of commuting signals}
In the main text, we have discussed two methods for simultaneously sensing multiple commuting signals: the quadratic Ramsey protocol and the tilted Ramsey protocol. 
In this section, we provide additional information about both procedures, including the details of their corresponding estimators, a more rigorous analysis of the sample complexity and the worst-case error, and a discussion of the impact of readout error and bias. 
To be more concrete, we consider signals generated by strings of Pauli $Z$ operators,
\begin{align}\label{eq:defcom}
   Z_a \equiv \otimes_{k=1}^{N} Z^{a_k},
\end{align}
where each 
$a = a_1 \cdots a_N$ is an $N$-bit bitstring used to uniquely label the Pauli string operators. Signals are imprinted on the system as
\begin{align}\label{eq:defcom2}
    \prod_{a} \exp(-i \theta_a Z_a),
\end{align}
and our goal is to determine each $\theta_a$ from local measurements. 

\subsection{Quadratic Ramsey protocol}\label{sec:quadram}

As shown in Fig. 2(a) of the main text, the quadratic Ramsey procedure starts from the $\ket{+}^{\otimes N}$ state on $N$ qubits. After accumulating signal described by Eq. \eqref{eq:defcom2}, a layer of Hadamard gates $\mathrm{H}^{\otimes N}$ is applied, followed by measurement in the computational basis.

This protocol has the following properties, the most important of which are stated in the main text:
\begin{enumerate}
\item The probability of measuring the bitstring $z=a$ is $p(z=a)\approx A\theta^2_a$, where $A = \prod_{a}\cos^2(\theta_{a})$. Thus, the estimator is given by $\hat \theta^2_a = \frac{\hat N_a}{AM}$, with $M$ the number of samples and $\hat N_a$ the number of observations of bitstring $a$. \label{num:quad_prob}
\item The typical sample complexity is given by $\Var(\hat \theta_a) \approx \frac{1}{4AM}$. \label{num:quad_sm}
\item Neglecting bias, a worst-case error $\max_a |\hat \theta_a-|\theta_a|| \leq \epsilon$ can be achieved with probability at least $1-\delta$ by taking $M\geq O(\frac{\log(K/\delta)}{\epsilon^2})$, where $K$ is the number of signals.\label{num:quad_bound}
\item The presence of readout error $\gamma_r$, corresponding to the probability of a local bit-flip error during measurement, initially leads to an error scaling worse than SQL, $\epsilon\sim M^{-1/4}$ for signals $\theta_a$ corresponding to low-weight $a$. The scaling transitions to $\sim M^{-1/2}$ for large $M$ at $M^* \sim O\bigg(\frac{\gamma_r\exp(\gamma_rN)}{\theta^4_a}\bigg)$. \label{num:quad_readout_err}
\end{enumerate}

An explicit derivation of these points remains to be shown. In the sections that follow, we justify each of these characteristics. Moreover, we comment on how the bias of this procedure---parametrically smaller than the statistical error arising from finite sampling in the regimes considered here---scales with $K$ and $N$.

\subsubsection{Estimator}

In this section, we present a constructive derivation of the quadratic sensing estimator (Property \ref{num:quad_prob} under Sec. \ref{sec:quadram}). We begin our analysis of this protocol by computing the effects of signal $\vec{\theta}$ on the output probability distribution $p(z|\vec{\theta})$ perturbatively. Now,
\begin{align}\label{eq:quadpz}
    p(z|\vec{\theta}) = |\braket{z|\mathrm{H}^{\otimes N}\prod_{a} \exp(-i \theta_a Z_a)\mathrm{H}^{\otimes N}|0}|^2
\end{align}
The first order term vanishes,
\begin{align}
    \partial_{\theta_a}  p(z|\vec{\theta})|_{\vec{\theta}=0}= -i \braket{z|\mathrm{H}^{\otimes N}Z_a\mathrm{H}^{\otimes N}|0} \braket{0|z} + h.c. \\
    \equiv  -i \braket{z|X_a|0} \braket{0|z} + h.c.= 0.
\end{align}
Here, we define $X_a$ analogously to $Z_a$ in Eq. \eqref{eq:defcom}, and we use the fact that $ \braket{z|X_a|0} \braket{0|z} $ is real in the second line. Meanwhile, the second order term is
\begin{align}
    \partial_{\theta_a}\partial_{\theta_b} p(z|\vec{\theta})|_{\vec{\theta}=0} = -2\braket{z|X_a X_b|0}\braket{0|z} + 2\braket{z|X_a|0}\braket{0|X_b|z}.
\end{align}
We note that this vanishes when $a\neq b$. As $X_a\ket{0} =\ket{a}$, $\braket{z|X_aX_b|0}\braket{0|z}$ is nonzero only when $z = 0$ and $z = a\oplus b$, which is only simultaneously satisfied when $a = b$. Similarly, $\braket{z|X_a|0}\braket{0|X_b|z}$ is nonzero only when both $z = a$ and $z = b$. 
Thus, the expansion of $p(z|\vec{\theta})$ is

\begin{align}
    p(z|\vec{\theta}) &\approx (1 + \sum_a\theta_a\partial_{\theta_a} + \sum_a \theta^2_a\frac{\partial^2_{\theta_a}}{2})p(z|\vec{\theta})\big|_{\vec{\theta}=0} + O(\theta^3) \\
    &= (1-\sum_a\theta^2_a)|\braket{z|0}|^2 + \sum_a\theta^2_a |\braket{z|X_a|0}|^2+ O(\theta^3) \\
    &= (1-\sum_a\theta^2_a)\delta_{z, 0} + \sum_a\theta^2_a \delta_{z, a} + O(\theta^3).\label{eq:quadpzexp}
\end{align}
From Eq.~\ref{eq:quadpzexp}, we see that for a single nonzero $\theta_a$, one
measures $z=0$ with probability $1-\theta_a^2$ and $z=a$ with probability
$\theta_a^2$ to second order. However, this expansion is taken about
$\vec{\theta}=0$ and therefore does not capture the suppression of probabilities arising from other nonzero amplitudes
$\theta_{b\neq a}$. To understand how background signals affect the leading
$\theta_a$-dependence of $p(z|\vec{\theta})$, we consider a local expansion in
$\theta_a$ about the point
\[
\vec{\theta}^{(a)} \equiv (\theta_a=0,\;\theta_b=\theta_b^{*}\ \forall\, b\neq a),
\]
i.e., we vary $\theta_a$ while holding the remaining coordinates fixed. The
leading contribution takes the form
\begin{align}
\partial^2_{\theta_{a}}p(z|\vec{\theta})\big|_{\vec{\theta}=\vec{\theta}^{(a)}}
&= -2|\braket{z\big|\prod_{b\neq a}e^{-i\theta^*_b X_b}\big|0}|^2
+ 2|\braket{z\big| X_{a}
\prod_{b\neq a}e^{-i\theta^*_b X_b}\big|0}|^2 \\
&\approx -2\prod_{b\neq a}\cos^2(\theta_b^*)\,\delta_{z,0}
+ 2\prod_{b\neq a}\cos^2(\theta_b^*)\,\delta_{z,a},
\end{align}
where we have neglected terms proportional to $\sin(\theta_b^*)$, as they
contribute to the probabilities of measuring bitstrings with support beyond
$\{0,a\}$ (e.g., $z=a\oplus b$). Since we are primarily concerned with how
background signals suppress the probabilities of the no-signal and
single-signal outcomes identified in the expansion about $\vec{\theta}=0$, we
omit these contributions. Consequently, in the perturbative regime we obtain
\begin{align}
p(z|\vec{\theta}) \approx
\Big(1-\sum_a\theta_a^2\Big)\delta_{z,0}
+ \sum_a \Big(1-\sum_{b\neq a}\theta_b^2\Big)\theta_a^2\,\delta_{z,a}
+ O(\theta^3),
\end{align}
where we used $\prod_{b\neq a}\cos^2(\theta_b)=1-\sum_{b\neq a}\theta_b^2+O(\theta^4)$.
Equivalently, before expanding the cosine factors,
\begin{equation}\label{eq:quadA}
p(z=a|\vec{\theta}) \approx \theta_a^2\prod_{b\neq a}\cos^2(\theta_b)
= A\,\frac{\theta_a^2}{\cos^2\theta_a}
\approx A\,\theta_a^2,
\end{equation}
where $A\equiv p(z=0|\vec{\theta})=\prod_b\cos^2\theta_b$, and the final
approximation uses $\cos^2\theta_a\simeq 1$ for small $\theta_a$. We therefore
approximate the output distribution as
\begin{align}\label{eq:probdistquad}
p(z=0)=A,\qquad p(z=a)\approx A\,\theta_a^2,
\end{align}
for $a\neq 0$. Physically, $A$ is the probability of measuring the `no signal' bitstring $z = 0$. In other words, $A$ is the \textit{signal fidelity}, defined as the fidelity of the experiment with respect to the no-signal dynamics. Values $A \approx 1$ correspond to the perturbative regime of weak signals, while $A \ll 1$ indicates strong, non-perturbative signals.

We therefore obtain an estimator
\begin{align}\label{eq:quadest}
\hat A = \frac{\hat N_0}{M}, \;\; \hat \theta^2_a = \frac{\hat N_a}{\hat AM}. 
\end{align}
Here, $\hat N_a$ is the number of counts where bitstring $a$ is measured, and $M$ is the total number of samples. We note that $\hat N_z$ is a random variable drawn from the multinomial distribution $\text{Multinomial}(M, p(z))$, where $p(z)$ is given in Eq. \eqref{eq:probdistquad}.

The estimator $\hat \theta_a^2$ is unbiased up to higher-order corrections in $\theta_a$. Moreover, $ \hat \theta_a \equiv \sqrt{\hat\theta^2_a}$ has bias determined by $\text{Var}(\hat \theta^2_a)$. We can show this by first defining $g(y) \equiv \sqrt{y}$ and $Y = \hat \theta_a^2$, with $\mathbb{E}(Y) = \theta_a^2$. Then, Taylor-expanding $g(Y)$ about $\mathbb{E}(Y) = \theta_a^2$ yields
\begin{align}
    g(Y) - g(\theta_a^2) &\approx  g'(\theta_a^2)\big(Y - \theta_a^2\big) + \frac{1}{2}g''(\theta_a^2)\big(Y -\theta_a^2\big)^2 \\
    \implies \mathbb{E}\sqrt{\hat \theta_a^2} - |\theta_a| &\approx -\frac{\Var(\hat \theta^2_a)}{8|\theta_a|^3}.
\end{align}
We will show in the section below that $\Var(\hat \theta^2_a) \approx \frac{\theta^2_a}{AM}$ (Eq. \eqref{eq:varthetsq}). In practice, we can threshold small signals to zero, a form of statistical regularization known as hard thresholding~\cite{Wainwright2019} (see Sec.~\ref{sec:thres}), setting a lower bound on $\theta_a$. The bias of $\hat \theta_a$ therefore scales as $\sim \frac{1}{M}$, which is smaller than the standard deviation $\sqrt{\text{Var}(\hat \theta_a)} \sim M^{-1/2}$. We thus neglect bias for the rest of our analysis, operating in a regime in which bias is negligible compared to statistical error from finite $M$. Corrections to this assumption can be obtained by including contributions from the bias in the error of the estimator, and a discussion of the scaling behavior of the bias---as well as how the bias may be reduced---is given in Sec.~\ref{sec:biasqram}.
\subsubsection{Typical sample complexity}\label{sec:qsampcomp}
After obtaining the estimator Eq. \eqref{eq:quadest}, we now compute its variance (Property \ref{num:quad_sm} under Sec. \ref{sec:quadram}). The variance of our estimator can be obtained by error propagation, 
\begin{align}
|\hat \theta_a-|\theta_a||^2 =\Var(\hat \theta_a) &= \frac{\Var(\hat \theta^2_a)}{(\partial_{\theta_a} \hat \theta^2_a)^2},
\end{align}
where we note that the quadratic estimator can only estimate the magnitudes $|\theta_\alpha|$. Throughout this work, the variance of our estimators is evaluated over measurement outcomes, which follow a multinomial distribution.

We can compute $\Var(\hat \theta^2_a)$ using the following approximate formula~\cite{SeltmanNotes},
\begin{align}\label{eq:varfrac}
    \Var\bigg(\frac{Y}{X}\bigg) = \frac{\Var(Y)}{\mathbb{E}(X)^2} + \frac{\mathbb{E}(Y)^2 \Var(X)}{\mathbb{E}(X)^4} - 2 \frac{\mathbb{E}(Y) \text{Cov}(X, Y)}{\mathbb{E}(X)^3}.
\end{align}
The approximation follows from a delta-method expansion of $Y/X$ about $(\mathbb{E}(X), \mathbb{E}(Y))$ and is valid provided the denominator remains bounded away from zero; in our perturbative regime, our $X = \hat A =\frac{\hat N_0}{M}$ is strictly positive. Now, as we work in the perturbative regime, we expect $A \approx 1$. Furthermore, we drop terms involving $\mathbb{E}(Y)^2 = A^2 \theta^4_a$ and $\mathbb{E}(Y) \text{Cov}(X, Y) = -A\theta^2_a\frac{A^2 \theta^2_a}{M}$, which are higher-order in $\theta_a$. So,
\begin{align}
    \Var(\hat \theta^2_a) &\approx \frac{1}{A^2}\Var\bigg(\frac{\hat N_a}{M}\bigg) \\ 
    &= \frac{p(a)(1-p(a))}{A^2M}  \\
    &\approx \frac{A \theta^2_a}{A^2M} \label{eq:varthetsq}
\end{align}
where we again drop higher order terms in $\theta_a$ in the last line. So, we finally obtain
\begin{align}
\Var(\hat \theta_a) &= \frac{\Var(\hat \theta^2_a)}{(\partial_{\theta_a} \hat \theta^2_a)^2} \\
&\approx \frac{A \theta^2_a}{A^2M} \times \frac{1}{4 \theta^2_a}  \\
&= \frac{1}{4A M} \label{eq:quadestsm}.
\end{align}
Thus, we expect $E_a|\hat \theta_a-|\theta_a| |= \frac{1}{2\sqrt{AM}}$.

\subsubsection{Bounds on sample complexity for worst-case error}\label{sec:quadbounds}
We can also obtain a rigorous bound on the worst-case error neglecting bias, $\text{max}_a |\hat \theta_a-|\theta_a||^2$, leading to the result in Property \ref{num:quad_bound}. The proof proceeds in two steps. First, Lemma~\ref{lem:tyscalquad} establishes the sample-complexity scaling for estimating a single signal $\theta_a$ to accuracy $\epsilon$, based on tail bounds of our estimator. Then, using a union bound over all signals, Theorem~\ref{thm:boundquad} extends this result to the worst-case error across all signals.

\begin{lemma}\label{lem:tyscalquad}
    In the perturbative regime where 
    $\frac{p(a)}{p(0)} <1$ and neglecting bias, estimating a single nonzero $\theta_a$ with precision $|\hat \theta_a -|\theta_a||\leq \epsilon$ and probability $1-\delta$ requires samples $M= O(\frac{1}{\epsilon^2}\log\frac{1}{\delta})$.
\end{lemma}

\textit{Proof. } 
We first focus on bounding $|\hat \theta^2_a - \theta^2_a|$ = $|\frac{\hat N_a}{\hat N_0} - \frac{p(0)}{p(a)}|$. By Hoeffding's inequality,
\begin{align}
    \text{Pr}\bigg(\bigg|\frac{\hat N_a}{M} -p(a)\bigg| \geq \epsilon\bigg) \leq 2\exp(-2 M \epsilon^2 ) \label{eq:hofquad1}\\
    \text{Pr}\bigg(\bigg|\frac{\hat N_0}{M} -p(0)\bigg| \geq \epsilon'\bigg) \leq 2\exp(-2 M \epsilon'^2 ). \label{eq:hofquad2}
\end{align}
Now, we can write
\begin{align}
    \frac{\hat N_a}{\hat N_0} = \frac{p(a) + (\frac{\hat N_a}{M}-p(a))}{p(0) + (\frac{\hat N_0}{M}-p(0))} = \frac{p(a) + \Delta n_a}{p(0) + \Delta n_0},
\end{align}
setting $\Delta n_a \equiv \frac{\hat N_a}{M}-p(a)$ and $\Delta n_0 \equiv \frac{\hat N_0}{M}-p(0)$. 

We work in a regime where $\frac{\Delta n_0}{p(0)} \leq r$, $\frac{\Delta n_a}{p(0)} \leq r$, where $r < 1$ is a fixed parameter that controls the relative difference of our measurements from their expected values. By Eqs. \eqref{eq:hofquad1} and \eqref{eq:hofquad2}, this occurs with probability $1-\delta_1$ given $M\geq  \frac{1}{2p(0)^2r^2}\log{\frac{4}{\delta_1}}$. 
Taylor expanding around small $\Delta n_0$, we have
\begin{align}
\frac{1}{p(0) +\Delta n_0} =  \frac{1}{p(0)} \bigg(1 - \frac{\Delta n_0}{p(0)} + c\bigg),
\end{align}
such that
\begin{align}
    \frac{\hat N_a}{\hat N_0} &= \frac{p(a) + \Delta n_a}{p(0)}\bigg(1 - \frac{\Delta n_0}{p(0)} +  c\bigg) \\ &= \frac{p(a) + \Delta n_a}{p(0)} - \frac{p(a)}{p(0)^2}\Delta n_0 - \frac{\Delta n_a \Delta n_0}{p(0)^2} + c\bigg(\frac{p(a) + \Delta n_a}{p(0)}\bigg),
\end{align}
and $c$ is the remainder term of the expansion, which we will bound below.

So, we have
\begin{align}
    \bigg|\frac{\hat N_a}{\hat N_0} - \frac{p(a)}{p(0)}\bigg| \leq \bigg| \frac{\Delta n_a}{p(0)}\bigg| + \frac{p(a)}{p(0)^2}|\Delta n_0| + \frac{|\Delta n_a| |\Delta n_0|}{p(0)^2} + |c|\frac{p(a) + |\Delta n_a|}{p(0)}
\end{align}
where we have used the Triangle inequality. Now, we note that
\begin{align}
    |c| &= \bigg|\bigg(\frac{\Delta n_0}{p(0)}\bigg)^2 -\bigg(\frac{\Delta n_0}{p(0)}\bigg)^3 \cdots \bigg| \leq\sum_{\alpha = 2} \bigg|\frac{\Delta n_0}{p(0)}\bigg|^\alpha \\
    &= \bigg(\frac{\Delta n_0}{p(0)}\bigg)^2\frac{1}{1-|\frac{\Delta n_0}{p(0)}|} \leq \bigg(\frac{\Delta n_0}{p(0)}\bigg)^2 \frac{1}{1-r} \equiv C\bigg(\frac{\Delta n_0}{p(0)}\bigg)^2.
\end{align}
Moreover, $\frac{\Delta n_0}{p(0)} \leq 1$, so $C\big(\frac{\Delta n_0}{p(0)}\big)^2 \leq C\frac{\Delta n_0}{p(0)}$ and $\frac{|\Delta n_a| |\Delta n_0|}{p(0)^2} \leq \frac{|\Delta n_a|}{p(0)}$.
Thus,
\begin{align}
    \bigg|\frac{\hat N_a}{\hat N_0} - \frac{p(a)}{p(0)}\bigg| \leq (C+2)\bigg| \frac{\Delta n_a}{p(0)}\bigg| + (C+1)\frac{p(a)}{p(0)^2}|\Delta n_0|.
\end{align}
We can now use a union bound,
\begin{align}
\text{Pr}\bigg(\bigg|\frac{\hat N_a}{\hat N_0} - \frac{p(a)}{p(0)}\bigg| \geq \epsilon\bigg) &\leq (C+2)\text{Pr}\bigg(\bigg| \frac{\Delta n_a}{p(0)}\bigg| \geq \frac{\epsilon}{2C+3}\bigg) + (C+1)\text{Pr}\bigg(\frac{p(a)}{p(0)^2}|\Delta n_0| \geq \frac{\epsilon}{2C+3}\bigg) \\
&\leq 2(C+2) \exp\bigg(-\frac{2}{(2C+3)^2}Mp(0)^2\epsilon^2\bigg) + 2(C+1) \exp\bigg(-\frac{2}{(2C+3)^2}M\frac{ p(0)^4}{p(a)^2}\epsilon^2\bigg) \\
&\leq (6C + 4) \exp\bigg(-\frac{2}{(2C+3)^2}Mp(0)^2\epsilon^2\bigg).
\end{align}
In the second line, we use Eqs. \eqref{eq:hofquad1} and \eqref{eq:hofquad2}. In the last line, we use the fact that $\frac{p(0)}{p(a)} \geq 1$, so  $\exp(-\frac{2}{(2C+3)^2}M\frac{ p(0)^4}{p(a)^2}\epsilon^2)  \leq \exp(-\frac{2}{(2C+3)^2}Mp(0)^2\epsilon^2)$.

Thus, 
\begin{align}\label{eq:quadbound}
   \text{Pr}( |\hat \theta^2_a - \theta^2_a| \geq \epsilon) = \text{Pr}\bigg(\bigg|\frac{\hat N_a}{\hat N_0} - \frac{p(a)}{p(0)}\bigg| \geq \epsilon\bigg) \leq (6C + 4) \exp\bigg(-\frac{2}{(2C+3)^2}Mp(0)^2\epsilon^2\bigg).
\end{align}

From here, we work in a regime where  $\bigg|\frac{\hat N_a}{\hat N_0} - \frac{p(a)}{p(0)}\bigg| \leq r'$, for some $r' < 1$. By Eq. \eqref{eq:quadbound}, this occurs with probability $1 - \delta_2$ for $M \geq \frac{(2C+3)^2}{2 A^2 r'^2}\log\frac{6C+4}{\delta_2}$.
Then, we can also Taylor expand $\sqrt{\frac{\hat N_a}{\hat N_0}}$ about $\sqrt{\frac{p(a)}{p(0)}}$,
\begin{align}
    \sqrt{\frac{\hat N_a}{\hat N_0}} = \sqrt{\frac{p(a)}{p(0)} }\bigg(1 + \frac{1}{2}\frac{\frac{\hat N_a}{\hat N_0} - \frac{p(a)}{p(0)}}{\frac{p(a)}{p(0)}} + d\bigg).
\end{align}
By Taylor's Remainder Theorem,
\begin{align}
    d = -\frac{1}{8 c^{3/2}}\bigg(\frac{\hat N_a}{\hat N_0} - \frac{p(a)}{p(0)}\bigg)^2 ,
\end{align}
where $c$ is some constant between $\frac{\hat N_a}{\hat N_0}$ and $\frac{p(a)}{p(0)}$. We have
\begin{align}
    |d| \leq \frac{1}{8 (\frac{p(a)}{p(0)}-r')^{3/2}}\bigg(\frac{\hat N_a}{\hat N_0} - \frac{p(a)}{p(0)}\bigg)^2\equiv C'\bigg(\frac{\hat N_a}{\hat N_0} - \frac{p(a)}{p(0)}\bigg)^2.
\end{align}
So, 
\begin{align}
    \bigg|\sqrt{\frac{\hat N_a}{\hat N_0}} -\sqrt{\frac{p(a)}{p(0)} }\bigg| \leq \bigg(\frac{1}{2}\sqrt{\frac{p(0)}{p(a)}} + \sqrt{\frac{p(a)}{p(0)}}C'\bigg)\bigg|\frac{\hat N_a}{\hat N_0} - \frac{p(a)}{p(0)}\bigg|, 
\end{align}
where we have used $|\frac{\hat N_a}{\hat N_0} - \frac{p(a)}{p(0)}|<1$. 
Thus,
\begin{align}
\text{Pr}\bigg( \bigg|\sqrt{\frac{\hat N_a}{\hat N_0}} -\sqrt{\frac{p(a)}{p(0)} }\bigg|\geq \epsilon\bigg) &\leq \text{Pr}\bigg(\bigg(\frac{1}{2}\sqrt{\frac{p(0)}{p(a)}} + \sqrt{\frac{p(a)}{p(0)}}C'\bigg)\bigg|\frac{\hat N_a}{\hat N_0} - \frac{p(a)}{p(0)}\bigg| \geq \epsilon\bigg) \\
&\leq (6C + 4)\exp\bigg(-\frac{2A^2}{(2C+3)^2(\frac{1}{2\theta_a}+\theta_aC')^2}M\epsilon^2\bigg) \label{eq:quadsqrtbound},
\end{align}
where we have used Eq. \eqref{eq:quadbound} and $p(a) = A\theta_a^2$, $p(0) = A$. 

By Eq. \eqref{eq:quadsqrtbound}, an error $\epsilon$ can be achieved by $M\geq\frac{(2C+3)^2(\frac{1}{2\theta_a}+\theta_aC')^2}{2 A^2 \epsilon^2}\log\frac{6C+4}{\delta_3} $ for a failure probability $\delta_3$.

Thus, putting Eqs. \eqref{eq:hofquad1}, \eqref{eq:hofquad2}, \eqref{eq:quadbound}, and \eqref{eq:quadsqrtbound} together, we can achieve
$\bigg|\sqrt{\frac{\hat N_a}{\hat N_0}} -\sqrt{\frac{p(a)}{p(0)} }\bigg|\geq \epsilon$ with probability $1 - \delta_1 -\delta_2 -\delta_3$ by taking $M\geq \text{max}\bigg(\frac{1}{2p(0)^2r^2}\log{\frac{4}{\delta_1}}, \frac{(2C+3)^2}{2 A^2 r'^2}\log\frac{6C+4}{\delta_2}, \frac{(2C+3)^2(\frac{1}{2\theta_a}+\theta_aC')^2}{2 A^2 \epsilon^2}\log\frac{6C+4}{\delta_3}\bigg)$. We can set $\delta_1 = \delta_2 = \delta_3 = \frac{\delta}{3}$, such that $\bigg|\sqrt{\frac{\hat N_a}{\hat N_0}} -\sqrt{\frac{p(a)}{p(0)} }\bigg|\geq \epsilon$ with probability $1-\delta$ by $M$ scaling as $ O(\frac{1}{\epsilon^2}\log (\frac{\mathcal{C}}{\delta}))$ for some constant $\mathcal{C}$. We note that this scaling may depend on the strength of the signal as $\frac{1}{\theta_a}$, which could be large for very small signals. In practice, we can treat all signals below a certain threshold $\theta_{min}$ as effectively zero (see Sec.~\ref{sec:thres}), thus setting a lower bound on $\theta_a$. $\qed$

\begin{theorem}\label{thm:boundquad}
 Assume the perturbative regime where $A = p(0)$ is close to 1, such that $\frac{p(a)}{p(0)} <1$, and neglect bias. In the presence of $K$ total signals $\theta_a$, a precision $\max_a|\hat \theta_a -|\theta_a||\leq \epsilon$ can be achieved with probability $1-\delta$ by taking $M= O(\frac{1}{\epsilon^2}\log\frac{K}{\delta})$.
\end{theorem}
\textit{Proof.} We can take $\delta \rightarrow \frac{\delta}{K}$ in Lemma \ref{lem:tyscalquad}. Applying the union bound gives the statement of the Theorem. $\qed$

\subsubsection{Impact of readout error}\label{sec:qreadout}
Here, we explore the effect of readout error on the quadratic sensing procedure. We model classical readout error $\gamma_r$ as an independent bit-flip error on each qubit with probability $\gamma_r$. Ultimately, we find that the sensitivity to signals $\theta_a$ with low-weight $a$ is affected the most, with the scaling becoming worse than SQL $\sim M^{-1/4}$ in regimes of low $M$ (Property \ref{num:quad_readout_err}).

Now, if we run the quadratic sensing procedure as-is in the presence of readout error, our estimator Eq. \eqref{eq:quadest} will be biased. Moreover, the factor $A$ will also be rescaled as  $A \rightarrow A(1-\gamma_r)^N$. To correct for this, we first construct the confusion matrix~\cite{Hashim2025}
\begin{align}
C = \bigotimes_{i = 1}^{N} \begin{pmatrix}
    1 - \gamma_r & \gamma_r\\
    \gamma_r & 1-\gamma_r
\end{pmatrix}.
\end{align}
We can then use modified estimators
\begin{align}\label{eq:quadesterr}
    \hat A = \sum_j C^{-1}_{0j}\frac{\tilde{N}_j}{M}, \;\; \hat \theta^2_a = \sum_jC^{-1}_{aj}\frac{\tilde{N}_j}{\hat AM}, \;\;\hat \theta_a =\sqrt{\hat \theta^2_a},
\end{align}
where $\tilde{N_i}$ indicates the number of counts for bitstring $i$ \textit{after} readout error. 
We now analyze the performance of our estimators in Eq. \eqref{eq:quadesterr}. To do so, we first note that in the presence of classical readout noise, 
\begin{align}\label{eq:varperr}
    \Var\bigg(\frac{\tilde{N}_j}{M}\bigg) = \sigma^2_{j, \text{ quantum}} + \sigma^2_{j, \text{ classical}},
\end{align}
where $\sigma^2_{j, \text{ quantum}} = \frac{p(j)(1-p(j))}{M}$ is our familiar quantum projection noise. $\sigma^2_{j, \text{ classical}}$ is the classical noise from readout error \cite{Degen2017}, which can be intuitively written as
\begin{align}
\sigma^2_{j, \text{ classical}} &= \frac{p(j)}{M}\sum_{i\neq j}[\text{variance from confusing $j$ with $i\neq j$}] + \sum_{i\neq j}\frac{p(i)}{M}[\text{variance from confusing $i\neq j$ with $j$}].
\end{align}
For example, if bitstring $j$ is confused with another bitstring $i\neq j$ with probability $\alpha$, then, conditional on a measurement being reported as $j$, the variance this confusion contributes is $\alpha(1-\alpha)$.
We note that here, $p(j)$ are the clean probabilities, not including the effect of readout error. 
For simplicity of analysis, we assume that all signals are small with equal magnitude $p(z\neq0) = p_k$.
Then,
\begin{align}
   \sigma^2_{j, \text{ classical}} &= \frac{p_k}{M}\sum_{m=1}
   ^N \begin{pmatrix}
       N \\m
   \end{pmatrix}\alpha_m(1-\alpha_m) + \frac{p_k}{M}\sum_{m=1} \begin{pmatrix}
       N \\m
   \end{pmatrix}\alpha_m(1-\alpha_m) + \frac{A-p_k}{M}\alpha_s(1-\alpha_s).
\end{align}
Here, $\alpha_m$ is the confusion probability between two bitstrings with Hamming distance $m$,
\begin{align}
    \alpha_m = \gamma_r^m(1-\gamma_r)^{N-m},
\end{align}
and $s$ is the Hamming distance between $j$ and $0$, $s = \text{wt}(j)$. We intentionally separate out the term proportional to $A = p(0)$ from the second sum. Now, the sum $\sum_{m = 1}^N \begin{pmatrix}
       N \\m
   \end{pmatrix}\alpha_m(1-\alpha_m)$ can be directly computed. First, we note
\begin{align}
    \sum_{m=0}
   ^N \begin{pmatrix}
       N \\m
   \end{pmatrix}\alpha_m(1-\alpha_m)  =  \sum_{m=0}
   ^N \begin{pmatrix}
       N \\m
   \end{pmatrix}\alpha_m -\sum_{m=0}
   ^N \begin{pmatrix}
       N \\m
   \end{pmatrix}\alpha^2_m = 1  - \sum_{m=0}
   ^N \begin{pmatrix}
       N \\m
   \end{pmatrix}\alpha^2_m,
\end{align}
where we have used the binomial theorem to simplify 
\begin{align}
    \sum_{m=0}
   ^N \begin{pmatrix}
       N \\m
   \end{pmatrix}\gamma_r^m(1-\gamma_r)^{N-m} = (1-\gamma_r +\gamma_r)^N = 1.
\end{align}
Now, for the second term,
\begin{align}
    \sum_{m=0}
   ^N \begin{pmatrix}
       N \\m
   \end{pmatrix}\alpha^2_m &= \sum_{m=0}
   ^N \begin{pmatrix}
       N \\m
   \end{pmatrix}\gamma_r^{2m}(1-\gamma_r)^{2N-2m} \\
   &= (1-\gamma_r)^{2N}\sum_{m=0}
   ^N \begin{pmatrix}
       N \\m
   \end{pmatrix}\bigg(\frac{\gamma_r}{1-\gamma_r}\bigg)^{2m}1^{N-m}\\
   &= (1-\gamma_r)^{2N} \bigg(1 
   + \bigg(\frac{\gamma_r}{1-\gamma_r}\bigg)^2\bigg)^N\\
   &= (1-2\gamma_r + 2\gamma^2_r)^N,
\end{align}
where we have used the Binomial theorem again in the second-to-last line. Subtracting off the $m=0$ terms, 
\begin{align}
    \sum_{m = 1}^N \begin{pmatrix}
       N \\m
   \end{pmatrix}\alpha_m(1-\alpha_m) &= 1 -  (1-2\gamma_r + 2\gamma^2_r)^N - (1-\gamma_r)^N + (1-\gamma_r)^{2N} \\
    &= \gamma N  -  \frac{\gamma^2}{2}N(N+1) + O(\gamma^3).
\end{align}
Thus, from Eq. \eqref{eq:varperr}, 
\begin{align} \label{eq:varnerr}
\Var\bigg(\frac{\tilde{N}_{j\neq 0}}{M}\bigg)   &\approx \frac{p_k(1-p_k)}{M} +\frac{2p_k}{M}(\gamma N  -  \frac{\gamma^2}{2}N(N+1) ) + \frac{A-p_k}{M} \alpha_s(1-\alpha_s) \\
\Var\bigg(\frac{\tilde{N}_{0}}{M}\bigg)   &\approx \frac{A(1-A)}{M} +\frac{p_k+A}{M}(\gamma N  -  \frac{\gamma^2}{2}N(N+1) )
\end{align}
Now, the second term in $\Var\big(\frac{\tilde{N}_{j\neq 0}}{M}\big)$ is $O(p_k\gamma) = O(A\theta^2 \gamma)$, while the third term is $O(A \gamma^s)$. Thus, recalling that $A$ is $O(1)$ in the perturbative regime, the dominant contribution to the variance will come from the third term. When $s = \text{wt}(j)$ is large, $\Var(\frac{\tilde{N}_j}{M})$ will be minimally affected. On the other hand, when $s$ is small, as for single-body and two-body signals, $\Var(\frac{\tilde{N}_j}{M})$ will be greatly impacted. This is illustrated in Fig. 2(b) of the main text.

Now, we can analyze the variance of the estimator Var$(\hat \theta^2_a)$ in the presence of readout error, and hence its sample complexity:
\begin{align}
    \Var(\hat \theta^2_a) = \Var\bigg( \sum_j C^{-1}_{aj}\frac{\tilde{N}_j}{\hat AM}\bigg) \approx \sum_j (C^{-1}_{aj})^2\Var\bigg(\frac{\tilde{N}_j}{AM}\bigg),
\end{align}
where we have neglected covariances, which will be higher-order in $p_k$ and $\gamma_r$. By similar logic as in Section \ref{sec:qsampcomp}, we also make the approximation of subsituting $\hat A$ with its expectation in the denominator. Finally, for simplicity of analysis, we also approximate all $\Var\big(\frac{\tilde{N}_{j\neq 0}}{AM}\big)$ to be of the same size. By Eq. \eqref{eq:varnerr}, this is valid if all signals are small.
\begin{align}
    \Var(\hat \theta^2_a) &\approx  \Var\bigg(\frac{\tilde{N}_a}{AM}\bigg)\sum_{j=0}^{2^N} (C^{-1}_{aj})^2 + \bigg(\Var\bigg(\frac{\tilde{N}_0}{AM}\bigg)-\Var\bigg(\frac{\tilde{N}_a}{AM}\bigg)\bigg)(C^{-1}_{a0})^2 \\
    &= \Var\bigg(\frac{\tilde{N}_a}{AM}\bigg) \underbrace{\frac{1}{(2\gamma_r-1)^{2N}}\sum_{m=0}^N\begin{pmatrix}
        N \\m
    \end{pmatrix} \gamma_r^{2m}(\gamma_r-1)^{2N-2m}}_{=\frac{(1-2\gamma_r + 2\gamma^2_r)^N}{(2\gamma_r-1)^{2N}}} + \bigg(\Var\bigg(\frac{\tilde{N}_0}{AM}\bigg)-\Var\bigg(\frac{\tilde{N}_a}{AM}\bigg)\bigg) \gamma_r^{2s}(\gamma_r-1)^{2N-2s} \\
    &\approx \frac{1}{(1-2\gamma_r)^N}\Var\bigg(\frac{\tilde{N}_a}{AM}\bigg) \label{eq:quadinvexp}.
\end{align}
In the last line, we have approximated $(1-2\gamma_r + 2\gamma^2_r)^N \approx (1-2\gamma_r)^N$ and dropped the term proportional to $\gamma_r^{2s}$, where as before, $s = \text{wt}(a)$. Thus, we see that our corrected estimators diverge exponentially in $N$ in the presence of readout error $\gamma_r$. Moreover, as discussed previously, signals $a$ with low weight will be affected the most by readout error according to Eq. \eqref{eq:varnerr}.

We can also look at the sample complexity scaling of our corrected estimators. 
\begin{align}\label{eq:sampcerr}
    \Var(\hat \theta_a) = \frac{\Var(\hat \theta^2_a)}{(\partial_{\theta_a} \hat \theta^2_a)^2} &\approx \frac{1}{(1-2\gamma_r)^N}\bigg(\frac{p(a)}{4 A^2M \theta^2_a} + \frac{A}{4A^2 M\theta^2_a}\gamma^s_r \bigg) \\
    &= \frac{1}{(1-2\gamma_r)^N}\bigg(\frac{1}{4 A M } + \frac{1}{4 A M\theta^2_a}\gamma^s_r \bigg) 
\end{align}
When $\theta_a$ is indistinguishable from $0$, the denominator of the formula above diverges and is inapplicable. Instead, we can approximate $\Var(\hat \theta_a) = \sqrt{\Var(\hat \theta^2_a)}$ \cite{Degen2017}, so 
\begin{align}
\Var(\hat \theta_a)  \approx \sqrt{\frac{1}{(1-2\gamma_r)^N}\bigg(\frac{\theta^2_a}{ AM} + \frac{\gamma^s_r}{A M} \bigg) }.
\end{align}
This is worse than SQL scaling. However, for finite $\theta_a$, SQL scaling as given by Eq. \eqref{eq:sampcerr} is restored. The transition from worse than SQL to SQL is given by the number of samples $M^*$ at which the statistical error is smaller than $\theta_a$---when $\theta_a$ becomes effectively nonzero. For single-body signals where $s =1$, this occurs at
\begin{align}
    &\sqrt{\Var(\hat \theta^2_a)} \sim \theta^2_a \\
    \implies &\frac{1}{(1-2\gamma_r)^N}\frac{\gamma_r}{A M^*} \sim \theta^4_a \\
    \implies &M^* \sim O\bigg(\frac{\gamma_r\exp(\gamma_rN)}{\theta^4_a}\bigg),
\end{align}
where in the second line, we have assumed that $\gamma_r$ is around the same magnitude as $\theta_a$ so $\gamma_r \gg \theta^2_a$. This is indeed the scaling of the transition shown in Fig. 2(c) of the main text.

\subsubsection{Bias scaling}\label{sec:biasqram}
Here, we discuss how the bias of the quadratic Ramsey estimator (Eq.~\ref{eq:quadest}) scales with the system size $N$, the total number of signals $K$, and the signal fidelity $A$. We show that the bias is suppressed with increasing $N$, grows slowly with decreasing $A$, and never exceeds the scaling of the statistical error with $K$. As a result, there exists a broad range of $M$ in which the statistical error dominates over the bias.

We define the squared bias as
\begin{align}
    (\hat \theta^*_a - \theta_a)^2, 
\end{align}
where $\theta_a$ is the true value of the signal, and $\hat \theta^*_a$ is the estimator Eq.~\ref{eq:quadest} evaluated using the \textit{true} probability distribution $p(z|\vec{\theta})$ instead of the empirical distribution $\hat p(z|\vec{\theta}) = \hat N_z/M$. Since $p(z|\vec{\theta})$ corresponds to the infinite-sample limit, $\hat \theta^*_a$ is free of statistical fluctuations due to finite sampling. Because the bias can vary in both magnitude and sign across different signals, we focus on the squared bias. In particular, we consider both the mean squared bias across all signals,
\begin{align}\label{eq:meansqb}
    \mathbb{E}_a[(\hat \theta^*_a - \theta_a)^2],
\end{align} 
as well as the worst-case squared bias,
\begin{align}\label{eq:wcasesqb}
    \max_a[(\hat \theta^*_a - \theta_a)^2].
\end{align}

\begin{figure}
    \centering
    \includegraphics[width=1\columnwidth]{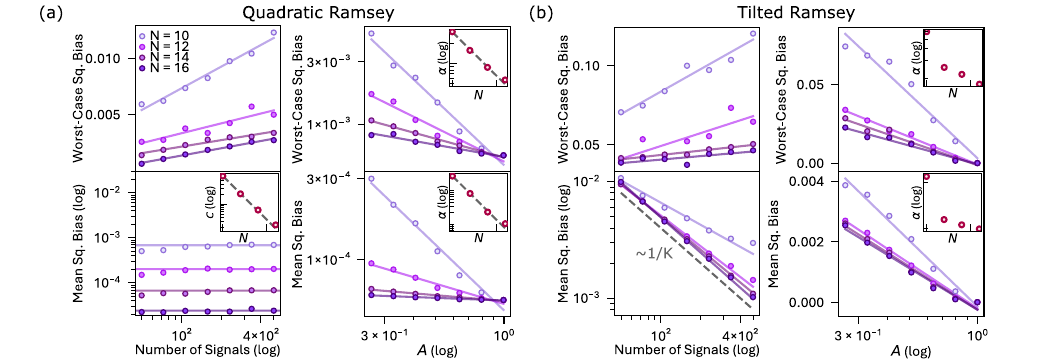}
    \caption{\textbf{(a)} Scaling of the worst-case squared bias (Eq.~\ref{eq:wcasesqb}, top panel) and mean squared bias (Eq.~\ref{eq:meansqb}, bottom panel) in the quadratic Ramsey protocol as function of the total number of signals $K$ and the signal fidelity $A$. Signals are drawn randomly from single, two, three, and four-body Pauli $Z$ strings of the form Eq.~\ref{eq:defcom}, with $A = 0.91$.  Each data point is obtained by averaging over 30 random signal instances. (Left) Scaling of bias with respect to $K$. The worst-case squared bias scales approximately as $\log(K)$; solid lines show best fits to $\alpha \log(K) +\beta$. The mean squared bias saturates as $K$ increases; solid lines indicate the corresponding saturation values $c$. The inset demonstrates an exponential decrease of $c$ with increasing $N$, with the dotted line indicating the best exponential fit. (Right) Scaling of bias with respect to $A$, for $K = 100$. The worst-case squared bias decreases with $\log(A)$ as $A$ approaches 1; solid lines indicate best fits to $-\alpha \log(A) +\beta$. As shown in the inset, $\alpha$ decreases exponentially with increasing $N$, with the best exponential fit marked by the dotted line. The mean squared bias follows similar scaling behavior. \textbf{(b)} Scaling of bias in the tilted Ramsey protocol as a function of $K$ and $A$, for the same setting as in (a). (Left) The worst-case squared bias increases approximately as $\log(K)$ with the number of signals $K$, with solid lines showing best fits to $\alpha \log(K) +\beta$. The mean-squared bias decreases as $K^{-\gamma}$ with increasing $K$, with $\gamma$ approaching $1$ at large $N$. Solid lines indicate best fits to $aK^{-\gamma}$, while the dotted line demonstrates $K^{-1}$ scaling. (Right) The worst-case squared bias decreases with $\log(A)$ as $A$ approaches $1$. Solid lines show best fits to to $-\alpha \log(A) +\beta$. As demonstrated in the inset, $\alpha$ decreases with increasing $N$. The mean squared bias exhibits similar behavior.}
    \label{fig:biasram}
\end{figure}

We investigate the bias of the quadratic Ramsey procedure numerically, with results shown in Fig.~\ref{fig:biasram}(a). The mean squared bias saturates to a constant $c$ with respect to $K$, with $c\sim e^{-N}$ decreasing exponentially with increasing $N$ over the range shown. This exponential suppression can be understood by noting that in the quadratic Ramsey protocol, bias arises primarily from higher-order effects that perturb the measurement probabilities of bitstrings associated with the desired signals. As the Hilbert space dimension grows exponentially with $N$, these higher-order perturbations are spread over an exponentially larger space of bitstrings,  thereby suppressing the probability of overlap with signal bitstrings and reducing the bias. Meanwhile, the worst-case squared bias scales as $\log(K)$ with the total number of signals $K$, in agreement with the scaling predicted by a union bound argument (see Sec.~\ref{sec:quadbounds}). Indeed, this follows the same scaling behavior as the worst-case statistical error, as proven in Sec.~\ref{sec:quadbounds}. 

In the perturbative regime, the worst-case squared bias decreases logarithmically with $A$ (Eq.~\ref{eq:quadA}) as $A\rightarrow 1$, scaling as $\sim -\alpha \log A$. Moreover, the prefactor $\alpha$ also decreases exponentially $\sim e^{-N}$ with $N$. The mean squared bias exhibits a similar logarithmic dependence on $A$. Such dependence indicates slow growth of the bias as $A$ decreases. 

Overall, the worst-case bias grows asymptotically as $\log K$, no faster than the statistical error. Since the worst-case statistical error scales as $\epsilon^2 \sim \log(K)/M$, while the worst-case bias scales as $\epsilon_{\text{bias}}^2\sim \log(K)$, statistical error dominates for sample sizes $M$ below a scale set by $1/{\epsilon^2_{\text{bias}}}$, which defines the regime relevant for our analyses. As shown in Fig.~\ref{fig:biasram}(a), when averaged over randomly drawn signal instances, both the mean and worst-case bias are further suppressed exponentially with increasing $N$. Bias may therefore be reduced both by increasing $N$ and through post-processing, as described in Sec.~\ref{sec:highorder}.


\subsection{Tilted Ramsey}\label{sec:tiltedram}
As shown in Fig. 2(a) of the main text, the tilted Ramsey procedure also starts from the $\ket{+}^{\otimes N}$ state. After accumulating signal described by Eq. \eqref{eq:defcom2}, a layer of $X$ rotations $X(\phi)^{\otimes N}$ is applied, followed by measurement in the computational basis. 

This protocol has the following properties, which are highlighted in the main text:
\begin{enumerate}
\item The presence of a signal $\theta_a$ changes the probability $p(z)$ of measuring bitstring $z$ in a manner linear in $\theta_a$. As the response of $p(z)$ is linear in the signals, approximate linear inversion can be can be used to construct estimators $\hat \theta_a$ from the empirical probabilities $\frac{\hat N_z}{M}$, which are unbiased up to higher-order corrections, in a computationally efficient manner. \label{num:tram_prob}
\item The typical sample complexity of this procedure is given by $\Var(\hat \theta_a) \approx  \frac{1}{4\sin^2(s_a\phi)M}$. \label{num:tram_sm}
\item Neglecting bias, a worst-case error $\max_a |\hat \theta_a -\theta_a|\leq \epsilon$ can be attained with probability $1-\delta$ when $ M \geq O(\frac{\log(K/\delta)}{\epsilon^2})$ when all signals $a$ are of bounded weight. When considering all signals $a$ with weight up to $\text{wt}(a) = N$, $ M \geq O(\mathrm{poly}(N)\frac{\log(K/\delta)}{\epsilon^2})$.
\label{num:tram_bound}
 
\item The presence of nonzero readout error will not change the SQL scaling of error with $M$.\label{num:tram_readout}
\item Though tilted Ramsey can sense commuting signals with unentangled resources, the protocol is not effective for non-commuting or time-dependent signals, and exhibits poor error scaling with the weight of the Pauli $Z_a$ terms to be sensed.  \label{num:tram_neg}
\end{enumerate}
In this section, we explicitly derive points \ref{num:tram_prob}, \ref{num:tram_sm}, and \ref{num:tram_bound}. We additionally elaborate on \ref{num:tram_readout} and \ref{num:tram_neg}. Finally, we discuss how the bias of this procedure, parametrically smaller than the statistical error from finite sampling in the regimes considered here, scales with $K$ and $N$.
\subsubsection{Estimator}
In this section, we present a constructive derivation of the tilted Ramsey estimator (Property \ref{num:tram_prob} under Sec. \ref{sec:tiltedram}) Specifically, we first analyze the protocol by computing the output probability distribution $p(z|\vec{\theta})$, and we then demonstrate how it leads to an estimator $\hat \theta_a$ (Eq. \eqref{eq:linest}). We further show that this estimator is efficient to compute via Theorem \ref{thm:diagestlin}.

We first calculate $p(z|\vec{\theta}) = |\braket{z|X(\phi)^{\otimes N} \Pi_{a} \exp(-i \theta_a Z_a) H^{\otimes N} |0}|^2$. Now, the first order terms are
\begin{align}
    \partial_{\theta_a} p(z|\vec{\theta})|_{\vec \theta =0} &= - i  \braket{z|X(\phi)^{\otimes N} Z_a \mathrm{H}^{\otimes N} |0}  \braket{0|\mathrm{H}^{\otimes N}X(-\phi)^{\otimes N}|z}  + h.c.\\
    &= 2 \im(\braket{z|X(\phi)^{\otimes N} Z_a \mathrm{H}^{\otimes N} |0}  \braket{0|\mathrm{H}^{\otimes N}X(-\phi)^{\otimes N}|z})\\
    & = \frac{(-1)^{n_{z,a}}}{2^{N-1}}\sin(s_a\phi), \label{eq:linforder}
\end{align}
where $s_a = \text{wt}(a)$ and $n_{z,a}= \text{wt}(z \odot a)$, and $\odot$ indicates bitwise multiplication. The first-order term is nonzero for any $\phi$ such that $s\phi \neq\pi$ for $s\in \mathbb{Z}$.

We also check the second-order terms,

\begin{align}
     \partial_{\theta_a}\partial_{\theta_b} p(z|\vec{\theta})|_{\vec \theta =0} &= -   \braket{z|X(\phi)^{\otimes N} Z_aZ_b \mathrm{H}^{\otimes N} |0}  \braket{0|\mathrm{H}^{\otimes N}X(-\phi)^{\otimes N}|z}  \notag\\
     &\quad + \braket{z|X(\phi)^{\otimes N} Z_a\mathrm{H}^{\otimes N} |0}  \braket{0|\mathrm{H}^{\otimes N}Z_bX(-\phi)^{\otimes N}|z} + h.c.\\
     &= -2\text{Re}(\braket{z|X(\phi)^{\otimes N} Z_aZ_b \mathrm{H}^{\otimes N} |0}  \braket{0|\mathrm{H}^{\otimes N}X(-\phi)^{\otimes N}|z} ) \notag\\ 
     &\quad + 2\text{Re}(\braket{z|X(\phi)^{\otimes N} Z_a \mathrm{H}^{\otimes N} |0}  \braket{0|\mathrm{H}^{\otimes N}Z_bX(-\phi)^{\otimes N}|z}) \\
     &= \frac{(-1)^{n_{z, a\oplus b}}}{2^{N-1}} \big(-\cos(s_{a\oplus b} \phi) + \cos((s_{a}-s_{b}) \phi)\big)\label{eq:linsorder},
\end{align}
where $n_{z, a\oplus b} = \text{wt}(z \odot (a \oplus b))$ and $s_{a \oplus b} = \text{wt}(a\oplus b)$, with $\oplus$ indicating bitwise $\text{XOR}$.
Although this will generally result in some complex pattern of bitstrings, the case when $a = b$ is of note. Then,
\begin{align}\label{eq:linsorder2}
\partial^2_{\theta_a}p(z|\vec{\theta})|_{\vec \theta =0}  = -2|\braket{z|X(\phi)^{\otimes N}\mathrm{H}^{\otimes N} |0}|^2 + 2|\braket{z|X(\phi)^{\otimes N} Z_aZ \mathrm{H}^{\otimes N} |0} |^2.
\end{align}
The first term is precisely the $\theta^2_a$ correction to the normalization of zero-th order term, $p(z|\vec{\theta} = 0)=|\braket{z|X(\phi)^{\otimes N}H^{\otimes N} |0}|^2$. In our quadratic estimator, we had absorbed the normalization into a constant $A$. Here, by Eq. \eqref{eq:linsorder} however, this term is precisely canceled in the second-order,
\begin{align}
    \partial^2_{\theta_a}p(z|\vec{\theta})|_{\vec \theta =0} =  \frac{(-1)^{n_{z, a\oplus b}}}{2^{N-1}} (-1 + 1) = 0.
\end{align}
Therefore, we do not include the factor $A$ in the tilted Ramsey procedure, 
\begin{align}\label{eq:problin}
    p(z|\vec{\theta})= p_0(z) + \sum_a \theta_a \delta p_a(z) + O(\theta_a\theta_b), \;\; p_0(z)  = \frac{1}{2^N}, \;\; \delta p_a(z) = \frac{(-1)^{n_{z,a}}}{2^{N-1}}\sin(s_a\phi).
\end{align}

Conceptually, our estimator performs least-squares regression on Eq. \eqref{eq:problin}. 
Specifically, thinking of $p_0(z)$ and $\delta p_a$ as $2^N$ dimensional vectors over bitstrings $z$, we can concatenate them together to form a $2^N \times (K_c+1)$ dimensional matrix $V$. The columns of $V$ are given by $V_{z, 0}=p_0(z)$ and $V_{z, a} = \delta p_a(z)$. Then, we see that $p(z|\vec{\theta}) = V_{z, 0} +\sum_{a}V_{z, a}\theta_a$. An estimator for $\theta_a$ can be obtained by inverting this equation, replacing $p(z|\vec{\theta})$ with the empirical distribution \cite{Hastie2001}: 
\begin{align}
    \hat \theta_a = \sum_{i, z}(V^TV)^{-1}_{a, i} V^{T}_{i, z}\frac{\hat N_z}{M}.
\end{align}

In practice, we do not need to explicitly store the $2^N \times (K_c+1)$ dimensional matrix $V$. Due to the structure of $\delta p_a(z)$ (Eq. \eqref{eq:linforder}), $V^TV$ is actually diagonal, as we show below. 

We begin by establishing several preliminary results that characterize the behavior of $\delta p_a(z)$. In Lemma~\ref{lem:lindpa}, we show that $\delta p_a(z)$ is negative for half of all bitstrings $z$. Next, Lemma~\ref{lem:lindpab} demonstrates that $\delta p_a(z)\delta p_b(z)$ is negative for half of all bitstrings when $a \neq b$. These results let us evaluate entries of $V^TV$, which have the form $\sum_z p_0(z) \delta p_a(z)$ and $\sum_z \delta p_a(z) \delta p_b(z)$. The resulting expression for $V^TV$, shown to be diagonal, is derived explicitly in Theorem~\ref{thm:diagestlin}.

\begin{lemma}\label{lem:lindpa}
    The number of bitstrings $z$ where $\delta p_a(z)$ is negative $|\{z\; |\; \sgn(\delta p_a(z)) = -1\}| = 2^{N-1}$.
\end{lemma}
\textit{Proof. } We analyze  $n_{z,a}= \text{wt}(z \odot a)$, which determines the sign $(-1)^{n_{z, a}}$ of $\delta p_a(z)$. Consider the qubit positions at which $a \in \{0, 1\}^N$ is 1. We call this the support of $a$, $\text{supp}(a) \equiv \{i \; | \; a_i = 1\}$.  The bitstrings $z$ which have $z_j=1$ on an even number of qubits in $\text{supp}(a)$ (bitstrings $z$ which satisfy $|\{j | z_j= 1 \text{ and } j \in\text{supp}(a)\}| = 0 \mod 2$) take $\sgn(\delta p_a(z))=1$, while the bitstrings $z$ which have $z_j=1$ on an odd number of qubits in $\text{supp}(a)$  take $\sgn(\delta p_a(z))=-1$.

Let $s_a = \text{wt}(a)$. Then, the number of bitstrings $z$ which have $z_j=1$ on an even number of qubits $j$ in $\text{supp}(a)$ is 
\begin{align}
    \sum_{\text{even } k}\begin{pmatrix}
        s_a \\
        k
    \end{pmatrix}2^{N-s_a} = \frac{1}{2} 2^{s_a}2^{N-s_a} = 2^{N-1}.
\end{align}
In the second line, we have used the Binomial theorem, as well as the identity
\begin{align}
    \sum_{\text{even } k}\begin{pmatrix}
        s_a \\
        k
    \end{pmatrix} = \sum_{\text{odd } k}\begin{pmatrix}
        s_a \\
        k
    \end{pmatrix}.
\end{align}
So, $2^{N-1}$ bitstrings $z$ have $\sgn(\delta p_a(z))=1$, and $2^{N-1}$ bitstrings $z$ have $\sgn(\delta p_a(z))=-1$. $\qed$

\begin{lemma}\label{lem:lindpab}
    For $a \neq b$, the number of bitstrings $z$ where $\delta p_a(z) \delta p_b(z)$ is negative $|\{z\; |\; \sgn(\delta p_a(z)\delta p_b(z)) = -1\}| = 2^{N-1}$.
\end{lemma}
\textit{Proof. } We have $\sgn(\delta p_a(z) \delta p_b(z)) =  (-1)^{n_{z,a}+n_{z,b}}$. First, we note that since 
\begin{align}
    n_{z,a}+n_{z,b} = \sum_i z_i(a_i+b_i) = 2\sum_{\{i |a_i=b_i\}}z_ia_i +\sum_{\{i |a_i\neq b_i\}}z_i,
\end{align}
\begin{align}
    \sgn(\delta p_a(z) \delta p_b(z))=(-1)^{n_{z,a}+n_{z,b}} = (-1)^{2\sum_{\{i |a_i=b_i\}}z_ia_i +\sum_{\{i |a_i\neq b_i\}}z_i}=(-1)^{\sum_{\{i |a_i\neq b_i\}}z_i}.
\end{align}

Therefore, the bitstrings $z$ which have $z_j=1$ on an odd number of qubits in the set $\{i |a_i\neq b_i\}$ correspond to $\sgn(\delta p_a(z) \delta p_b(z)) =-1$. 

Now, we assert that for $a$ and $b$ which have bits that differ at any $k$ positions, $|\{z\; |\; \sgn(\delta p_a(z)\delta p_b(z)) = -1\}| = 2^{N-1}$. We prove this by induction. For $a$ and $b$ differing at a single bit $i$, half of all bitstrings will have $z_i=1$, leading to $2^{N-1}$ bitstrings with $\sgn(\delta p_a(z) \delta p_b(z)) = -1$. For $a$ and $b$ differing at $k-1$ positions, assume $2^{N-1}$ bitstrings take $\sgn(\delta p_a(z) \delta p_b(z)) =-1$. WLOG, suppose we then obtain $a'$ by flipping a single bit at position $j$ in $a$. $a'$ and $b$ now differ at $k$ positions. 

Now, there are $2^{N-2}$ bitstrings $z$ where  $z_j=0$ and $\sgn(\delta p_{a} (z)\delta p_b (z)) = -1$, and $2^{N-2}$ bitstrings $z$ where $z_j=0$ and $\sgn(\delta p_{a} (z)\delta p_b (z)) = 1$. This is true as we can consider simply truncating bit $j$ from our system, which takes $N \rightarrow N-1$. In this case, our inductive hypothesis states that there are $2^{N-2}$ bitstrings with $\sgn(\delta p_{a} (z)\delta p_b (z)) = -1$. Adding bit $j$ back in, $\sgn(\delta p_{a} (z)\delta p_b (z))$ for bitstrings with $z_j=0$ will be unchanged, as $a_j = b_j$. We also note that $\sgn(\delta p_a (z)\delta p_b (z)) = \sgn(\delta p_{a'} (z)\delta p_b (z))$ for $z$ where $z_j = 0$.  Thus, there are also $2^{N-2}$  bitstrings $z$ where  $z_j=0$ and $\sgn(\delta p_{a'} (z)\delta p_b (z)) = -1$, as well as $2^{N-2}$  bitstrings $z$ where  $z_j=0$ and $\sgn(\delta p_{a'} (z)\delta p_b (z)) = 1$.

Meanwhile, $\sgn(\delta p_{a'} (z)\delta p_b (z))=-\sgn(\delta p_a (z)\delta p_b (z))$ for bitstrings $z$ where $z_j = 1$. This again leads to $2^{N-2}$ bitstrings $z$ where  $z_j=1$ and $\sgn(\delta p_{a'} (z)\delta p_b (z)) = -1$, as well as $2^{N-2}$ bitstrings $z$ where  $z_j=1$ and $\sgn(\delta p_{a'} (z)\delta p_b (z)) = 1$. In total, we still obtain $|\{z\; |\; \sgn(\delta p_{a'}(z)\delta p_b(z)) = -1\}| = 2\times 2^{N-2}=2^{N-1}$. $\qed$

Now, we have enough information to show that $V^TV$ is actually a diagonal matrix. 
\begin{theorem}\label{thm:diagestlin}
    $V^TV$, where $V_{z,0}=p_0(z)$ and $V_{z, a} = \delta p_a(z)$ (Eq. \eqref{eq:problin}), is a diagonal matrix: 
    \begin{align}
       (V^TV)_{0, 0}=\frac{1}{2^N},\;\; (V^TV)_{a, a}= \frac{1}{2^{N-2}}\sin^2(s_a\phi),\;\; (V^TV)_{l, m} = 0 \text{ for } l\neq m
    \end{align}
\end{theorem}
\textit{Proof. } We first compute the diagonal terms. Note that $\delta p_a(z)^2 =  \frac{(-1)^{2n_{z,a}}}{2^{2N-2}}\sin^2(s_a\phi) = \frac{\sin^2(s_a\phi)}{2^{2N-2}}$, so $ (V^TV)_{a,a}= \sum_z\delta p_a(z)^2 = 2^N\frac{\sin^2(s_a\phi)}{2^{2N-2}} = \frac{1}{2^{N-2}}\sin^2(s_a\phi)$. Meanwhile, $(V^TV)_{0,0}= \sum_z\frac{1}{2^{2N}} = \frac{1}{2^N}$.

Now, we consider the off-diagonal terms. For the first row and column, $(V^TV)_{a,0} = (V^TV)_{0,a} = \sum_z \frac{\delta p_a(z)}{2^N}$. By Lemma \ref{lem:lindpa},   
$\sum_z \delta p_a(z) = 0$, so $(V^TV)_{a,0} = (V^TV)_{0,a} = 0$, For the rest of the terms, 
\begin{align*}
    (V^TV)_{a,b} = \frac{\sin(s_a \phi)\sin(s_b \phi)}{2^{N-2}} \sum_z (-1)^{n_{z, a} + n_{z, b}},
\end{align*} where $a\neq b$. By Lemma \ref{lem:lindpab}, $\sum_z  (-1)^{n_{z,a}+n_{z,b}} = 2^{N-1}-2^{N-1} =0$. Thus, $(V^TV)_{a,b} = 0$ for $a \neq b$.
$\qed$

From Theorem \ref{thm:diagestlin}, we can then write our estimator as 
\begin{align}\label{eq:linest}
    \hat \theta_a = \frac{2^{N-2}}{ \sin^2(s_a\phi)M} \sum_{m=1}^{M} \delta p_a(z_m),
\end{align}
where $\{z_m\}$ are the bitstrings measured in experiment. Since we can choose to only compute $\delta p_a(z_m)$ for the bitstrings we measure using Eq. \eqref{eq:problin}, the classical complexity of this estimator is $O(MN)$, where $O(N)$ time complexity is required to naively calculate $s_a$ and $n_{z, a}$. This estimator is also unbiased up to higher-order corrections in $\theta_a$, and for the rest of our analysis, we assume $\hat \theta_a$ is unbiased, instead focusing on the regime where statistical error arising from finite $M$ is much larger than bias. Corrections to this assumption can be obtained by including contributions from the bias in the error of the estimator, and a discussion of the behavior of this bias, as well as how it may be suppressed, is given in Sec.~\ref{sec:biastram}.

To better understand the behavior of our estimator (Eq.~\eqref{eq:linest}), 
we analyze the magnitude  $|\sin(s_a \phi)|$. 
We define $F(S)$ as the minimum of $|\sin(s_a \phi)|$ over all 
$s_a \in [1, S]$, after maximizing over $\phi$. 
Since the sample complexity of estimating $\hat{\theta}_a$ 
scales as $1 / \sin^2(s_a \phi) $, $F(S)$ characterizes the 
worst-case value of $|\sin(s_a \phi)|$ when signals with weights from 
\(1\) to \(S\) are considered under the optimal choice of $\phi$. 
In Lemma~\ref{lem:sinmag}, we show that $F(S) = \Theta(1/S)$.

\begin{lemma}\label{lem:sinmag} Let
\[
F(S)\;:=\;\sup_{\phi\in\mathbb{R}}\;\min_{1\le s\le S} \bigl|\sin(s\phi)\bigr|,
\qquad S\in\mathbb{N}.
\]
Then there exist absolute constants $c,C>0$ such that
\[
\frac{2c}{S}\;\le\;F(S)\;\le\;\frac{C}{S}.
\]
In particular, $F(S)=\Theta(S^{-1})$.

\end{lemma}

\textit{Proof.}
Write $\mathrm{dist}_\pi(x):=\min_{k\in\mathbb{Z}}|x-k\pi|\in[0,\pi/2]$ for the
distance to the lattice $\pi\mathbb{Z}$. For all $x\in\mathbb{R}$ one has the
standard bounds
\begin{equation}\label{eq:sin-bounds}
\frac{2}{\pi}\,\mathrm{dist}_\pi(x)\;\le\;|\sin x|\;\le\;\mathrm{dist}_\pi(x).
\end{equation}
(Indeed, by periodicity and symmetry it suffices to consider $x\in[0,\pi/2]$,
where $\sin x\le x$ and $\sin x\ge \tfrac{2}{\pi}x$.)

\emph{Upper bound.} Fix $\phi\in\mathbb{R}$ and set $\alpha=\phi/\pi$. By
Dirichlet’s approximation theorem, for each integer $S\ge1$ there exists
$1\le s\le S$ such that $\|s\alpha\|:=\min_{m\in\mathbb{Z}}|s\alpha-m|
\le 1/S$. Hence,
\[
\min_{1\le s\le S}\mathrm{dist}_\pi(s\phi)
=\min_{1\le s\le S}\pi\,\|s\frac{\phi}{\pi}\|\;=\min_{1\le s\le S}\pi\,\|s\alpha\|\;\le\;\frac{\pi}{S}.
\]
Using the right inequality in \eqref{eq:sin-bounds} gives
$\min_{1\le s\le S}|\sin(s\phi)|\le \pi/S$, and since this holds for every
$\phi$, we have $F(S)\le \pi/S$. Thus we may take $C=\pi$.

\emph{Lower bound.} Choose $\alpha\in\mathbb{R}$ that is \emph{badly
approximable}~\cite{Wagon2002} (e.g., a quadratic irrational such as the golden-ratio
conjugate), so there exists $c>0$ such that for all rationals $\frac{p}{q}$, $|\alpha - \frac{p}{q}| \geq \frac{c}{q^2}$, or equivalently $\|q\alpha\|\ge c/q$ for all integers $q\ge1$. Set $\phi=\pi\alpha$.
Then for all $1\le s\le S$,
\[
\mathrm{dist}_\pi(s\phi)=\pi\,\|s\alpha\|\;\ge\;\pi\,\frac{c}{s}\;\ge\;\pi\,\frac{c}{S}.
\]
Applying the left inequality in \eqref{eq:sin-bounds} yields
\[
\min_{1\le s\le S}|\sin(s\phi)|\;\ge\;\frac{2}{\pi}\,\mathrm{dist}_\pi(s\phi)
\;\ge\;\frac{2c}{S}.
\]
Therefore $F(S)\ge (2c)/S$.  $\qed$

Thus, by Lemma~\ref{lem:sinmag}, if we include signals $a$ with weights from $1$ to $S$, then even with optimized $\phi$, the worst case $\min_a|\sin(s_a \phi)|$ scales as $S^{-1}$. As we shall see in the following section, this implies from Eq.~\eqref{eq:linest} that if signals $a$ with weights $1$ through $N$ are included, then the worst-case sample complexity for $\hat \theta_a$ acquires a prefactor of $N^2$. Here, the weight of $a$ determines the size of the support of the corresponding Pauli operator $Z_a$ (Eq.~\eqref{eq:defcom}).

\subsubsection{Typical sample complexity}

After obtaining the estimator Eq. \eqref{eq:linest}, we now compute its variance (Property \ref{num:tram_sm} under Sec. \ref{sec:tiltedram}). 
Now, 
\begin{align}
    \Var(\hat \theta_a) = \frac{2^{2N-4}}{ \sin^4(s_a \phi)M^2}\Var(\sum_z \delta p_a(z) \hat N_z), 
\end{align}
where $\hat N_z \sim \text{Multinomial}(M, p(z))$. It can be straightforwardly shown that
\begin{align}\label{eq:wmultvar}
    \Var(\sum_z \delta p_a(z) \hat N_z) = M\sum_z\delta p^2_a(z) p(z)-M\bigg(\sum_z \delta p_a(z)  p(z)\bigg)^2.
\end{align}
We approximate $p(z) = p_0(z) = \frac{1}{2^N}$ and ignore higher-order terms in $\theta_a$. So,
\begin{align}
    \Var(\sum_z \delta p_a(z) \hat N_z)\approx M\sum_z\delta p^2_a(z) p(z) \approx  \frac{M}{2^{2N-2}}\sin^2(s_a\phi).
\end{align}
This yields
\begin{align}
    \Var(\hat \theta_a) \approx \frac{2^{2N-4}}{ \sin^4(s_a \phi)M^2} \frac{M}{2^{2N-2}}\sin^2(s_a\phi)= \frac{1}{4\sin^2(s_a\phi)M}.
\end{align}
Thus, we expect $|\hat \theta_a-\theta_a| = \frac{1}{2|\sin(s_a\phi)|\sqrt{M}}$. By Lemma~\ref{lem:sinmag}, $\min_a|\sin(s_a\phi)| \sim N^{-1}$ if signals $a$ with weights $1$ up to $N$ are considered.

\subsubsection{Bounds on sample complexity for worst-case error}\label{sec:tiltedbounds}
We can obtain a bound on the worst-case error, $\text{max}_a |\hat \theta_a-\theta_a|^2$, which we state below in Theorem \ref{thm:boundtram} (Property \ref{num:tram_bound}). The proof relies on tail bounds for the estimator $\hat \theta_a$, obtained by mapping the multinomial distribution of the bitstring measurements to a Poisson model and analyzing the probability of large deviations. This result provides the sample complexity scaling in terms of the worst-case error.

To proceed, we recall some standard properties of the Poisson distribution, which will be useful in the ensuing analysis of our estimator $\hat \theta_a$. These results follow from standard references (e.g., Ref.~\cite{Wainwright2019}) and are reproduced here for completeness.

\begin{lemma}\label{lem:poissub}
    For a random variable $X\sim \text{Poisson}(\lambda)$, $X$ is sub-exponential with parameters $(\sqrt{e\lambda}, 1)$.
\end{lemma}
\textit{Proof. } First, we note the definition of sub-exponential \cite{Wainwright2019}. A random variable $X$ with mean $\mu$ is sub-exponential if there are non-negative
parameters $(\nu, \alpha)$ such that 
\begin{align}
    \mathbb{E}(e^{t(X-\mu)}) \leq e^{\frac{\nu^2 t^2}{2}} \text{ for all $|t| \leq \frac{1}{\alpha}$}
\end{align}
Now, for $X\sim \text{Poisson}(\lambda)$,
\begin{align}
    \mathbb{E}(e^{t(X-\lambda)})=e^{-\lambda t}\mathbb{E}(e^{tX}) = \exp(\lambda (-t + e^t-1)).
\end{align}
Now, 
\begin{align}
    e^t-t-1 &= \frac{t^2}{2} + \frac{t^3}{3!} +\cdots \\
    &\leq \frac{e^c}{2}t^2 \text{ for some c between 0 and $t$} \\
    &\leq \frac{e}{2}t^2 \text{ for $|t| \leq 1$},
\end{align}
where we have Taylor expanded $e^t$ in the first line and used Taylor's Remainder Theorem in the second. This implies
\begin{align}
     \exp(\lambda (e^t-t-1))\leq\exp(\frac{e}{2}t^2 \lambda)  \text{ for $|t| \leq 1$}.
\end{align}
So, $X$ is sub-exponential with $\nu = \sqrt{e\lambda}$, $\alpha = 1$. $\qed$

\begin{corollary}\label{cor:rpois}
     $aX$ is sub-exponential with parameters $(a\sqrt{e\lambda}, |a|)$.
\end{corollary}
\textit{Proof. } The rescaled variable $aX$ has mean $a \lambda$. We can compute
\begin{align}
    \mathbb{E}(e^{at(X-\lambda)}) = \exp(\lambda(e^{at}-at-1)).
\end{align}
Using Taylor's Remainder Theorem again,
\begin{align}
    e^{at}-at-1 &\leq \frac{e^{ac}}{2}a^2t^2 \text{ for some c between 0 and $t$}\\
    &\leq \frac{e}{2}a^2t^2 \text{ for $|t|\leq 1/|a|$},
\end{align}
such that 
\begin{align}
    \exp(\lambda(e^{at}-at-1)) \leq \exp(\lambda\frac{e}{2}a^2t^2 ) \text{ for $|t|\leq 1/|a|$}.
\end{align}
So, $aX$ is sub-exponential with $\nu = a\sqrt{e\lambda}$, $\alpha = |a|$. $\qed$

We also recall Corollary D.13 from Ref.~\cite{Canonne2020}, restated here for convenience.
\begin{lemma}\label{lem:poisbound}
    For any $M\geq 20$ and $X\sim \text{Poisson}(2M)$, 
    \begin{align}
        \Pr(X\notin [M, 3M]) \leq \frac{2}{\sqrt{M}}\left(\frac{8e}{27}\right)^M.
    \end{align}
\end{lemma}
\textit{Proof. } See Ref. \cite{Canonne2020}. 

Now, we show our main result.
\begin{theorem}\label{thm:boundtram}
 Assume the perturbative regime of small signals $\sum_a \theta^2_a \ll 1$, and neglect bias. In the presence of $K$ total signals $\theta_a$ (and no a priori information about their weights), a precision $\max_a|\hat \theta_a -\theta_a|\leq \epsilon$ can be achieved with probability $1-\delta$ by taking $M= O(\max\{\frac{\log(\frac{K}{\delta})}{\epsilon^2}, \frac{\log(\frac{K}{\delta})}{\epsilon}\})$, where $\frac{1}{\epsilon^2}$ controls the sample complexity for small $\epsilon$.
\end{theorem}
\textit{Proof.} Using our estimator,
\begin{align}\label{eq:linestbound}
     |\hat \theta_a-\theta_a| = \frac{2^{N-2}}{ \sin^2(s_a\phi)} \bigg|\sum_{z} \delta p_a(z)\frac{\hat N_z}{M} - \mathbb{E}\bigg(\sum_{z} \delta p_a(z)\frac{\hat N_z}{M}\bigg)\bigg|.
\end{align}
Now, $\hat N_{z} \sim \text{Multinomial}(M, p(z))$. However, we can conduct our analysis by first considering a process where we take $M'\sim \text{Poisson}(M)$ samples, and each of our counts $\hat N_z\sim \text{Poisson}(Mp(z))$, with $\hat N_{z}$ and $N_{z'}$ independent for $z \neq z'$. By Lemma \ref{lem:poisbound}, this \textit{Poissonization} process incurs at most constant factors in our sample complexity analysis.

With poissonized random variables $\hat N_z$, we can use the fact that $\delta p_a(z)\hat N_z$ is sub-exponential with $\nu^2 = \delta p_a(z)^2eMp(z)$ and $\alpha =|\delta p_a(z)|$ by Lemma \ref{lem:poissub} and Corollary \ref{cor:rpois}. 
Since summations of independent sub-exponential random variables remain sub-exponential, $\sum_{z} \delta p_a(z)\hat N_z$ is sub-exponential with effective $\nu' = \sqrt{
\sum_z \delta p_a(z)^2Mp(z) e}$ and $\alpha' = \max_z |\delta p_a(z)|$ \cite{Wainwright2019}. Now, let's examine $\nu'$ more carefully:
\begin{align}
    (\nu')^2 = e \sum_z \delta p_a(z)^2Mp(z) .
\end{align}
$\sum_{z} \delta p_a(z)^2 = \frac{\sin^2(s_a\phi)}{2^{N-2}}$ using Eq. \eqref{eq:problin}, and $p(z) \approx \frac{1}{2^N}$ for all $z$ in the perturbative regime. We can thus write
\begin{align}
    (\nu')^2 =M \frac{ \mathcal{C}_a}{2^{2N-2}}
\end{align}
for some $O(1)$ constant $\mathcal{C}_a$. Meanwhile, $\alpha' = \frac{|\sin(s_a\phi)|}{2^{N-1}}$ by Eq. \eqref{eq:problin}. So, using the sub-exponential tail bound,
\begin{align}
    \text{Pr}\bigg(\bigg|\sum_{z} \delta p_a(z)\hat N_z -\sum_{z} M\delta p_a(z)p(z)\bigg| \geq \epsilon \bigg) &\leq 
    \begin{cases}
        2\exp\bigg(-\frac{2^{2N-3}\epsilon^2}{\mathcal{C}_aM}\bigg) \text{ if $\epsilon \leq \frac{ \mathcal{C}_aM}{2^{N-1}|\sin(s_a\phi)|}$} \\
        2\exp\big(-\frac{2^{N-2}\epsilon}{|\sin(s_a\phi)|}\big) \text{ if $\epsilon \geq \frac{ \mathcal{C}_aM}{2^{N-1}|\sin(s_a\phi)|}$} 
    \end{cases} \\
    \implies \text{Pr}\bigg(\bigg|\sum_{z} \delta p_a(z)\frac{\hat N_z}{M} -\sum_{z} \delta p_a(z)p(z)\bigg| \geq \epsilon \bigg) &\leq 
     \begin{cases}
        2\exp\bigg(-\frac{2^{2N-3}M\epsilon^2}{\mathcal{C}_a}\bigg) \text{ if $\epsilon \leq \frac{ \mathcal{C}_a}{2^{N-1}|\sin(s_a\phi)|}$} \\
        2\exp\big(-\frac{2^{N-2}M\epsilon}{|\sin(s_a\phi)|}\big) \text{ if $\epsilon \geq \frac{ \mathcal{C}_a}{2^{N-1}|\sin(s_a\phi)|}$} 
    \end{cases}\\
      \implies  \text{Pr}(|\hat \theta_a-\theta_a|\geq \epsilon) &\leq
      \begin{cases}
        2\exp\bigg(-\frac{2\sin^4(s_a \phi) M\epsilon^2}{\mathcal{C}_a}\bigg) \text{ if $\epsilon \leq \frac{ \mathcal{C}_a}{2|\sin(s_a\phi)|^3}$} \\
        2\exp\big(-|\sin(s_a\phi)|M\epsilon\big) \text{ if $\epsilon \geq \frac{ \mathcal{C}_a}{2|\sin(s_a\phi)|^3}$} 
    \end{cases}.
\end{align}
In the last line, we have used Eq. \eqref{eq:linestbound}.
So, to achieve $|\hat \theta_a-\theta_a|\leq \epsilon$ with probability $1-\delta$, we want $M\geq \max\{\frac{\log(\frac{2}{\delta})}{2\sin^4(s_a \phi)\epsilon^2}\mathcal{C
}_a, \frac{\log(\frac{2}{\delta})}{|\sin(s_a \phi)|\epsilon}\mathcal{C
}_a\}$, with the $\frac{1}{\epsilon^2}$ scaling controlling behavior for small $\epsilon \leq \frac{ \mathcal{C}_a}{2|\sin(s_a\phi)|^3}$. To achieve $\text{max}_a|\hat \theta_a-\theta_a|\leq \epsilon$ with probability $1-\delta$, we use the union bound. This therefore requires $M\geq \max_a\{\frac{\log(\frac{2K}{\delta})}{2\sin^4(s_a \phi)\epsilon^2}\mathcal{C
}_a, \frac{\log(\frac{2K}{\delta})}{|\sin(s_a \phi)|\epsilon}\mathcal{C
}_a\}$, where $K$ is the total number of signals. Moreover, we have the $\frac{1}{\epsilon^2}$ scaling  controlling behavior for small $\epsilon \leq \min_a\frac{ \mathcal{C}_a}{2|\sin(s_a\phi)|^3}$. 

We can translate this analysis back to a deterministic number of samples rather than a probabilistic number $M'\sim \text{Poisson}(M)$. By Lemma \ref{lem:poisbound}, 
\begin{align}
    \text{Pr}\left(M' \notin [\frac{M}{2}, \frac{3M}{2}]\right) \leq \frac{2\sqrt{2}}{\sqrt{M}}\bigg(\frac{8e}{27}\bigg)^{M/2} \leq \frac{2\sqrt{2}}{\sqrt{40}}\bigg(\frac{8e}{27}\bigg)^{M/2} 
\end{align}
for $M \geq 40$, noting that $M$ here corresponds to $2M$ in Lemma~\ref{lem:poisbound}. Thus, to achieve a probability of $1-\delta'$ that the true number of samples $M'$ lies within $[\frac{M}{2}, \frac{3M}{2}]$, we can take $M  \geq \max\{{2\frac{\log(\frac{\delta'}{\sqrt{5}})}{\log(\frac{8e}{27})}}, 40\}$. So, with probability $1-\delta$, $\text{max}_a|\hat \theta_a-\theta_a|\leq \epsilon$ when we take a number of samples $M$ within a constant interval $M/2$ of $\max\big\{\frac{\log(\frac{4K}{\delta})}{2\epsilon^2}\max_a\frac{\mathcal{C
}_a}{\sin^4(s_a \phi)}, \frac{\log(\frac{4K}{\delta})}{\epsilon}\max_a\frac{\mathcal{C
}_a}{|\sin(s_a \phi)|}, {2\frac{\log(\frac{\delta}{2\sqrt{5}})}{\log(\frac{8e}{27})}}, 40\big\}$.
$\qed$

\subsubsection{Impact of readout error}\label{sec:linreadout}
Here, we justify Property \ref{num:tram_readout}. Recall the variance of our empirical probabilities in the presence of readout error, Eq. \eqref{eq:varnerr}. For the tilted Ramsey estimator, $p_k = A \approx \frac{1}{2^N}$, so $\Var(\frac{\tilde N_z}{M})-\Var(\frac{\hat N_z}{M})\sim \frac{\gamma_r}{2^N}$, which is subdominant compared to quantum projection noise. We thus expect the sample complexity to maintain SQL scaling, albeit with a factor $\Var(\hat \theta_a) \rightarrow \frac{1}{(1-2\gamma_r)^N}\Var(\hat \theta_a)$ that comes from correcting the bias introduced by readout error (see Eq. \eqref{eq:quadinvexp}).

\subsubsection{Bias scaling}\label{sec:biastram}
Here, we discuss how the bias of the tilted Ramsey estimator (Eq.~\ref{eq:linest}) scales with the system size $N$, the total number of signals $K$, and the signal fidelity $A$. Specifically, we show that bias is suppressed with increasing $N$, grows slowly with decreasing $A$, and never exceeds the scaling of the statistical error with $K$. Consequently, there is a broad regime of small $M$ in which statistical error dominates over the bias. We analyze both the mean squared bias (Eq.~\ref{eq:meansqb}) and the worst-case squared bias (Eq.~\ref{eq:wcasesqb}). The results of our numerical study are shown in Fig.~\ref{fig:biasram}(b) and summarized below.

The mean squared bias decreases as $K^{-\gamma}$ with increasing $K$. 
Although this behavior may appear counterintuitive, it follows directly from the fact that increasing the total number of \textit{candidate} signals $K$ at a fixed signal fidelity $A$ makes the signal set increasingly sparse. In this regime, most candidate signals are zero, and the bias associated with zero signals is smaller than that of nonzero signals, approaching zero as the system size $N$ increases. As a result, the mean squared bias scales as $K^{-1}$ in the limit of large $N$. This scaling behavior is not a generic feature of all protocols---for example, the mean squared bias of the quadratic Ramsey protocol, shown in  Fig.~\ref{fig:biasram}(a), remains constant with respect to $K$, reflecting the fact that the bias on zero and nonzero signals is comparable.

As expected from the same arguments as in Sec.~\ref{sec:tiltedbounds}, the worst-case squared bias scales with the total number of signals $K$ as $\log(K)$---the same scaling behavior as the statistical error (Sec.~\ref{sec:tiltedbounds}). The worst-case squared bias also decreases logarithmically with $A$ as $-\alpha\log(A)$, where the prefactor $\alpha$ is suppressed with the system size $N$; however, this suppression does not exhibit a clear polynomial or exponential scaling. The reduction in $\alpha$ arises because the bias originates from higher-order contributions to the probability distribution $p(z | \vec{\theta})$ that have finite overlap with the first-order perturbation directions $\delta p_a(z)$ associated with the signals $\theta_a$ (Eq.~\ref{eq:problin}). As the Hilbert space dimension grows exponentially with $N$, these higher-order perturbations become increasingly orthogonal to the signal directions on average, leading to a reduction of the bias. The mean squared bias exhibits similar scaling behavior with respect to $A$. The bias thus grows slowly as $A$ decreases. 

Importantly, the worst-case bias grows as $\epsilon_{\text{bias}}^2\sim \log(K)$, the same rate as the worst-case statistical error $\epsilon^2\sim  \log(K)/M$. Thus, for $M$ smaller than a scale set by $\frac{1}{\epsilon_{\text{bias}}^2}$, the statistical error is larger than the bias---the regime relevant for our work. Moreover, for a fixed $A$ and $K$, both the average-case and worst-case bias are suppressed on average with increasing $N$ (Fig.~\ref{fig:biasram}). Thus, bias can be reduced both by working with a larger system size and by post-processing, as discussed in Sec.~\ref{sec:highorder}.

\subsubsection{Drawbacks}

First, our tilted Ramsey estimator is not sensitive some commuting signals. In particular, the protocol is insensitive to coherent signals generated by Pauli $X$ terms. In this case, the analogue of Eq. \eqref{eq:linforder} is 
\begin{align}
    \partial_{\theta_a} p(z|\vec{\theta})|_{\vec \theta =0} &= 2 \im(\braket{z|X(\phi)^{\otimes N} X_a \mathrm{H}^{\otimes N} |0}  \braket{0|\mathrm{H}^{\otimes N}X(-\phi)^{\otimes N}|z})\\ &= 2 \im(\braket{z|X(\phi)^{\otimes N}\mathrm{H}^{\otimes N} Z_a|0}  \braket{0|\mathrm{H}^{\otimes N}X(-\phi)^{\otimes N}|z}) = 2 \im(|\braket{z|X(\phi)^{\otimes N}\mathrm{H}^{\otimes N} Z_a|0}|^2)=0.
\end{align}
Second, the tilted Ramsey procedure is also not sensitive to incoherent signals $\gamma_a$, even when they are Pauli $Z$ errors. As incoherent signals take $\rho \rightarrow (1-\gamma_a) \rho + Z_a \rho P_a$, the first order term is
\begin{align}
    \partial_{\gamma_a} p(z|\vec{\theta})|_{\vec \theta, \vec \gamma =0} = |\braket{z|X(\phi)^{\otimes N} Z_a \mathrm{H}^{\otimes N} |0}|^2=|\braket{z|X(\phi)^{\otimes N}\mathrm{H}^{\otimes N} |0}|^2= p_0(z),
\end{align}
as we discussed in Eq. \eqref{eq:linsorder2}. Therefore, all first order terms for incoherent signals do not modify the probability distribution. When these incoherent signals coexist with detectable coherent signals, they only contribute to a renormalization of $p_0(z)$ (Eq. \eqref{eq:problin}) and do not give any information about the magnitude of $\gamma_a$. 

Additionally, if we consider all signals $\theta_{a}$ up to weight $s_{a} = N$, then by Lemma \ref{lem:sinmag}, our sample complexity becomes $M= O(\max\{\frac{N
^4\log(K)}{\epsilon^2}, \frac{N\log(\frac{K}{\delta})}{\epsilon}\})$ for a worst-case error $\max_a|\hat \theta_a -\theta_a|\leq \epsilon$. Although we can still sense exponentially many signals $K$, this incurs an additional polynomial overhead in the system size $N$.

Finally, this procedure does not resolve time-dependent signals, as it can only yield an estimate for a single time step.

%



\subsection{Correction for higher-order contributions}\label{sec:highorder}

Although we have neglected higher-order terms in our analysis, in practice it may be useful to compute and explicitly subtract higher-order contributions. 

For example, consider sensing one and two-body signals with the quadratic Ramsey protocol.
In this case, for $N = 2$ qubits, we might expect the two-body signal $Z_{11} \equiv Z_1 \otimes Z_2$ to have smaller magnitude than the single-body signals $Z_{01}\equiv \id \otimes Z_2$, $Z_{10}\equiv Z_1 \otimes \id$, where bitstring subscripts indicate the support of the Pauli $Z$ operator. Thus, we can calculate the effect of higher-order terms in $\theta_{01}$, $\theta_{10}$ on the estimate for $\theta_{11}$. Recall that our estimator  $\hat \theta^2_{11} \propto \frac{N_{11}}{M}$ (Eq.~\eqref{eq:quadest}), where $N_{11}$ denotes the number of measurements of the bitstring $z=11$ out of $M$ total shots. We can compute the  $\theta_{01}^2\theta_{10}^2$ component of the output probability $p(z|\vec{\theta})$ (Eq.~\eqref{eq:quadpz}), given by 
\begin{align}
\frac{(\partial_{\theta_{01}}\partial_{\theta_{10}})^2}{4} p(z|\vec{\theta}).
\end{align} Most terms at this order do not contribute to the probability of measuring $z=11$, which determines our estimate of $\theta_{11}^2$. However, one term does:  $|\braket{z|X_{01}X_{10}|0}|^2 = |\braket{z|X_{11}|0}|^2$. This is precisely the same form as the contribution of $\theta^2_{11}$, as seen from Eq.~\eqref{eq:quadpzexp}. Thus, we see that 
\begin{align}
     p(z = 11)=\mathbb{E}\frac{N_{11}}{M}   \propto  \theta^2_{11} + \theta^2_{10}\theta^2_{01}.
\end{align}
To correct this contribution, we can modify our estimator $\hat \theta^2_{11}$ by
\begin{align}
    \hat \theta^2_{11} \rightarrow \hat \theta^2_{11}-\hat \theta^2_{10}\hat \theta^2_{01}.
\end{align}
An analogous procedure follows for more general signals.

We can correspondingly correct for higher-order contributions in the tilted Ramsey protocol. Indeed, Eq.~\eqref{eq:linsorder} captures the second-order contributions arising from any pair of signals $\theta_a$ and $\theta_b$. Comparing this expression to Eq.~\eqref{eq:linforder}, we see that $\theta_a$ and $\theta_b$ contribute to the estimate of $\theta_c$ at second order whenever $a \oplus b = c$. These higher-order effects can be explicitly calculated using Eq.~\eqref{eq:linsorder} and then subtracted appropriately. For example, single-body signals $\theta_{01}$ and $\theta_{10}$ will affect the estimate for a two-body signal $\theta_{11}$ at order $\theta_{10}\theta_{01}$. To correct this contribution, we can modify our estimator $\hat \theta_{11}$ by
\begin{align}
    \hat \theta_{11} \rightarrow \hat \theta_{11}- (\tan\phi)\hat \theta_{10}\hat \theta_{01},
\end{align}
where the coefficient of $\hat \theta_{10}\hat \theta_{01}$, $\frac{2\sin^2{\phi}}{\sin{2\phi}} = \tan\phi$, is determined using Eq.~\eqref{eq:linsorder}.

\newpage

\section{Multiparameter sensing of non-commuting signals}\label{sec:multiparam}
In this section, we consider sensing coherent and incoherent signals generated by sets of $K_c$ and $K_{ic}$ Pauli operators up to weight $N$, respectively. Specifically, coherent signals $\theta_\alpha$ couple to the system as $e^{-i \sum_\alpha\theta_\alpha P_\alpha}$, where $P_\alpha$ is a Pauli operator, while incoherent signals $\gamma_\alpha$ act on the system as $\rho \rightarrow (1-\gamma_\alpha)\rho + P_\alpha \rho P_\alpha$. Our signals are also allowed to be time-dependent. To handle time-dependent signal, we consider discrete time steps $1 \cdots T$, and we estimate $\theta_\alpha(t)$  and $\gamma_\alpha(t)$ for each $t \in [1, \dots, T].$ 

We first discuss our multiparameter sensing protocol with global Clifford unitaries in Sec. \ref{sec:glob_cliff}. We then elaborate on how our methods can be extended to local systems---specifically, local Clifford circuits (Sec. \ref{sec:loc_cliff}), local Haar-random unitary circuits (Sec. \ref{sec:lruc}), and ergodic Hamiltonian evolution (Sec. \ref{sec:ham_ev}).

\subsection{Global Clifford protocol}\label{sec:glob_cliff}
Our multiparameter estimation procedure using global Clifford unitaries is illustrated in Fig. 3(a) of the main text. Starting from the $\ket{0}^{\otimes N}$ state, we alternate between evolving under global Clifford unitaries $C_t$ and accumulating signal. This discretizes our signal into $T$ time steps, in which each time step corresponds to a layer of accumulated signal. The last layer of Clifford unitaries applied prior to measurement is $\big(\prod_{t=1} ^{T}C_{t}\big)^{\dagger}$, which ensures that in the presence of no signal, the output measurement distribution is deterministically $p(0)=1$ when measuring in the computational basis. 

In practice, we must choose several sets of Clifford circuits $\{C_t\}_{t=1}^T$, each specifying a different circuit to be run. The number of sets $n_c$ scales logarithmically with the number of signals. To detect incoherent signals, $\frac{M}{n_c}$ measurements in the $z-$basis are taken for each circuit. To detect coherent signals, $\frac{M}{n_c}$ measurements are performed in the $x-$basis for each circuit--- specifically, Hadamard gates are applied before the standard computational basis measurement. As we introduce in the main text, the output probability distribution of this protocol in the presence of coherent signals $\vec{\theta}$ and incoherent signals $\vec{\gamma}$ is given by 
\begin{align}
p(z|\vec{\theta}, \vec{\gamma}) 
&\approx p_0(z)
 +  \sum_{\alpha,t}\theta_\alpha(t)\,\delta p_{\alpha,t}(z) + \sum_{\alpha,t}\gamma_\alpha(t)\,k_{\alpha,t}(z) + O(\gamma_\alpha^2, \theta_\alpha^2),
\end{align}
where \begin{align}
    \delta p_{\alpha, t}(z) &\equiv \partial_{\theta_\alpha(t)} p(z|\vec{\theta}, \vec{\gamma}) \vert_{\vec{\theta}=0, \vec{\gamma}=0},  \label{eq:dp} \\
    k_{\alpha, t}(z) &\equiv \partial_{\gamma_\alpha(t)} p(z|\vec{\theta}, \vec{\gamma}) \vert_{\vec{\theta} = 0, \vec{\gamma}=0} \label{eq:k},
\end{align}
are the perturbations imparted by the presence of each signal.

This multiparameter estimation protocol has the following properties, which are summarized in the main text:
\begin{enumerate}
    \item Each incoherent signal $\gamma_\alpha(t)$ leads to a perturbation $k_{\alpha,t}(z)$ supported on a single bitstring.\label{num:gcliff_probinc} 
    \item For a given circuit, a coherent signal $\theta_\alpha(t)$ leads to a linear nonzero  perturbation $\delta p_{\alpha, t}(z)$ with probability $\frac{1}{2}$. Otherwise, $\theta_\alpha(t)$ will not affect $p(z)$ to first-order---$\delta p_{\alpha, t}(z) = 0$---with probability $\frac{1}{2}$.  \label{num:gcliff_probcoh}
    \item The number of circuits $n_c$ required to achieve a probability $1-\delta$ of being sensitive to each incoherent signal scales as $n_c \sim \log(K_{ic}^2/\delta)/N$, and  $n_c \sim \log(K_{c}/\delta)$ for coherent signals. $K_{ic} (K_c)$ is the total number of incoherent (coherent) signals.\label{num:gcliff_nc}
    \item As $p(z)$ responds linearly to $\gamma_\alpha(t)$  ($\theta_\alpha(t)$) in the incoherent (coherent) protocol, least-squares regression can be used to determine estimators $ \hat \gamma_\alpha(t)$  ($\hat \theta_\alpha(t)$), which are unbiased up to higher-order corrections and require only polynomial classical resources. \label{num:gcliff_est}
    
    \item The sensing procedure for incoherent signals has a typical sample complexity $\Var(\hat \gamma_\beta(t)) \approx \frac{\gamma_\beta(t)}{AM}$, where $A$ is the fidelity of our experiment to the signal-free dynamics. Meanwhile, the sensing procedure for coherent signals has $\Var(\hat \theta_\alpha(t)) \approx \frac{1}{2M A^2}$. \label{num:gcliff_sm}
    \item Neglecting bias, the incoherent sensing protocol achieves $\max_{\alpha ,t}|\hat \gamma_\alpha(t) -\gamma_\alpha(t)| \leq \epsilon$ with probability $1-\delta$ by taking $M= O(\frac{\log(K_{ic}/\delta)}{\epsilon^2})$. Meanwhile, the coherent sensing protocol achieves $\max_\alpha |\hat \theta_\alpha(t) -\theta_\alpha(t)|\leq \epsilon$ with probability $1-\delta$ for $M= O(\frac{\log (K_{c}/\delta)}{\epsilon^2})$. \label{num:gcliff_bound}
    \item In the incoherent sensing scheme, active error correction can be used to counteract readout error at rate $\gamma_r$ per qubit by correcting measured bitstrings to the corresponding signal bitstring closest in Hamming distance. Error correction is possible when $ \gamma_r <\frac{d_{min}}{2N}$, where $d_{min}$ is the minimum Hamming distance between pairwise signal bitstrings--- our ``codewords." With random Clifford circuits, our signal bitstrings will statisfy $\frac{d_{min}}{N} > \alpha$ with probability at least $1-\delta$ for $N \sim \log \big(\frac{K_{ic}^2}{\delta}\big)/(1-H(\alpha))$, where  $H(\cdot)$ is the binary entropy. Here, $\alpha <\frac{1}{2}$ denotes a target relative distance, or the minimum fraction of bits by which any two distinct signal bitstrings differ. Thus, for large system sizes $N$ scaling as $\log(K_{ic}^2)$, our protocol can correct with high probability error rates up to $ \gamma_r <\frac{\alpha}{2}$, at the cost of increasing qubit overhead. \label{num:gcliff_rob}
\end{enumerate}
In this section, we derive these characteristics explicitly. We further elaborate on several aspects of our protocol which are touched upon in the main text. Specifically, we discuss how a thresholding procedure can be used on estimated signals $\hat \theta_{\alpha}(t), \hat \gamma_{\beta}(t)$ to obtain sparse solutions. We further comment on how to address situations in which the Pauli generators for coherent signals $\{\hat P_\alpha\}$ and incoherent signals $\{\hat P_\beta\}$ overlap, discuss how the bias of our protocol---parametrically smaller than the statistical error arising from finite sampling in settings considered here---behaves as a function of $K_{c}$, $K_{ic}$, and $N$, and give more details about the numerical results presented in the main text. 



\subsubsection{Estimator} 
Here, we analyze the probability distribution of the resulting measurements for both the coherent and incoherent procedures, thus showing Properties \ref{num:gcliff_probinc} (stated in Theorem \eqref{eq:Acliff}) and \ref{num:gcliff_probcoh} (stated in Theorem \ref{thm:cliffcohdp}) listed above. With this understanding, we then demonstrate Property \ref{num:gcliff_nc}, that the number of circuits $n_c$ required for an effective protocol scales logarithmically with the number of signals, $K_{ic}$ or $K_{c}$ (Theorems \ref{thm:failurencincoh}, \ref{thm:failurecoh}). We use our analysis to construct estimators $\hat \gamma_\beta(t)$ (Eq. \eqref{eq:estincoh1cliff}) and $\hat \theta_\alpha(t)$  (Eq. \eqref{eq:estcohcliff}) and show that they can be computed efficiently (Theorems \ref{thm:gcliffcompincoh},  \ref{thm:gcliffcompcoh}), thereby validating Property \ref{num:gcliff_est}. 


\noindent\textbf{Incoherent signals---}
When measuring in the $z-$ basis, we have the first order terms
\begin{align}\label{eq:cliffzforder}
    \partial_{\theta_\alpha(t)} p(z|\vec{\theta}, \vec{\gamma})|_{\vec{\theta} = 0, \vec{\gamma}=0} &= 2\im\big[\bra{z}C^{
    \dagger}_1\cdots C^{\dagger}_{t}P_\alpha C_t \cdots C_1\ket{0} \braket{0|z}\big]
    \\
     \partial_{\gamma_\alpha(t)} p(z|\vec{\theta}, \vec{\gamma})|_{\vec{\theta} = 0, \vec{\gamma}=0} & = \lvert \bra{z} C_1^\dagger...C^\dagger_{t} P_\alpha  C_{t}... C_1 \ket{0}\rvert^2 - \lvert \braket{z|0}\rvert^2 \label{eq:cliffzfordergam}
\end{align}
as well as the second order term (assuming $t' \geq t$ WLOG)
\begin{align}\label{eq:cliffzsorder}
    \partial_{\theta_\alpha(t), \theta(t')_\beta} p(z|\vec{\theta}, \vec{\gamma})|_{\vec{\theta} = 0, \vec{\gamma}=0} &= - 2\text{Re}\big[\bra{z}C^{
    \dagger}_1\cdots C^\dagger_{t'} P_\beta C_{t'} \cdots C_{t+1} P_\alpha C_t \cdots C_1\ket{0} \braket{0|z}\big]  \notag \\
    &+ 2\text{Re}\big[\bra{z}C^{
    \dagger}_1\cdots C^\dagger_{t} P_\alpha C_t \cdots C_1\ket{0} \braket{0|C_1\cdots C_{t'} P_\beta C^\dagger_{t'} \cdots C^\dagger_1|z}\big].
\end{align} 
Now, we can examine the structure of these terms in order to determine how they behave for random Clifford circuits. We note that in all of our work below, we work with Pauli operators $P_\alpha$ that are assumed to be tensor products of local Pauli operators---there is no global phase. 

We begin by establishing several results that allow us to explicitly evaluate Eqs.~\eqref{eq:cliffzforder} and~\eqref{eq:cliffzsorder} for random Clifford circuits. Specifically, Lemma~\ref{lem:cliffz1} below provides a probabilistic result that allows us to show that the first-order term in $\theta_\alpha(t)$ (Eq.~\eqref{eq:cliffzforder}) vanishes with exponentially high probability. Similarly, Lemmas~\ref{lem:cliffz2} and~\ref{lem:cliffz3} below provide supporting results used to demonstrate that the second-order term (Eq.~\eqref{eq:cliffzsorder}) is zero with exponentially high probability whenever $(\alpha, t) \neq (\beta, t')$.

\begin{lemma}\label{lem:cliffz1}
    For a random Clifford $C$ and a Pauli operator $P_\alpha \neq \id$, 
    \begin{align}
    \Pr(\braket{0|C^\dagger P_\alpha C|0}=0) = 1-\frac{1}{2^N}.
    \end{align}
\end{lemma}

\textit{Proof. } We note that $C^\dagger P_\alpha C$ = $(\pm 1)\times P'_\alpha$, where $P'_\alpha$ is another Pauli string, and the phase is $\pm 1$. We note that since $P_\alpha$ does not have a global phase, the phase after conjugation cannot be imaginary. Therefore, as $P'_\alpha\ket{0} = \ket{z'}$, $|\braket{0|C^\dagger P_\alpha C|0}|= |\braket{0|z'}|=\delta_{0, z'}$. The magnitude of the overlap is either $0$ or $1$. Now, $|\braket{0|C^\dagger P_\alpha C|0}| = 1$ when $P'_\alpha$ consists of only Pauli $Z$ or $\id$ on each of the qubits. Thus, there are a total of $2\times 2^{N}$ possibilities for $C^\dagger P_\alpha C$ such that $|\braket{0|C^\dagger P_\alpha C|0}| = 1$. Meanwhile, the space of all possible Pauli group elements with phase $\pm 1$ is $2\times4^{N}$. Thus, the probability $\text{Pr}(|\braket{0|C^\dagger P_\alpha C|0}| = 1)=\frac{2\times 2^N}{2\times 2^{2N}} = \frac{1}{2^N}$. So the probability $\text{Pr}(\braket{0|C^\dagger P_\alpha C|0} = 0) = 1- \text{Pr}(|\braket{0|C^\dagger P_\alpha C|0}| = 1) = 1-\frac{1}{2^N}$. $\qed$

\begin{lemma}\label{lem:cliffz2}
    For randomly selected Clifford circuits $\{C_t\}_{t = 1}^T$ and Pauli operators $P_\alpha$, $P_\beta$, assuming that $P_\alpha = P_\beta $ and $t = t'$ are not simultaneously true,
    \begin{align}
        \Pr(\bra{0}C^{
    \dagger}_1\cdots C^\dagger_{t'} P_\beta C_{t'}\cdots C_{t+1} P_\alpha C_t \cdots C_1\ket{0} = 0) \geq 1 - \frac{1}{2^N} - \frac{1}{2^{2N}}.
    \end{align}
\end{lemma}
\textit{Proof. } If $t = t'$ and $P_\alpha \neq P_\beta$, the result follows from Lemma \ref{lem:cliffz1}.

For $t \neq t'$, we can rewrite $\bra{0}C^{
    \dagger}_1\cdots C^\dagger_{t'} P_\beta C_{t'}\cdots C_{t+1} P_\alpha C_t \cdots C_1\ket{0}$ = $\bra{0}C^{
    \dagger}_1\cdots C^\dagger_{t'} P_\beta 
 \tilde{P}_\alpha C_{t'}\cdots C_1\ket{0}$, where $\tilde{P}_\alpha = C_{t'}\cdots C_{t+1} P_\alpha C^{\dagger}_{t+1}\cdots C^\dagger_{t'}$. By the same logic as Lemma \ref{lem:cliffz1}, $$\text{Pr}(\bra{0}C^{
    \dagger}_1\cdots C^\dagger_{t'} P_\beta 
 \tilde{P}_\alpha C_{t'}\cdots C_1\ket{0} = 0) = 1- \text{Pr}(|\bra{0}C^{
    \dagger}_1\cdots C^\dagger_{t'} P_\beta 
 \tilde{P}_\alpha C_{t'}\cdots C_1\ket{0}|=1).$$ Now, $|\bra{0}C^{
    \dagger}_1\cdots C^\dagger_{t'} P_\beta 
 \tilde{P}_\alpha C_{t'}\cdots C_1\ket{0}|=1$ only if $\tilde{P}_\alpha \propto P_\beta$, such that $\tilde{P}_\alpha P_\beta \propto \id$, or if $\tilde{P}_\alpha \not\propto P_\beta$ but $C^{
    \dagger}_1\cdots C^\dagger_{t'} P_\beta 
 \tilde{P}_\alpha C_{t'}\cdots C_1$ stabilizes $\ket{0}$. There are 2 Pauli group operators $\propto P_\beta$, so the probability that $\tilde{P}_\alpha \propto P_\beta$ is $\frac{1}{2^{2N}}$. Meanwhile, if $\tilde{P}_\alpha \not\propto P_\beta$, the probability $|\bra{0}C^{
    \dagger}_1\cdots C^\dagger_{t'} P_\beta 
 \tilde{P}_\alpha C_{t'}\cdots C_1\ket{0}|=1$ is $\frac{1}{2^N}$ by Lemma \ref{lem:cliffz1}. So,
$$ \text{Pr}(|\bra{0}C^{
    \dagger}_1\cdots C^\dagger_{t'} P_\beta 
 \tilde{P}_\alpha C_{t'}\cdots C_1\ket{0}|=1) = \frac{1}{2^{2N}} + (1-\frac{1}{2^{2N}})\frac{1}{2^N}= \frac{1 + \frac{1}{2^N} - \frac{1}{2^{2N}}}{2^N},$$
 and $\text{Pr}(\bra{0}C^{
    \dagger}_1\cdots C^\dagger_{t'} P_\beta C_{t'}\cdots C_{t+1} P_\alpha C_t \cdots C_1\ket{0} = 0) \geq 1 - \frac{1}{2^N} - \frac{1}{2^{2N}}.$ $\qed$
    
\begin{lemma}\label{lem:cliffz3}
    For randomly selected Clifford circuits $\{C_t\}_{t = 1}^T$ and Pauli operators $P_\alpha$, $P_\beta$, assuming that $P_\alpha = P_\beta $ and $t = t'$ are not simultaneously true,
    \begin{align}
        \Pr(\bra{z}C^{
    \dagger}_1\cdots C^\dagger_{t} P_\alpha C_t \cdots C_1\ket{0} \braket{0|C_1\cdots C_{t'} P_\beta C^\dagger_{t'} \cdots C^\dagger_1|z} = 0) \geq 1 - \frac{1}{2^N}.
    \end{align}
\end{lemma}
\textit{Proof. } By the same logic as in Lemma \ref{lem:cliffz1}, we can write $\braket{0|C_1\cdots C_{t'} P_\beta C^\dagger_{t'} \cdots C^\dagger_1|z}  = \delta_{z, z'}$, where $z'$ satisfies $C^\dagger_1\cdots C^\dagger_{t'} P_\beta C_{t'} \cdots C_1\ket{0} \propto \ket{z'}$. As before, we first compute $\text{Pr}(|\bra{z}C^{
    \dagger}_1\cdots C^\dagger_{t} P_\alpha C_t \cdots C_1\ket{0} \braket{0|C_1\cdots C_{t'} P_\beta C^\dagger_{t'} \cdots C^\dagger_1|z}| = 1) $. This only occurs if $C^{
    \dagger}_1\cdots C^\dagger_{t} P_\alpha C_t \cdots C_1\ket{0} \propto \ket{z'}$. Now, there are $2\times 2^{N}$ elements $\tilde{P}$ of the Pauli group for which $\tilde{P}\ket{0} = \ket{z'}$ (omitting those with imaginary phase, as they cannot be reached by Clifford conjugation from Pauli strings). Specifically, to construct $\tilde P$, we choose $\id$ or $Z$ for the qubits $j$ where $z'_j=0$, either $X$ or $Y$ for the qubits $j$ where $z'_j=1$, and one of two phases $\pm 1$. Thus, $\text{Pr}(|\bra{z}C^{
    \dagger}_1\cdots C^\dagger_{t} P_\alpha C_t \cdots C_1\ket{0} \braket{0|C_1\cdots C_{t'} P_\beta C^\dagger_{t'} \cdots C^\dagger_1|z}| = 1)  = \frac{2 \times 2^N}{2 \times 2^{2N}} = \frac{1}{2^N}$. So, $\text{Pr}(\bra{z}C^{
    \dagger}_1\cdots C^\dagger_{t} P_\alpha C_t \cdots C_1\ket{0} \braket{0|C_1\cdots C_{t'} P_\beta C^\dagger_{t'} \cdots C^\dagger_1|z} = 0) =1-\frac{1}{2^N}$. $\qed$

Based on these results, we can now establish the derivatives 
\begin{align}
    \partial_{\theta_\alpha(t)} p(z|\vec{\theta}, \vec{\gamma})|_{\vec{\theta} = 0,
    \vec{\gamma}=0}  = 0, \label{eq:cliffincoh1st} \\
    \partial_{\theta_\alpha(t), \theta(t')_\beta} p(z|\vec{\theta}, \vec{\gamma})|_{\vec{\theta} = 0, \vec{\gamma}=0} &= \big(- 2|\braket{0|z}|^2 + 2|\braket{0|C_1\cdots C_{t} P_\alpha C^\dagger_{t} \cdots C^\dagger_1|z}|^2\big)\delta_{\alpha, \beta}\delta_{t, t'} \label{eq:cliffincoh2nd}
\end{align}
with exponentially high probability. Specifically, the first equation comes using Lemma \ref{lem:cliffz1} with Eq. \eqref{eq:cliffzforder}, while the second comes from applying Lemmas \ref{lem:cliffz2} and \ref{lem:cliffz3} to Eq. \eqref{eq:cliffzsorder}.

Defining
\begin{align}\label{eq:kalphacliff}
    k_{\alpha, t}(z) &\equiv \lvert \bra{z} C_1^\dagger...C^\dagger_{t} P_\alpha  C_{t}... C_1 \ket{0}\rvert^2, \\
    p_0(z) &\equiv |\braket{0|z}|^2,
\end{align} we see that we can approximate $p(z|\vec{\theta}, \vec{\gamma})$ perturbatively as
\begin{align}
    p(z|\vec{\theta}, \vec{\gamma}) &\approx A\big(p_0(z) + \sum_{\beta, t}\frac{\gamma_\beta(t)}{1-\gamma_\beta(t)}k_{\beta, t}(z) + \sum_{\alpha , t}\theta_\alpha(t)^2 k_{\alpha, t}(z)\big) \\
    &\approx A\big(p_0(z) + \sum_{\beta, t}\frac{\gamma_\beta(t)}{1-\gamma_\beta(t)}k_{\beta, t}(z)\big),\label{eq:cliffpincoh}
\end{align}
where we drop higher-order terms in $\theta^2_\alpha$ and  
\begin{align}\label{eq:Acliff}
    A = \prod_{\alpha, t}\cos^2(\theta_\alpha(t))\prod_{\beta, t'}(1-\gamma_\beta(t'))
\end{align} is the probability that no signal occurs. Note that in our expansion above, $A \approx 1 - \sum_{\alpha, t} \theta_{\alpha}(t)^2 - \sum_{\beta, t} \gamma_\beta(t)$. The first term in Eq. \eqref{eq:cliffincoh2nd}, for example, contributes to $Ap_0(z)$. For a more extensive discussion of the renormalization $A$, see Section \ref{sec:quadram}. Here, we have assumed coherent signals couple to the system via operators 
$\{P_\alpha\}$, while  incoherent signals couple to the system via different operators $\{P_\beta\}$ for simplicity. However, our analysis can be easily extended to when they are present with the same Pauli generators (see Sec. \ref{sec:ovlapcliff}).

We also establish that $k_{\beta, t}(z)$ is supported on a single bitstring, as stated in the main text. This property will play an important role in demonstrating the robustness of our approach.
\begin{theorem}\label{thm:thm1}
    For any set Clifford circuits $\{C_t\}_{t = 1}^T$ drawn uniformly at random and a Pauli operator $P_\alpha$, $$k_{\alpha, t}(z) = \lvert \bra{z} C_1^\dagger...C^\dagger_{t} P_\alpha  C_{t}... C_1 \ket{0}\rvert^2$$ is supported on a single bitstring.
\end{theorem}
\textit{Proof---} $C_1^\dagger...C^\dagger_{t} P_\alpha  C_{t}... C_1 \ket{0} \propto \ket{z'}$ for some bitstring $z'$. Thus, $\lvert \bra{z} C_1^\dagger...C^\dagger_{t} P_\alpha  C_{t}... C_1 \ket{0}\rvert^2 = \delta_{z, z'}$. This is a distribution which is supported only on $z'$, $k_{\alpha, t}(z') = 1$. $\qed$

Now, if each $k_{\beta, t}(z)$ are supported on different bitstrings, then it is easy to see from Eq. \eqref{eq:cliffpincoh} that all signals $\gamma_\beta(t)$ will be distinguishable from the measured distribution. Indeed, this is the same principle we used in quadratic sensing (Section \ref{sec:quadram}). For a single set of Clifford operators $\{C_t\}^T_{t=1}$ on an $N$-qubit system, we can constrain the probability of failure of our protocol---when two or more signals map to the same bitstring. The following theorem shows that a successful protocol requires $N = O(\log K_{ic}^2)$.

\begin{theorem}
For a single set of Clifford operators $\{C_t\}_{t=1}^T$ drawn uniformly at random and $K_{ic} > 1$ incoherent Pauli signals on an $N-$qubit system, the probability of a collision (when two or more signals are indistinguishable), defined as
\[
\mathsf{Coll}:=\Big\{\exists\,(\alpha,t)\neq(\beta,t')\ \text{such that}\ k_{\alpha,t}(z)=k_{\beta,t'}(z)  \,\forall z\Big\},
\]
satisfies
\[
\Pr(\mathsf{Coll}) \;\le\; \binom{K_{ic}}{2}\,2^{-N}
\;\le\; \frac{(K_{ic})^2}{2^{N+1}}.
\]
\end{theorem}

\textit{Proof---}
For any fixed pair $(\alpha,t)\neq(\beta,t')$, uniform randomness of the support of $k_{\alpha,t}(z)$ and $k_{\beta,t'}(z)$  implies
\[
\Pr\!\left[k_{\alpha,t}(z)=k_{\beta,t'}(z)\right]=2^{-N}.
\]
Applying a union bound over all $\binom{K_{ic}}{2}$ distinct pairs yields
\[
\Pr(\mathsf{Coll})
\le \sum_{(\alpha,t)\neq(\beta,t')} \Pr\!\left[k_{\alpha,t}(z)=k_{\beta,t'}(z)\right]
= \binom{K_{ic}}{2}\,2^{-N},
\]
which implies the stated bound. $\qed$

This means that to guarantee success probability at least $1-\delta$ (i.e., $\Pr(\mathsf{Coll})\le \delta$),
it suffices to choose $N$ such that
\[
\binom{K_{ic}}{2}\,2^{-N}\le \delta,
\qquad\text{e.g.,}\qquad
N \ge \log\bigg(\frac{K_{ic}(K_{ic}-1)}{2\delta}\bigg).
\]
In particular, achieving success probability $1-\delta$ requires $N$
to grow logarithmically with $K_{ic}^2$.


Although we can suppress the probability of error by increasing the system size $N$, it is also useful to be able to ensure accuracy at a fixed  system size. We therefore allow selecting multiple sets of Clifford circuits to run in our protocol. Theorem~\ref{thm:failurencincoh} shows that, in this case, a sucessful protocol requires $n_c = O(\log(K_{ic}^2)/N)$ sets of Clifford circuits.

\begin{theorem}\label{thm:failurencincoh}
Fix $n_c$ sets of Clifford circuits drawn uniformly randomly, denoted
$\mathrm{C}_n \equiv \{C^{(n)}_t\}_{t=1}^T$ for $n=1,\dots,n_c$. Let $k^{(n)}_{\alpha,t}(z)$ be defined as in Eq.~\eqref{eq:kalphacliff} calculated for the circuit $\mathrm{C}_n$. Then, the probability of a collision over all $n_c$ circuits, such that there exists
$(\alpha, t) \neq (\beta, t')$ such that
\[
k^{(n)}_{\alpha,t}(z)=k^{(n)}_{\beta,t'}(z)
\quad \text{for all } n\in\{1,\dots,n_c\},
\]
is bounded as
\begin{equation}
\Pr\!\left[
\exists\;(\alpha, t) \neq (\beta, t')
\text{ s.t. }
k^{(n)}_{\alpha,t}(z)=k^{(n)}_{\beta,t'}(z)\ \forall n
\right]
\;\le\;
\binom{K_{ic}}{2}\,2^{-N n_c}
\;=\;
\frac{K_{ic}(K_{ic}-1)}{2}\,2^{-N n_c}.
\end{equation}
\end{theorem}

\textit{Proof---} For $n_c$ circuits, failure occurs when two or more signals are indistinguishable in every circuit. For any given two signals $\gamma_{\alpha}(t)$, $\gamma_{\beta}(t)$, the probability that they are indistinguishable in a single circuit is $\frac{1}{2^N}$. Thus, the probability $\Pr(\text{failure})$ that there is indistinguishability across any two signals for all $n_c$ circuits is $\Pr(\text{failure}) \leq \begin{pmatrix}
    K_{ic} \\2
\end{pmatrix} \frac{1}{2^{Nn_c}}$ = $\frac{K_{ic}(K_{ic}-1)}{2}\frac{1}{2^{Nn_c}}$, where we have used a union bound over all pairs of signals. $\qed$

This implies that to guarantee success probability at least $1-\delta$, it suffices to choose $n_c$ such that 
\begin{align}
    \frac{K_{ic}(K_{ic}-1)}{2}\,2^{-N n_c} \leq \delta, \qquad\text{e.g.,}\qquad
    n_c \ge \log\left(\frac{K_{ic}(K_{ic}-1)}{2\delta}\right)/N.
\end{align}

To control the failure probability $\delta$ of incoherent multiparameter estimation for a given $N$, we thus choose the number of circuits $n_c$ based on Theorem \ref{thm:failurencincoh}. Now, when we run $n_c$ circuits, we will have $n_c$ different relations of the form Eq. \eqref{eq:cliffpincoh}. To construct our estimator for $\gamma_{\beta}(t)$, we simply invert all of these equations simultaneously using linear regression \cite{Hastie2001}.

Specifically, we think of $p^{(n)}_0(z)$ and $k^{(n)}_{\beta, t}(z)$ each as $2^N$ dimensional vectors over bitstrings $z$, where the superscript indicates the circuit iteration considered. Note that for every circuit $\mathrm{C}_n$, $p^{(n)}_0(0)=1$. We then concatenate of these vectors vertically across circuits to form $n_c \times 2^N$ dimensional vectors:
\begin{align}\label{eq:kall}
    k^{all}_{\beta,t} (\tilde z) \equiv \begin{bmatrix} k^{(1)}_{\beta, t} (z^{(1)}) \\ \vdots \\k^{(n_c)}_{\beta,t} (z^{(n_c)}) \end{bmatrix}, \,\, p^{all}_{0}(\tilde z) \equiv \begin{bmatrix} p^{(1)}_{0}(z^{(1)}) \\ \vdots \\ p^{(n_c)}_{0}(z^{(n_c)})\end{bmatrix}.
\end{align}
Here, $z^{(n)}$ indicates the bitstring $z$ drawn from the circuit $\mathrm{C}_n$, while  $\tilde z$ represents bitstrings from any circuit. After forming these super vectors, we can concatenate them horizontally to form a $n_c 2^N \times (K_{ic} +1)$ dimensional matrix $V$. Here, $K_{ic}$ is the number of incoherent signals $\gamma_\beta(t)$. The columns of $V$ are given by $V_{\tilde z, 0}=p^{all}_0(\tilde z)$, and $V_{z, \beta t} = k^{all}_{\beta, t}(\tilde z)$,
where we combine $\beta$ and $t$ into a single index.
Then, we see that $p(\tilde z|\vec{\theta}, \vec{\gamma}) = AV_{\tilde z, 0} +A\sum_{\beta t}V_{\tilde z, \beta t}\frac{\gamma_\beta(t)}{1-\gamma_\beta(t)}$,  where we implicitly assume $\beta t$ runs from $1$ to $K_{ic}$.
An estimator for $\gamma_{\beta}(t)$ can then be obtained by inverting this equation using least-squares regression, replacing $p(\tilde z|\vec{\theta}, \vec{\gamma})$ with the empirical distribution: 
\begin{align}\label{eq:estincoh1cliff}
    v_{{ic},{\beta t}}= \sum_{i, \tilde z}(V^TV)^{-1}_{\beta t, i} V^{T}_{i, \tilde z}\frac{\hat N_{\tilde z}}{M/n_c}, \;\; \hat \gamma_\beta(t) = \frac{v_{{ic},{\beta t}}}{v_{{ic},{\beta t}} + v_{{ic},{0}}}, \;\;  \hat A = v_{ic, 0}
\end{align}
$\hat N_{\tilde z}$ is again formed by concatenating across circuits,
\begin{align}\label{eq:ntild}
    \hat N_{\tilde z} = \begin{bmatrix}
        \hat N_{z^{(1)}} \\
        \vdots \\
         \hat N_{z^{(n_c)}}
    \end{bmatrix},
\end{align}
where $\hat N_{z^{(n)}}$ indicates the number of times bitstring $z$ is measured for circuit $n$. So, for instance, $\sum_{\tilde z}k^{all}_{\beta ,t}(\tilde z)\frac{\hat N_{\tilde z}}{M/n_c} = \sum_{z, n}k^{(n)}_{\beta, t}(z)\frac{\hat N_{z^{(n)}}}{M/n_c}$. Note that we sample $\frac{M}{n_c}$ times for each circuit, for a total of $M$ samples. This estimator is unbiased up to higher-order corrections in the signals, and for the rest of our analysis, we assume $\hat \gamma_\beta(t)$ is indeed unbiased, instead operating in the regime where statistical error dominates over bias---relevant for small $M$. Corrections to this assumption can be obtained by including contributions from both bias and variance in the error of the estimator, and further discussion of the behavior of the bias is given in Sec.~\ref{sec:biascliff}.

Now, although our estimator is defined in terms of a $n_c 2^N \times (K_{ic}+1)$ dimensional matrix $V$, we do not ever need to construct this matrix in practice. Instead, we can directly compute the elements of the $(K_{ic} + 1) \times (K_{ic} + 1)$ matrix $V^TV$, which can be done efficiently. Consequently, the overall data-analysis procedure is polynomial in all relevant parameters, as shown in Theorem ~\ref{thm:gcliffcompincoh}.

\begin{theorem}\label{thm:gcliffcompincoh}
    The classical complexity of computing $\hat \gamma_\beta (t)$ (Eq. \eqref{eq:estincoh1cliff}) is $O(K_{ic}n_cN^2 +K_{ic}^2n_c + K_{ic}^3  +  K_{ic} M)$.
\end{theorem}
\textit{Proof. } We first compute each of the $n_c\times K_{ic}$ perturbations $k^{(n)}_{\beta, t}(z)$. By Theorem \ref{thm:thm1}, each is supported on a single bitstring, so we just need to compute $C_1^\dagger...C^\dagger_{t} P_\beta  C_{t}... C_1 $ in the stabilizer formalism and find the non-$Z, I$ term. This costs $n_c\times K_{ic} \times N^2$, ignoring $\log N$ factors for search. 

We then construct $V^TV$. Each entry of $V^TV$ is given by the inner product of two sparse vectors with $n_c$ nonzero entries. Thus, this costs $n_c(K_{ic} +1)^2$. Inverting the $(K_{ic} +1)\times (K_{ic} +1)$ dimensional matrix $V^TV$ costs $(K_{ic} +1)^3$.

Finally, we only compute $V^{T}_{i, \tilde z}\frac{\hat N_{\tilde z}}{M/n_c}$ for each $\tilde z$ that we have sampled. As we have $M$ total samples and  $K_{ic} + 1$ total rows $i$ of $V^T$, this costs time $ K_{ic} M$. $\qed$

\noindent\textbf{Coherent signals---} Recall that to determine 
coherent signals using the Clifford procedure, we must measure in the $x-$basis at the end. This ensures that the first-order perturbation in $\theta_\alpha(t)$ is not zero, as is the case in the incoherent protocol (c.f. Eq. \eqref{eq:cliffincoh1st}). This results in first-order terms 
\begin{align}
    \partial_{\theta_\alpha(t)} p(z|\vec{\theta}, \vec{\gamma})|_{\vec{\theta} = 0, \vec{\gamma}=0} &= 2\im\big[\bra{z}\mathrm{H}^{\otimes N}C^{
    \dagger}_1\cdots C^{\dagger}_{t}P_\alpha C_t \cdots C_1\ket{0} \braket{0|\mathrm{H}^{\otimes N}|z}\big] \equiv \delta p_{\alpha, t}(z) \label{eq:cliffcohforder1}
    \\
     \partial_{\gamma_\alpha(t)} p(z|\vec{\theta}, \vec{\gamma})|_{\vec{\theta} = 0, \vec{\gamma}=0} & = \lvert \bra{z} \mathrm{H}^{\otimes N}C_1^\dagger...C^\dagger_{t} P_\alpha  C_{t}... C_1 \ket{0}\rvert^2 - \lvert \braket{z|\mathrm{H}^{\otimes N}|0}\rvert^2 = 0. \label{eq:cliffcohforder2}
\end{align}
As before, we can examine the structure of these terms. We can rewrite Eq. \eqref{eq:cliffcohforder2} as $\lvert \bra{z} \mathrm{H}^{\otimes N}\ket{z'}\rvert^2 - \lvert \braket{z|\mathrm{H}^{\otimes N}|0}\rvert^2$, where $C_1^\dagger...C^\dagger_{t} P_\alpha  C_{t}... C_1 \ket{0} \propto \ket{z'}$. In this form, we see that both terms will lead to a uniform distribution over all bitstrings $z$, so they cancel to zero: the coherent protocol is insensitive to incoherent signals to first order. Meanwhile, Eq. \eqref{eq:cliffcohforder1} may sometimes evaluate to zero, as we show below.
\begin{theorem}\label{thm:cliffcohdp}
    For a set of Clifford circuits $\{C_t\}^T_{t=1}$ selected uniformly randomly and a fixed signal indicated by $(\alpha, t)$, $\delta p_{\alpha, t}(z) = 0$ for all bitstrings $z$  with probability $p = \frac{1}{2}$. 
\end{theorem}
\textit{Proof---} We want to analyze when $\bra{z}\mathrm{H}^{\otimes N}C^{
    \dagger}_1\cdots C^{\dagger}_{t}P_\alpha C_t \cdots C_1\ket{0} \braket{0|\mathrm{H}^{\otimes N}|z}$ is completely real. Now, we denote $\tilde P_\alpha \equiv C^{
    \dagger}_1\cdots C^{\dagger}_{t}P_\alpha C_t \cdots C_1 $. We thus want to know when
    $\bra{z}\mathrm{H}^{\otimes N}\tilde P_\alpha\ket{0} \braket{0|\mathrm{H}^{\otimes N}|z}$ is real. $ \braket{0|\mathrm{H}^{\otimes N}|z}$ is always real, so it depends on $\bra{z}\mathrm{H}^{\otimes N}\tilde P_\alpha\ket{0}$. Now,  $ \braket{0|\mathrm{H}^{\otimes N}\tilde P_\alpha|z}$ is real when $\tilde P_\alpha$ contains an odd number of Pauli $Y$. This happens with probability $\frac{1}{2}$ in the limit of large $N$. $\qed$

Thus, we can approximate the output of the coherent estimation protocol for a single circuit as
\begin{align}\label{eq:cliffpcoh}
    p(z|\vec{\theta}, \vec{\gamma}) \approx p_0(z) + A \sum_{\alpha, t}\theta_\alpha(t)\delta p_{\alpha, t}(z),
\end{align}
where $A = \prod_{\alpha, t}\cos^2(\theta_\alpha(t))\prod_{\beta, t'}(1-\gamma_\beta(t'))$ and  $p_0(z)$ is the uniform distribution over all bitstrings $z$. We note that $A$ does not renormalize the coefficient of $p_0(z)$ here as the terms which contribute to this renormalization cancel (see Eq. \eqref{eq:cliffcohforder2}, which is equal to the term $\propto \theta^2_\alpha$ and vanishes). 

Now, due to Theorem \ref{thm:cliffcohdp}, we cannot just take one set of Clifford operators $\{C_t\}^T_{t=1}$, as there is a probability of $\frac{1}{2}$ for each signal that they have $\delta p_{\alpha, t}(z) = 0$---the circuit is not sensitive to the signal. To rectify this issue, we again consider multiple sets of Clifford circuits. Theorem~\ref{thm:failurecoh} shows that a successful protocol requires $n_c = O(\log K_c)$ sets of Clifford circuits.

\begin{theorem}\label{thm:failurecoh}
Fix $n_c$ sets of Clifford circuits drawn uniformly randomly, denoted
$\mathrm{C}_n \equiv \{C^{(n)}_t\}_{t=1}^T$ for $n=1,\dots,n_c$.
Let $\delta p^{(n)}_{\alpha,t}(z)$ be defined as in
Eq.~\eqref{eq:cliffcohforder1}, calculated for the circuit $\mathrm{C}_n$.
Then the probability that there exists ($\alpha$, $t$) such that
\[
\delta p^{(n)}_{\alpha,t}(z)=0
\quad \text{for all } n\in\{1,\dots,n_c\},
\]
is given by
\begin{equation}
\Pr\!\left[
\exists\;\alpha,t\ \text{s.t.}\ 
\delta p^{(n)}_{\alpha,t}(z)=0\ \forall n
\right]
=
1-\big(1-2^{-n_c}\big)^{K_c},
\end{equation}
where $K_c$ is the total number of coherent signals $\theta_\alpha(t)$.
\end{theorem}

\textit{Proof.}
Failure occurs if there exists a coherent signal $\theta_\alpha(t)$ whose
first-order contribution $\delta p^{(n)}_{\alpha,t}(z)$ vanishes for every
one of the $n_c$ circuits.
For a fixed signal $\alpha$ and time $t$, the probability that
$\delta p^{(n)}_{\alpha,t}(z)=0$ for a single circuit is $1/2$.
Independence of the circuits implies that the probability this occurs for
all $n_c$ circuits is $2^{-n_c}$.

Therefore, the probability that a given signal is detected (i.e., is
nonzero for at least one circuit) is $1-2^{-n_c}$.
Requiring this to hold for all $K_c$ coherent signals yields
\[
\Pr(\text{success}) = \big(1-2^{-n_c}\big)^{K_c},
\]
and hence the failure probability is
\[
\Pr(\text{failure}) = 1-\big(1-2^{-n_c}\big)^{K_c}.
\] \hfill$\square$

This implies that to guarantee a success probability of at least $1-\delta$, it suffices to
choose $n_c$ such that
\[
1-\big(1-2^{-n_c}\big)^{K_c} \le \delta.
\]
For small $\delta$, this condition implies that we need to take
\[
n_c \gtrsim \log_2\!\left(\frac{K_c}{\delta}\right),
\]
so that the number of circuit sets required grows logarithmically with
the number of coherent signals.


As before, we can control the failure probability by choosing $n_c$ according to Theorem \ref{thm:failurecoh}. When we run $n_c$ circuits, we will again have $n_c$ relations like Eq. \eqref{eq:cliffpcoh}. We can similarly invert them to obtain our estimator for $\theta_\alpha(t)$. This procedure is almost identical to the discussion surrounding Eq. \eqref{eq:kall}, so we shorten the explanation here for brevity. As in Eq. \eqref{eq:kall}, we can concatenate our $\delta p^{n}_{\alpha, t}(z)$ and $p_0^{(n)}(z)$ for each circuit $n$ vertically, to make $n_c\times 2^N$ dimensional vectors $p_0^{all}(\tilde z)$, $\delta p^{all}_{\alpha, t}(\tilde z)$. In this case, $p_0^{(n)}(z)$ is the uniform distribution over all bitstrings $z$ for each circuit $n$. We then join them together horizontally to make a $n_c 2^N \times (K_c + 1)$ dimensional matrix $V$. The columns of $V$ are $V_{\tilde z, 0}=p^{all}_0(\tilde z)$, and $V_{z, \alpha t} = \delta p^{all}_{\alpha, t}(\tilde z)$. Then, we see that $p(\tilde z|\vec{\theta}, \vec{\gamma}) = V_{\tilde z, 0} +A\sum_{\alpha t}V_{\tilde z, \alpha t}\theta_\alpha(t)$,  where we implicitly assume $\alpha t$ runs from $1$ to $K_{c}$. An estimator for $\theta_{\alpha}(t)$ can then be obtained by inverting this equation approximately: 
\begin{align}\label{eq:estcohcliff}
    \hat \theta_\alpha(t) = \frac{1}{\hat A}\sum_{i, \tilde z}(V^TV)^{-1}_{\alpha t, i} V^{T}_{i, \tilde z}\frac{\hat N_{\tilde z}}{M/n_c}
\end{align}
where $\hat A$ is obtained from the incoherent estimator Eq. \eqref{eq:estincoh1cliff} and $\hat N_{\tilde z}$ is defined in Eq. \eqref{eq:ntild}. This estimator is unbiased up to higher-order corrections in the signals, and for the rest of our analysis, we assume $\hat \theta_\alpha(t)$ is indeed unbiased, instead working in the regime where statistical error dominates. Corrections to this assumption can be obtained by including contributions from both bias and variance in the error of the estimator, and further discussion of the behavior of the bias is given in Sec.~\ref{sec:biascliff}.

Now, although our estimator is defined in terms of an $n_c 2^N \times (K_c +1)$ dimensional matrix $V$, we do not ever need to construct this matrix in practice. Instead, we can directly compute the entries of the $(K_{c} + 1) \times (K_{c} + 1)$ matrix $V^TV$. To do so, we first establish several properties of $\delta p^{(n)}_{\alpha, t}(z)$.  In Lemma~\ref{lem:imsign}, we show that $\delta p^{(n)}_{\alpha, t}(z)$ is negative for half of all bitstrings $z$. Lemma~\ref{lem:imsigndot} then characterizes the product $\delta p^{(n)}_{\alpha, t}(z)\delta p^{(n)}_{\beta, t'}(z)$. These results let us evaluate entries of $V^TV$, which take the form $\sum_{z, n} p^{(n)}_0(z) \delta p^{(n)}_{\alpha, t}(z)$ and $\sum_{z, n} \delta p^{(n)}_{\alpha, t}(z) \delta p^{(n)}_{\beta, t'}(z)$. Finally, in Theorem~\ref{thm:gcliffcompcoh}, we show that $V^TV$, and hence our estimator $\hat \theta_\alpha(t)$, can be constructed efficiently.

\begin{lemma}\label{lem:imsign}
    Assuming $\delta p^{(n)}_{\alpha, t}(z)$  $\neq 0$, the number of bitstrings $z$ where $\delta p^{(n)}_{\alpha, t}(z)$ is negative $|\{z\; |\; \sgn(\delta p^{(n)}_{\alpha, t}(z)) = -1\}| = 2^{N-1}$.
\end{lemma}
\textit{Proof---} Recall that $\delta p^{(n)}_{\alpha, t}(z) = 2\im\big[\bra{z}\mathrm{H}^{\otimes N}(C^{(n)}_{1})^{\dagger}\cdots (C^{(n)}_{t})^{\dagger}P_\alpha C^{(n)}_{t} \cdots C^{(n)}_1\ket{0} \braket{0|\mathrm{H}^{\otimes N}|z}\big]$. Now, we first analyze the sign of the $\im[]$ term.
Now, let $(C^{(n)}_{1})^{\dagger}\cdots (C^{(n)}_{t})^{\dagger}P_\alpha C^{(n)}_{t} \cdots C^{(n)}_1=\phi P'_\alpha$, where $\phi = \pm 1$. We want to examine the sign of
\begin{align} \label{eq:imsign}
    \phi \bra{z}\mathrm{H}^{\otimes N}P'_\alpha \ket{0} \braket{0|\mathrm{H}^{\otimes N}|z}.
\end{align}
By Theorem \ref{thm:cliffcohdp}, we know if $\delta p^{(n)}_{\alpha, t}(z) \neq 0$, then this is precisely the sign of $\delta p^{(n)}_{\alpha, t}(z)$.

To proceed, let $a = a_1 \cdots a_N$ be a bitstring defined as follows:
\begin{align}\label{eq:astab}
a_i =
\begin{cases}
1, & \text{if } (P'_\alpha)_i \in \{X, Y\} \\
0, & \text{if } (P'_\alpha)_i \in \{Z, I\}.
\end{cases}
\end{align}
In other words, $a$ is a bitstring denoting the positions where $P'_\alpha$ is either $X$ or $Y$. From examining Eq. \eqref{eq:imsign}, it is straightforward the see that $\sgn( \phi \bra{z}\mathrm{H}^{\otimes N}P'_\alpha \ket{0} \braket{0|\mathrm{H}^{\otimes N}|z}) = \sgn(\phi (-1)^{n_{z, a}} i^{n_y})$, where $n_{z, a} = \text{wt}(z\odot a)$ and $n_y$ is the number of Pauli $Y$ in $P'_\alpha$. As $\phi$ and $i^{n_y}$ are constant across all bitstrings $z$, we just need to show that $(-1)^{n_{z, a}}$ is negative on $2^{N-1}$ bitstrings and positive on the other $2^{N-1}$ bitstrings in order to prove our statement. This is precisely shown in Lemma \ref{lem:lindpa}. $\qed$

From our above discussion, we can also conclude the following formula,
\begin{align}
    \delta p^{(n)}_{\alpha, t}(z) = \frac{\im[\phi (-1)^{n_{z, a}} i^{n_y}]}{2^{N-1}},
\end{align}
where the definitions of $\phi$, $n_{z, a}$, and $n_y$ can be found in the above Lemma. Now, we establish some more properties of $\delta p^{(n)}_{\alpha, t}(z)$ that allow us to analyze the entries of $V^TV$.
\begin{lemma}\label{lem:imsigndot}
    Consider
    \begin{align*}
        \delta p^{(n)}_{\alpha, t}(z) &=2\im\big[\bra{z}\mathrm{H}^{\otimes N}(C^{(n)}_{1})^{\dagger}\cdots (C^{(n)}_{t})^{\dagger}P_\alpha C^{(n)}_{t} \cdots C^{(n)}_1\ket{0} \braket{0|\mathrm{H}^{\otimes N}|z}\big] \\
        &= 2 \im[\phi \bra{z}\mathrm{H}^{\otimes N}P'_\alpha \ket{0}\braket{0|\mathrm{H}^{\otimes N}|z}]
    \end{align*} 
    and
    \begin{align*}
        \delta p^{(n)}_{\beta, t'}(z) &=2\im\big[\bra{z}\mathrm{H}^{\otimes N}(C^{(n)}_{1})^{\dagger}\cdots (C^{(n)}_{t'})^{\dagger}P_\beta C^{(n)}_{t'} \cdots C^{(n)}_1\ket{0} \braket{0|\mathrm{H}^{\otimes N}|z}\big] \\
        &= 2 \im[\varphi \bra{z}\mathrm{H}^{\otimes N}P'_\beta \ket{0}\braket{0|\mathrm{H}^{\otimes N}|z}],
    \end{align*} 
    where $\phi, \varphi \in [\pm1]$ and $  (C^{(n)}_{1})^{\dagger}\cdots (C^{(n)}_{t(t')})^{\dagger}P_{\alpha(\beta)} C^{(n)}_{t(t')} \cdots C^{(n)}_1=\phi (\varphi)\times P'_{\alpha (\beta)}$. Let $a$ be the bitstring indicating the positions of $X$ and $Y$ in $P'_\alpha$, and $a'$ analogously for $P_\beta$ (Eq. \eqref{eq:astab}). Finally let $n_y(n'_y)$ be the number of Pauli $Y$s in $P_{\alpha (\beta)}$. The number of bitstrings $z$ where $\delta p^{(n)}_{\alpha, t}(z) \delta p^{(n)}_{\beta, t'}(z)$ is negative is 
    \begin{align}
        |\{z\; |\; \sgn(\delta p^{(n)}_{\alpha, t}(z) \delta p^{(n)}_{\beta, t'}(z)) = -1\}|=
        \begin{cases}
        2^{N-1} & \text{when } a\neq a', \\ 0 & \text{when } a=  a' \text{ and } \sgn(i^{n_y}\phi) = \sgn(i^{n'_y}\varphi)\\
        2^N, & \text{when }  a=  a' \text{ and } \sgn(i^{n_y} \phi) = -\sgn(i^{n'_y} \varphi).
   \end{cases}
    \end{align}
\end{lemma}
\textit{Proof---} We first consider the case when $a \neq a'$. By the logic of Lemma \ref{lem:imsign}, $\sgn(\delta p^{(n)}_{\alpha, t}(z) \delta p^{(n)}_{\beta, t'}(z))$ is determined by $\phi\varphi(-1)^{n_{z, a}+n_{z, a'}}i^{n_y + n'_y}$, where $n_{z, a} = \text{wt}(z\odot a)$. Now, as $i^{n_y + n'_y}\phi\varphi$ is constant across all bitstrings, we just need to show that $(-1)^{n_{z, a}+n_{z, a'}}$ switches sign on half of all bitstrings. Indeed, by Lemma \ref{lem:lindpab}, if $a \neq a'$, $\sgn((-1)^{n_{z, a}+n_{z, a'}}) = -1$ on $2^{N-1}$ bitstrings and $=1$ on the rest.

We now consider when $a = a'$ and $\sgn(i^{n_y} \phi) = \sgn(i^{n'_y} \varphi)$. This means that $n_{z, a}  = n_{z, a'}$. So, $\sgn(\delta p^{(n)}_{\alpha, t}(z) \delta p^{(n)}_{\beta, t'}(z))$ is determined by $\phi\varphi(-1)^{2n_{z, a}}i^{n_y + n'_y} = i^{n_y + n'_y}\phi\varphi$. By assumption, $\sgn(i^{n_y} \phi) = \sgn(i^{n'_y} \varphi)$, so $\sgn(i^{n_y + n'_y}\phi\varphi)$ $=1$. Thus, $\sgn(\delta p^{(n)}_{\alpha, t}(z) \delta p^{(n)}_{\beta, t'}(z))=1$ for all bitstrings.

Finally, we consider when $ a = a'$ and $\sgn(i^{n_y} \phi) = -\sgn(i^{n'_y} \varphi)$. As before, $\sgn(\delta p^{(n)}_{\alpha, t}(z) \delta p^{(n)}_{\beta, t'}(z))$ is determined by $i^{n_y + n'_y}\phi\varphi$ in this case. By our assumption, $\sgn(i^{n_y + n'_y}\phi\varphi)=-1$. Thus, $\sgn(\delta p^{(n)}_{\alpha, t}(z) \delta p^{(n)}_{\beta, t'}(z))=-1$ for all bitstrings. $\qed$

Using the above results, we can now explicitly compute $V^TV$, and thereby our estimator  $\hat \theta_\alpha (t)$ (Eq. \eqref{eq:estcohcliff}),  in a computationally efficient manner. This is established in the Theorem below.

\begin{theorem}\label{thm:gcliffcompcoh}
   The classical complexity of computing $\hat \theta_\alpha (t)$  (Eq. \eqref{eq:estcohcliff}) is $O(n_cK_cN^2 + n_cK_c^2N +K^3_c + K_c NM)$.
\end{theorem}

\textit{Proof---} We first need to compute $\phi$, $a$, and $n_y$ for each of the $n_c K_c$ perturbations $\delta p^{(n)}_{\alpha, t}(z)$, as defined in Lemma \ref{lem:imsigndot}. This involves evolving a Pauli string by a Clifford in the stabilizer formalism, which takes $O(N^2)$ time, for a total cost of $O(n_c K_c N^2)$.

We then compute the entries of $V^TV$. Now, $(V^TV)_{0, \alpha t}$  = $\sum_{\tilde z}p^{all}_{0}(\tilde z) \delta p^{all}_{\alpha, t}(\tilde z) = \sum_{n, z}p^{(n)}_{0}(z)\delta p^{(n)}_{\alpha, t}(z) = 0$ deterministically by Lemma~\ref{lem:imsign}. Meanwhile, $(V^TV)_{\alpha t, \beta t'}= \sum_{\tilde z}\delta p^{all}_{\alpha, t}(\tilde z) \delta p^{all}_{\beta, t'}(\tilde z) = \sum_{n, z}\delta p^{(n)}_{\alpha, t}(z)\delta p^{(n)}_{\beta, t'}(z)$. Now, $\delta p^{(n)}_{\alpha,t}(z)$ has magnitude $\frac{1}{2^{N-1}}$ for every bitstring $z$, with sign given by $\sgn(\phi (-1)^{n_{z, a}}i^{n_y})$. Thus, by Lemma \ref{lem:imsigndot}
\begin{align}\label{eq:entriescoh}
\sum_{ z}\delta p^{(n)}_{\alpha, t}(z)\delta p^{(n)}_{\beta, t'}(z) =
\begin{cases}
0 & \text{when } a\neq a'\\
        \frac{1}{2^{N-2}}, & \text{when } a= a'\text{ and } \sgn(i^{n_y} \phi) = \sgn(i^{n'_y} \varphi)\\
        -\frac{1}{2^{N-2}}, & \text{when } a= a' \text{ and } \sgn(i^{n_y} \phi) = -\sgn(i^{n'_y} \varphi).
\end{cases}
\end{align}
Computing this for each of $ (K_c+ 1)^2$ pairs costs $O(n_c (K_c+ 1)^2N)$. We note that this implies that $V^TV$ is approximately diagonal, only having off-diagonal terms where the transformed Pauli strings corresponding to signals $\alpha,t$ and $\beta,t'$ have Pauli operators $X$ and $Y$ at the exact same positions.

We then invert $V^TV$, which generally costs $O((K_c+1)^3)$. We must also calculate $V^T_{i, \tilde z}\hat N_{\tilde z}$ for all $K_c+1$ rows $i$ and every sample we measure. This can be done using the formula $\delta p^{(n)}_{\alpha, t}(z) = \frac{1}{2^{N-1}}\im[\phi(-1)^{n_{z, a}}i^{n_y }]$, which costs $O(K_cNM)$. $\qed$

\subsubsection{Invertibility}
Before we analyze the sample complexity of our estimators, we first highlight some important conclusions from our preceding analysis. Specifically, we justify why we expect $V^TV$ to be invertible, even with exponentially many signals.

First, we have calculated the entries of $V^TV$ for both incoherent and coherent signals, where the matrix $V$ is defined above Eq. \eqref{eq:estincoh1cliff} for incoherent signals and above Eq. \eqref{eq:estcohcliff} for coherent signals. Now, in both cases, our results show that $V^TV$ is approximately a diagonal matrix.

Specifically, for incoherent signals, the entries of $V^TV$ are given by:

\begin{align}
    (V^TV)_{0, 0} &= \sum_{n, z}(p_0^{(n)}(z))^2 = n_c, \\
    (V^TV)_{0, \beta t} &= \sum_{n, z}p_0^{(n)}(z) k_{\beta, t}^{(n)}(z) \approx 0, \\
    (V^TV)_{\alpha t', \beta t} &= \sum_{n, z}k_{\alpha, t'}^{(n)}(z) k_{\beta, t}^{(n)}(z) \approx n_c \delta_{\alpha,\beta} \delta_{t, t'}.
\end{align}
In all entries, $n$ runs from $1$ to $n_c$ and $z$ across all $2^N$ bitstrings. In the first equation, we have used the fact that $p_0^{(n)}(z) = \delta_{z, 0}$. In the second, we note that each $(V^TV)_{0, \beta t}$ is only nonzero if the signal $\gamma_\beta (t)$ maps exactly to $z=0$ in the sense of Theorem \ref{thm:thm1} for any circuit. This happens with exponentially small probability $\frac{n_c}{2^N}$, so we approximate $(V^TV)_{0, \beta t} \approx 0$. Similarly, in the third equation, when $\gamma_\alpha(t') = \gamma_\beta(t)$, then $k_{\alpha ,t'}^{(n)}(z)  = k_{\beta. t}^{(n)}(z)$ are both probability distributions supported on a single bitstring, and $(V^TV)_{\alpha t', \beta t} = n_c$. Otherwise, when $\gamma_\alpha(t') \neq \gamma_\beta(t)$, $(V^TV)_{\alpha t', \beta t} = 0$ with exponentially high probability for the same reason as the second equation.

For coherent signals, the entries of $V^TV$ are given by 

\begin{align}
    (V^TV)_{0, 0} &= \sum_{n, z}(\underbrace{p_0^{(n)}(z))^2}_{=\frac{1}{2^{2N}}} = \frac{n_c}{2^N}, \\
    (V^TV)_{0, \alpha t} &= \sum_{n, z}p_0^{(n)}(z) \delta p_{\alpha, t}^{(n)}(z) = 0, \\
    (V^TV)_{\alpha t', \beta t} &= \sum_{n, z} \delta p_{\alpha, t}^{(n)}(z) \delta p_{\beta, t'}^{(n)}(z) \approx 
    \frac{n_c}{2^{N-1}} \delta_{\alpha,\beta} \delta_{t, t'}.
\end{align}
In the first equation, we have used the fact that for coherent signals, $p_0^{(n)}(z)$ is the uniform distribution overall bitstrings $z$. In the second, we note that each $(V^TV)_{0, \alpha t}$ is zero by Lemma \eqref{eq:imsign}. In the third, when $\theta_\alpha(t') = \theta_\beta(t)$, then $\sum_{z} \delta p_{\alpha, t}^{(n)}(z) \delta p_{\beta, t'}^{(n)}(z) = \frac{1}{2^{N-2}}$ by Theorem \ref{thm:gcliffcompcoh}. Thus, using the fact that approximately $\frac{n_c}{2}$ $\delta p_{\alpha, t}^{(n)}(z)$ are nonzero over all circuits, $(V^TV)_{\alpha t, \alpha t} = \frac{n_c}{2}\frac{1}{2^{N-2}}$. Otherwise, when $\theta_\alpha(t') \neq \theta_\beta(t)$, $(V^TV)_{\alpha t', \beta t}$ is only nonzero when two random Pauli strings corresponding to the two signals have $X$ and $Y$ operators in the exact same qubit positions (see Theorem \ref{thm:gcliffcompcoh}). This happens with exponentially small probability $\frac{n_c}{2^N}$ over all circuits, so we approximate $(V^TV)_{\alpha t', \beta t}\approx 0$. Moreover, if an off-diagonal entry is indeed nonzero, its magnitude would be $a\frac{1}{2^{N-2}}$, where $a$ is the number of circuits in which two different signals happened to collide---this is, on average, much less than $\frac{n_c}{2}$.

Overall, we see that for both types of signal, $V^TV$ is approximately diagonal, with sparse, randomly distributed off-diagonal entries of smaller magnitude. As long as the diagonal entries are bounded away from zero (which we achieve in the coherent case with a sufficiently large $n_c$), we expect $V^TV$ to be invertible~\cite{Cook2018}. This justifies the use of least-squares linear regression (expressions Eq. \eqref{eq:estincoh1cliff} and \eqref{eq:estcohcliff}) for our estimators, which requires that the columns of $V$ are linearly independent and hence $V^TV$ is invertible.

\subsubsection{Typical sample complexity}\label{sec:sccliff}
We now compute the variances $\Var(\hat \gamma_{\beta}(t))$ and $\Var(\hat \theta_{\alpha}(t))$, with results given in  Eq. \eqref{eq:varcliffincoh} and Eq. \eqref{eq:varcliffcoh} and stated in Property \ref{num:gcliff_sm}.

\noindent\textbf{Incoherent Signals---} 
$V^TV$, as defined above in Eq.~\eqref{eq:estincoh1cliff}, is approximately a diagonal matrix, with off-diagonal entries nonzero only when two signals map to the same bitstring in a circuit, in the sense of Theorem \ref{thm:thm1}. Thus, for our calculation here, we approximate $V^TV = \text{diag}(\frac{1}{n_c}, \cdots, \frac{1}{n_c}) $. Then,
\begin{align}
    v_{ic,0} = \frac{1}{n_c} \sum_{z, n}p_0^{(n)}(z)\frac{\hat N_{z^{(n)}}}{M/n_c} = \sum_{\tilde z}p_0^{all}(\tilde z)\frac{\hat N_{\tilde z}}{M} ,\;\;v_{ic, \beta t} = \frac{1}{n_c} \sum_{z, n}k^{(n)}_{\beta t}(z)\frac{\hat N_{z^{(n)}}}{M/n_c}=\sum_{z}k^{(n)}_{\beta t}(\tilde z)\frac{\hat N_{\tilde z}}{M},
\end{align}
where $\hat N_{z^{(n)}}$ are the counts for bitstring $z$ measured for the $n$th circuit and $\hat N_{\tilde z}$ is defined in Eq. ~\eqref{eq:ntild}. We can use Eq. \eqref{eq:wmultvar} to calculate the variance of this, along with some modifications:
\begin{align}
    \Var( v_{ic, \beta t}) &\approx \frac{1}{M} \sum_{z, n}(k^{(n)}_{\beta t}(z))^2\frac{p^{(n)}(z|\vec{\theta}, \vec{\gamma})}{n_c} \\ &\approx \frac{1}{M}A\underbrace{\sum_{z, n}(k^{(n)}_{\beta t}(z))^2\frac{p^{(n)}_0(z)}{n_c}}_{=0} +\frac{1}{M} A\underbrace{\sum_{z, n}(k^{(n)}_{\beta t}(z))^2\sum_{\alpha t'} \frac{k^{(n)}_{\alpha t'}}{n_c}(z)}_{=\delta_{\alpha t', \beta t}}\frac{\gamma_\alpha(t')}{1-\gamma_\alpha(t')} \\
    &= \frac{A}{M}\frac{\gamma_\beta(t)}{1-\gamma_\beta(t)}.
\end{align}
In the first line, we must consider $\frac{p^{(n)}(z|\vec{\theta}, \vec{\gamma})}{n_c} $ as this is a properly normalized probability distribution: $\sum_{z, n} \frac{p^{(n)}(z|\vec{\theta}, \vec{\gamma})}{n_c}=1$. Moreover, $\hat N_{\tilde z}$ can be considered to be drawn from the distribution $\frac{p^{all}(\tilde z|\vec{\theta}, \vec{\gamma})}{n_c} $ with $M$ samples. We also drop higher-order terms in $\gamma_\beta(t)$. In the second line, we replaced $p^{(n)}(z|\vec{\theta}, \vec{\gamma})$ with Eq. \eqref{eq:cliffpincoh} and used the approximation that each signal maps to a unique bitstring for every circuit.

Thus, 
\begin{align}
    \Var(\hat \gamma_\beta (t))  = \Var\bigg(\frac{v_{ic, \beta t}}{v_{ic, 0} +v_{ic, \beta t}}\bigg) &\approx \frac{\Var(v_{ic, \beta t})}{(A+\frac{A\gamma_\beta(t)}{1-\gamma_\beta(t)})^2}\\
    &= \frac{A}{M}\frac{\gamma_\beta(t)}{1-\gamma_\beta(t)} \times \bigg(\frac{1-\gamma_\beta(t)}{A}\bigg)^2\\
    &\approx \frac{\gamma_\beta(t)}{AM}\label{eq:varcliffincoh}.
\end{align}
In the first line, we have used Eq. \eqref{eq:varfrac} and dropped higher-order terms in $\gamma_\beta (t)$. We also drop higher-order terms in the last line.

\noindent\textbf{Coherent Signals---} 
$V^TV$ here, as defined above Eq. \eqref{eq:estcohcliff}, is also approximately a diagonal matrix: an off-diagonal term is only nonzero when the second or third conditions of Eq. \eqref{eq:entriescoh} are satisfied. This happens if two random Pauli strings happen to have $X$ and $Y$ operators in the exact same qubit positions, which occurs with exponentially small probability. Thus, we approximate $V^TV = \text{diag}(\frac{2^N}{n_c}, \frac{2}{n_c}2^{N-2}, \cdots, \frac{2}{n_c}2^{N-2})$, where we use the fact that approximately $\frac{n_c}{2}$  $\delta p^{n}_{\alpha t}(z)$ are nonzero over all circuits. Then,
\begin{align}
    \hat \theta _\alpha(t) = \frac{2^{N-1}}{\hat A n_c}\sum_{z, n} \delta p^{n}_{\alpha t}(z)\frac{\hat N_{z^{(n)}}}{M/n_c} = \frac{2^{N-1}}{\hat A}\sum_{\tilde z} \delta p^{all}_{\alpha t}(z)\frac{\hat N_{\tilde z}}{M},
\end{align}
where $\hat N_{z^{(n)}}$ is the number of times bitstring $z$ is measured for circuit $n$, and $\hat N_{\tilde z}$ is defined in Eq. ~\eqref{eq:ntild}.
As before, $\hat N_{\tilde z}$ can be considered to be drawn from the distribution $\frac{p^{all}(\tilde z|\vec{\theta}, \vec{\gamma})}{n_c}$ with $M$ samples. Now,
\begin{align}
    \Var(\hat \theta _\alpha(t)) &\approx \frac{2^{2N-2}}{MA^2}\sum_{z, n}(\delta p^{(n)}_{\alpha t}(z))^2\frac{p^{(n)}(z|\vec{\theta}, \vec{\gamma})}{n_c} \\
    &\approx  \frac{2^{2N-2}}{MA^2}\underbrace{\sum_{z, n}(\delta p^{(n)}_{\alpha t}(z))^2\frac{p^{(n)}_0(z)}{n_c}}_{ = \frac{2^N n_c}{2} \times \frac{1}{2^{2N-2}}\times\frac{1}{2^N n_c}=\frac{1}{2^{2N-1}}} \\
    &= \frac{1}{2M A^2} \label{eq:varcliffcoh}.
\end{align}
As before, we used Eqs. \eqref{eq:varfrac} and \eqref{eq:wmultvar} while ignoring higher-order terms in $\theta_\alpha$. 
\subsubsection{Bounds on sample complexity for worst-case error}\label{sec:cliffbounds}
We can also bound the worst-case errors of our estimators $\max_{\beta, t} |\hat \gamma_\beta(t)-\gamma_\beta(t)|$  (given in Theorem \ref{thm:boundcliffincoh})  and $\max_{\alpha, t} |\hat \theta_\alpha(t)-\theta_\alpha(t)|$ (given in Theorem \ref{thm:boundcliffcoh}), thereby justifying Property \ref{num:gcliff_bound}. The proofs of Theorems \ref{thm:boundcliffincoh} and \ref{thm:boundcliffcoh} proceed by applying sub-exponential tail bounds of the Poisson distribution, which can be related to the multinomial statistics of the measured bitstrings, in order to constrain the error of our estimators. 

\begin{theorem}\label{thm:boundcliffincoh}
 Assume the perturbative regime of small signals, such that $A$ (defined in Eq. \eqref{eq:Acliff}) is $O(1)$, and neglect bias.  In the presence of $K_{ic}$ total incoherent signals $\gamma_\beta(t)$, a precision $\max_{\beta, t}|\hat \gamma_{\beta}(t) -\gamma_{\beta}(t)|\leq \epsilon$ for small $\epsilon$ can be achieved with probability $1-\delta$ by taking $M= O(\max\{\frac{\log(\frac{K_{ic}}{\delta})}{\epsilon^2}, \frac{\log(\frac{K_{ic}}{\delta})}{\epsilon}, \frac{\log(\frac{K_{ic}^2}{\delta})\log(\frac{1}{\delta})}{N}\})$, with the first term being dominant for small enough $\epsilon$.
\end{theorem}

\textit{Proof. } We want to bound the error of our estimator Eq. \eqref{eq:estincoh1cliff}. First, we take $n_c = O(\frac{K^2_{ic}}{\delta'}/N)$ large enough such that with high probability $1-\delta'$, the protocol succeeds (Theorem \ref{thm:failurencincoh}). We can rewrite our estimator as
\begin{align}\label{eq:incohrewrite}
     v_{{ic},l}= \sum_{z, n}w_{l,z^{(n)}}\frac{\hat N_{z^{(n)}}}{M/n_c}, \;\; \hat \gamma_\beta(t) = \frac{v_{{ic},{\beta t}}}{v_{{ic},{\beta t}} + v_{{ic},{0}}},
\end{align}
where 
\begin{align}
    w_{l,z^{(n)}} = \sum_{i}(V^TV)_{l, i}V^T_{i, z^{(n)}},
\end{align}
and $z^{(n)}$ indicates bitstring $z$ drawn from the $n$th circuit. We further denote $\mathbb{E}(v_{{ic},l}) = \bar{v}_{{ic},l}$.  

Now, $\hat N_{z^{(n)}} \sim \text{Multinomial}(M/n_c, p^{(n)}(z))$, where we take $M/n_c$ samples deterministically. We can consider an alternative process whereby $X \sim\text{Poisson}(M/n_c)$ samples are taken for each circuit, such that $\hat N_{z^{(n)}} \sim \text{Poisson}(Mp^{(n)}(z)/n_c)$ \cite{Canonne2020}. In this case, $\hat N_{z^{(n)}}$ for different $z$, $n$, are independent random variables. By Lemma \ref{lem:poisbound}, analyzing our procedure in this way incurs at most a constant error in the sample complexity. We therefore proceed by treating our variables as Poisson.

After Poissonization, $w_{l, z^{(n)}} \hat N_{z^{(n)}}$ is a sub-exponential random variable with parameters $\nu^2 = w^2_{l, z^{(n)}} e p^{(n)}(z) \frac{M}{n_c}$ and $\alpha = |w_{l, z^{(n)}}|$ (see Lemma \ref{lem:poissub} and Corollary \ref{cor:rpois}). Using the fact that summations of independent sub-exponential random variables are also sub-exponential, $\sum_{z, n} w_{l, z^{(n)}}\hat N_{z^{(n)}}$ is sub-exponential with effective $\nu_l^2 = \sum_{z, n} w^2_{l, z^{(n)}} e p^{(n)}(z) \frac{M}{n_c}$ and $\alpha_l =\max_{z, n} |w_{l, z^{(n)}}|$ \cite{Wainwright2019}. 

Let us now consider the cases $l= 0$ and $l = \beta t$ seperately. By the discussion in Section \ref{sec:sccliff}, $w_{\beta t, z^{(n)}} \approx \frac{1}{n_c}k^{(n)}_{\beta t}(z)$, where $k^{(n)}_{\beta t}(z)$ is defined for a fixed circuit $n$ in Eq. \eqref{eq:kalphacliff}. Similarly, $w_{\beta t, z^{(n)}} \approx \frac{1}{n_c}p^0_n(z)$. Recall that $p^0_n(z)$ is deterministic, $p^0_n(0)=1$. By Theorem \ref{thm:thm1}, $k^{(n)}_{\beta t}(z)$ is also deterministic: for each $n$, $k^{(n)}_{\beta t}(z^*)=1$ for some $z^*$. Using this information, we see that $\nu_{\beta t}^2 \approx \frac{eM}{n^3_c} \sum_n \gamma_{\beta}(t) =\frac{eM \gamma_{\beta}(t)}{n^2_c}$  and ${\nu_{0}}^2\approx \frac{e MA}{n^2_c}$, where $A$ is defined in Eq. \eqref{eq:Acliff}. Thus, we can write
\begin{align}
    \nu_l^2 =  \frac{M\mathcal{C}_{l}}{n^2_c}, \,\,  \alpha_l = \frac{\mathcal{A}_{l}}{n_c},
\end{align}
where $\mathcal{C}_{l}$ and $\mathcal{A}_{l}$ are $O(1)$ constants. Using sub-exponential tail bounds, 
\begin{align}
\text{Pr}\bigg(\bigg|\sum_{z, n} w_{\beta t, z^{(n)}}\hat N_{z^{(n)}} -\sum_{z}w_{\beta t, z^{(n)}}\frac{M}{n_c}p^{(n)}(z)\bigg| \geq \epsilon \bigg) &\leq 2\exp\bigg(-\min\bigg\{\frac{n^2_c\epsilon^2}{2M\mathcal{C}_{\beta t}}, \frac{n_c\epsilon }{2\mathcal{A}_{\beta t}}\bigg\}\bigg),\\
\text{Pr}\bigg(\bigg|\sum_{z, n} w_{0, z^{(n)}}\hat N_{z^{(n)}} -\sum_{z}w_{0, z^{(n)}}\frac{M}{n_c}p^{(n)}(z)\bigg| \geq \epsilon \bigg) &\leq 2\exp\bigg(-\min\bigg\{\frac{n^2_c\epsilon^2}{2M\mathcal{C}_{0}}, \frac{n_c\epsilon }{2\mathcal{A}_{0}}\bigg\}\bigg).
\end{align}
These bounds imply
\begin{align}
    \text{Pr}(|v_{ic, {\beta t}} - \bar v_{ic, {\beta t}}|\geq \epsilon)=\text{Pr}\bigg(\bigg|\sum_{z, n} w_{\beta t, z^{(n)}}\frac{\hat N_{z^{(n)}}}{M/n_c} -\sum_{z}w_{\beta t, z^{(n)}}p^{(n)}(z)\bigg| \geq \epsilon \bigg) &\leq 2\exp\bigg(-\min\bigg\{\frac{M\epsilon^2}{2\mathcal{C}_{\beta t}}, \frac{M\epsilon }{2\mathcal{A}_{\beta t}}\bigg\}\bigg), \label{eq:expincoh1}\\
    \text{Pr}(|v_{ic, {0}} - \bar v_{ic, {0}}|\geq \epsilon)=\text{Pr}\bigg(\bigg|\sum_{z, n} w_{0, z^{(n)}}\frac{\hat N_{z^{(n)}}}{M/n_c} -\sum_{z}w_{0, z^{(n)}}p^{(n)}(z)\bigg| \geq \epsilon \bigg) &\leq 2\exp\bigg(-\min\bigg\{\frac{M\epsilon^2}{2\mathcal{C}_{0}}, \frac{M\epsilon }{2\mathcal{A}_{0}}\bigg\}\bigg) \label{eq:expincoh2}.
\end{align}
We note that the Gaussian-like tail $e^{-\epsilon^2}$ occurs for small deviations about the mean, $\epsilon \leq \frac{\mathcal{C}_{l}}{\mathcal{A}_l}$.

Now, we further want to bound
\begin{align}
    \bigg|\frac{v_{{ic},{\beta t}}}{v_{{ic},{\beta t}} + v_{{ic},{0}}}-\frac{\bar{v}_{{ic},{\beta t}}}{\bar{v}_{{ic},{\beta t}} + \bar{v}_{{ic},{0}}}\bigg| = |\hat \gamma_\beta(t) - \gamma_\beta (t)|.
\end{align}

First, we work in a regime where $|v_{{ic},{0}}-\bar{v}_{{ic},{0}}| \leq r$ and $|v_{{ic},{\beta t}}-\bar{v}_{{ic},{\beta t}}| \leq r'$ for some chosen $r, r' < 1$. By Eqs. \eqref{eq:expincoh1} and \eqref{eq:expincoh2}, this can be achieved with probability $1-\delta_1-\delta_2$ by taking 

\begin{align}
M \geq \max\bigg\{\frac{2 \log(2/\delta_1)\mathcal{C}_0}{r^2},\frac{2 \log(2/\delta_1)\mathcal{A}_0}{r}, \frac{2 \log(2/\delta_2)\mathcal{C}_{\beta t}}{r'^2},\frac{2 \log(2/\delta_2)\mathcal{A}_{\beta t}}{r'} \bigg\}.
\end{align}

Now, we have chosen $r$ and $r'$ sufficiently small so that we can Taylor expand $\frac{v_{{ic},{\beta t}}}{v_{{ic},{\beta t}} + v_{{ic},{0}}}$ about its mean:
\begin{align}
    \frac{v_{{ic},{\beta t}}}{v_{{ic},{\beta t}} + v_{{ic},{0}}}-\frac{\bar{v}_{{ic},{\beta t}}}{\bar{v}_{{ic},{\beta t}} + \bar{v}_{{ic},{0}}} = \frac{\bar{v}_{{ic},{0}}}{(\bar{v}_{{ic},{\beta t}} + \bar{v}_{{ic},{0}})^2}(v_{{ic},{\beta t}}-\bar{v}_{{ic},{\beta t}}) - \frac{\bar{v}_{{ic},{\beta t}}}{(\bar{v}_{{ic},{\beta t}} + \bar{v}_{{ic},{0}})^2}(v_{{ic},{0}}-\bar{v}_{{ic},{0}}) + d.
\end{align}
$d$ is the error of the Taylor expansion, which can be quantified by Taylor's Remainder Theorem,
\begin{align}
    d = -\frac{v_1}{(v_1 +v_2)^3} (v_{{ic},{\beta t}}-\bar{v}_{{ic},{\beta t}})^2 + \frac{v_2}{(v_1 +v_2)^3} (v_{{ic},{0}}-\bar{v}_{{ic},{0}} )^2+\frac{v_2-v_1}{(v_1 +v_2)^3} (v_{{ic},{\beta t}}-\bar{v}_{{ic},{\beta t}}) (v_{{ic},{0}}-\bar{v}_{{ic},{0}} ).
\end{align}
The coefficients originate from the second-order derivatives of $\frac{v_{{ic},{\beta t}}}{v_{{ic},{\beta t}} + v_{{ic},{0}}}$ evaluated at some point $v_1 \in [v_{{ic},{0}}, \bar{v}_{{ic},{0}}]$ and $v_2\in[v_{{ic},{\beta t}}, \bar{v}_{{ic},{\beta t}}]$.  As $\bar{v}_{{ic},{0}} = A \approx 1$, there are no divergences in the second-order derivatives for small variations $|v_{{ic},{0}}-\bar{v}_{{ic},{0}}|\leq r$, and we are justified in treating their output as bounded.

Thus,
\begin{align}
     \bigg|\frac{v_{{ic},{\beta t}}}{v_{{ic},{\beta t}} + v_{{ic},{0}}}-\frac{\bar{v}_{{ic},{\beta t}}}{\bar{v}_{{ic},{\beta t}} + \bar{v}_{{ic},{0}}}\bigg| \leq \frac{\bar{v}_{{ic},{0}}}{(\bar{v}_{{ic},{\beta t}} + \bar{v}_{{ic},{0}})^2}|v_{{ic},{\beta t}}-\bar{v}_{{ic},{\beta t}}|+ \frac{\bar{v}_{{ic},{\beta t}}}{(\bar{v}_{{ic},{\beta t}} + \bar{v}_{{ic},{0}})^2}|v_{{ic},{0}}-\bar{v}_{{ic},{0}}| \notag \\
     +\frac{v_1}{(v_1 +v_2)^3}  |v_{{ic},{\beta t}}-\bar{v}_{{ic},{\beta t}}|^2 + \frac{v_2}{(v_1 +v_2)^3}  |v_{{ic},{0}}-\bar{v}_{{ic},{0}}|^2+ \frac{|v_2-v_1|}{(v_1 +v_2)^3} |v_{{ic},{0}}-\bar{v}_{{ic},{0}} ||v_{{ic},{\beta t}}-\bar{v}_{{ic},{\beta t}}|\\
     \leq \bigg(\frac{\bar{v}_{{ic},{0}}}{(\bar{v}_{{ic},{\beta t}} + \bar{v}_{{ic},{0}})^2}+C_1\bigg)|v_{{ic},{\beta t}}-\bar{v}_{{ic},{\beta t}}|+ \bigg(\frac{\bar{v}_{{ic},{\beta t}}}{(\bar{v}_{{ic},{\beta t}} + \bar{v}_{{ic},{0}})^2} + C_2\bigg)|v_{{ic},{0}}-\bar{v}_{{ic},{0}}|.\label{eq:boundincohcliff}
\end{align}
In the second line, we have used the fact that we choose $r$, $r'$ so that $\frac{|v_{{ic},{\beta t}}-\bar{v}_{{ic},{\beta t}}|}{(v_1+v_2)} \leq 1 $ and $\frac{|v_{{ic},{0}}-\bar{v}_{{ic},{0}}|}{(v_1+v_2)} \leq 1 $. Thus, $\frac{v_1}{(v_1 +v_2)^3}  |v_{{ic},{\beta t}}-\bar{v}_{{ic},{\beta t}}|^2 
 \leq C_1 |v_{{ic},{\beta t}}-\bar{v}_{{ic},{\beta t}}|$ for some $C_1$. Similarly, $\frac{v_2}{(v_1 +v_2)^3}  |v_{{ic},{0}}-\bar{v}_{{ic},{0}}|^2
 \leq C' |v_{{ic},{0}}-\bar{v}_{{ic},{0}}|$ for some $C'$, and $\frac{|v_2-v_1|}{(v_1 +v_2)^3} |v_{{ic},{0}}-\bar{v}_{{ic},{0}} ||v_{{ic},{\beta t}}-\bar{v}_{{ic},{\beta t}}|\leq C''|v_{{ic},{0}}-\bar{v}_{{ic},{0}} |$ for some $C''$. We define $C_2 \equiv C' + C''$.

Combining this with Eqs. \eqref{eq:expincoh1} and \eqref{eq:expincoh2}, we have
\begin{align}
    \text{Pr}\bigg( \bigg|\frac{v_{{ic},{\beta t}}}{v_{{ic},{\beta t}} + v_{{ic},{0}}}-\frac{\bar{v}_{{ic},{\beta t}}}{\bar{v}_{{ic},{\beta t}} + \bar{v}_{{ic},{0}}}\bigg|\geq \epsilon\bigg) &\leq \bigg(\frac{\bar{v}_{{ic},{0}}}{(\bar{v}_{{ic},{\beta t}} + \bar{v}_{{ic},{0}})^2}+C_1\bigg)\text{Pr}\bigg(|v_{{ic},{\beta t}}-\bar{v}_{{ic},{\beta t}}| \geq \frac{\epsilon}{D}\bigg) \notag \\
    &+ \bigg(\frac{\bar{v}_{{ic},{\beta t}}}{(\bar{v}_{{ic},{\beta t}} + \bar{v}_{{ic},{0}})^2} + C_2\bigg)\text{Pr}\bigg(|v_{{ic},{0}}-\bar{v}_{{ic},{0}}| \geq \frac{\epsilon}{D}\bigg) \\
    &\leq 2\bigg(\frac{\bar{v}_{{ic},{0}}}{(\bar{v}_{{ic},{\beta t}} + \bar{v}_{{ic},{0}})^2}+C_1\bigg) \exp\bigg(-\min\bigg\{\frac{M\epsilon^2}{2D^2\mathcal{C}_{\beta t}}, \frac{M\epsilon }{2D\mathcal{A}_{\beta t}}\bigg\}\bigg)\notag  \\ &+2\bigg(\frac{\bar{v}_{{ic},{\beta t}}}{(\bar{v}_{{ic},{\beta t}} + \bar{v}_{{ic},{0}})^2} + C_2\bigg)\exp\bigg(-\min\bigg\{\frac{M\epsilon^2}{2D^2\mathcal{C}_{0}}, \frac{M\epsilon }{2D\mathcal{A}_{0}}\bigg\}\bigg) \\
    &\leq 2D \exp\bigg(-\min\bigg\{\frac{M\epsilon^2}{2D^2\mathcal{C}_{\beta t}}, \frac{M\epsilon }{2D\mathcal{A}_{\beta t}}, \frac{M\epsilon^2}{2D^2\mathcal{C}_{0}}, \frac{M\epsilon }{2D\mathcal{A}_{0}}\bigg\}\bigg),
\end{align}
where $D= \frac{\bar{v}_{{ic},{\beta t}}+\bar{v}_{{ic},{0}}}{(\bar{v}_{{ic},{\beta t}} + \bar{v}_{{ic},{0}})^2} + C_2 +C_1$. Thus, to achieve $|\hat \gamma_\beta(t) - \gamma_\beta (t)| \leq \epsilon$ with probability $1-\delta_1-\delta_2-\delta_3$, we need 
\begin{align}
    M \geq \max&\bigg\{\frac{2 \log(\frac{2}{\delta_1})\mathcal{C}_0}{r^2},\frac{2 \log(\frac{2}{\delta_1})\mathcal{A}_0}{r}, \frac{2 \log(\frac{2}{\delta_2})\mathcal{C}_{\beta t}}{r'^2},\frac{2 \log(\frac{2}{\delta_2})\mathcal{A}_{\beta t}}{r'}, \notag \\ &\frac{2 D^2 \log(\frac{2D}{\delta_3})\mathcal{C}_0}{\epsilon^2},\frac{2 D \log(\frac{2D}{\delta_3})\mathcal{A}_0}{\epsilon}, \frac{2 D^2 \log(\frac{2D}{\delta_3})\mathcal{C}_{\beta t}}{\epsilon^2},\frac{2 D\log(\frac{2D}{\delta_3})\mathcal{A}_{\beta t}}{\epsilon} \bigg\}.
\end{align} Setting $\delta_1 =\delta_2=\delta_3 = \delta/3$, we see that we can achieve probability $1 - \delta$ with $M=O(\max\{\frac{\log(\frac{1}{\delta})}{\epsilon^2}, \frac{\log(\frac{1}{\delta})}{\epsilon}\})$. To guarantee $\max_{\beta, t}|\hat \gamma_\beta(t) - \gamma_\beta (t)| \leq \epsilon$ with probability $1-\delta$, we then need 
\begin{align}\label{eq:midincohsamp}
    M=O(\max\{\frac{\log(\frac{K_{ic}}{\delta})}{\epsilon^2}, \frac{\log(\frac{K_{ic}}{\delta})}{\epsilon}\}),
\end{align} where we take $\delta \to \frac{\delta}{K_{ic}}$ and union bound. 

We can translate this analysis back to a deterministic number of samples per circuit rather than a probabilistic number $X\sim \text{Poisson}(M/n_c)$. By Lemma \ref{lem:poisbound},
\begin{align}\label{eq:depoisincoh}
    \text{Pr}(X \notin [\frac{M}{2n_c}, \frac{3M}{2n_c}]) \leq \frac{2\sqrt{2n_c}}{\sqrt{M}}\bigg(\frac{8e}{27}\bigg)^{M/2n_c} \leq \frac{2\sqrt{2}}{\sqrt{40}}\bigg(\frac{8e}{27}\bigg)^{M/2n_c} 
\end{align}
for $M/n_c \geq 40$, noting that $M$ here corresponds to $2M$ in Lemma~\ref{lem:poisbound}. Thus, to achieve a probability of $1-\delta'-\delta ''$ that the protocol succeeds and the true number of samples for each circuit lies within $[\frac{M}{2n_c}, \frac{3M}{2n_c}]$, we can take $M  \geq \max\{{2n_c\frac{\log(\frac{\delta''}{\sqrt{5}})}{\log(\frac{8e}{27})}}, n_c40\}=O(\log(\frac{K_{ic}^2}{\delta'})\log(\frac{1}{\delta''})/N)$. Setting $\delta'=\delta''=\frac{\delta}{3}$ and combining with Eq. \eqref{eq:midincohsamp} with $\delta \rightarrow \frac{\delta}{3}$ using a union bound, we require 
\begin{align}
    M=O(\max\{\frac{\log(\frac{K_{ic}}{\delta})}{\epsilon^2}, \frac{\log(\frac{K_{ic}}{\delta})}{\epsilon}, \frac{\log(\frac{K_{ic}^2}{\delta})\log(\frac{1}{\delta})}{N}\})
\end{align} samples to ensure $\max_{\beta, t}|\hat \gamma_\beta(t) - \gamma_\beta (t)| \leq \epsilon$ with total probability $1-\delta$. For small $\epsilon$, the $\frac{1}{\epsilon^2}$ rate dominates as long as $N >\epsilon^2 \log(\frac{1}{\delta}) \approx 7 \epsilon^2$ for $\delta = 0.1\%$. Thus,  $M = O(\frac{\log(\frac{K_{ic}}{\delta})}{\epsilon^2})$ for relevant scenarios with small $\epsilon$. $\qed$

\begin{theorem}\label{thm:boundcliffcoh}
 Assume that the signal strengths are sufficiently small that we lie in the perturbative regime, i.e.  $\sum_{\beta ,t}\gamma_\beta(t) +\sum_{\alpha ,t}\theta_\alpha(t)^2 \ll 1$, and neglect bias. In the presence of $K_{c}$ total coherent signals $\theta_\alpha(t)$, a precision $\max_{\alpha, t}|\hat \theta_\alpha(t) -\theta_\alpha(t)|\leq \epsilon$ can be achieved with probability $1-\delta$ by taking $M=O(\max\{\frac{\log(\frac{K_{c}}{\delta})}{\epsilon^2}, \frac{\log(\frac{K_{c}}{\delta})}{\epsilon}, \log(\frac{K_{c}}{\delta})\log(\frac{1}{\delta})\})$, with the first term dominant for small $\epsilon$.
\end{theorem}

\textit{Proof. } We want to bound the error of our estimator Eq. \eqref{eq:estcohcliff}. First, we take $n_c = O(\frac{K_{c}}{\delta'})$ large enough such that with high probability $1-\delta'$, the protocol succeeds (Theorem \ref{thm:failurecoh}).  We rewrite our estimator as 
\begin{align}
    \hat \theta_\alpha (t) = \frac{1}{\hat A} \sum_{ z, n}w_{{\alpha t},z^{(n)}}\frac{\hat N_{z^{(n)}}}{M/n_c} \equiv \frac{v_{c, \alpha t}}{v_{ic, 0}}
\end{align}
where
\begin{align}
    w_{{\alpha t},z^{(n)}}= \sum_i (V^TV)_{\alpha t, i} V^T_{i, z^{(n)}}
\end{align}
and $z^{(n)}$ indicates bitstring $z$ drawn from the $n$th circuit. We have used the fact that $\hat A = v_{ic, 0}$ (Eq. \eqref{eq:estincoh1cliff}). 

Although $\hat N_{z^{(n)}} \sim \text{Multinomial}(M/n_c, p^{(n)}(z))$, we can effectively consider a process whereby $X \sim \text{Poisson}(M/n_c)$ samples are taken for each circuit. In this case, $\hat N_{z^{(n)}} \sim \text{Poisson}$($M/n_cp^{(n)}(z)$) and are independent for different $z, n$. Poissonizing in this manner will only incur up to constant errors in our sample complexity analysis (Lemma \ref{lem:poisbound}). 

We proceed with Poisson variables $\hat N_{z^{(n)}}$. By Corollary \ref{cor:rpois}, $w_{{\alpha t},z^{(n)}}\hat N_{z^{(n)}}$ is sub-exponential with parameters $\nu^2 = w^2_{{\alpha t},z^{(n)}}e \frac{M}{n_c}p^{(n)}(z)$ and $\alpha = |w_{{\alpha t},z^{(n)}}|$. As summations of sub-exponential random variables are also sub-exponential, $\sum_{ z, n}w_{{\alpha t},z^{(n)}}\hat N_{z^{(n)}}$ is sub-exponential with effective parameters $\nu_{\alpha t}^2 = \sum_{z, n} w^2_{{\alpha t},z^{(n)}}e \frac{M}{n_c}p^{(n)}(z)$ and $\alpha_{\alpha t} = \max_{z, n} |w_{{\alpha t},z^{(n)}}|$ \cite{Wainwright2019}. From the discussion in Sec. \ref{sec:sccliff}, we know that $w_{{\alpha t},z^{(n)}} \approx \frac{2^{N-1}}{n_c}\delta p^{(n)}_{\alpha t}(z)$, with $\delta p^{(n)}_{\alpha t}(z)$ defined for a single circuit $\mathrm{C}_{n}$ in Eq. \eqref{eq:cliffcohforder1}. Thus, 
\begin{align}
    \nu_{\alpha t}^2 \approx \frac{Me2^{2N-2}}{n^3_c} \sum_{z, n} \delta p^{(n)}_{\alpha t}(z)^2 p^{(n)}(z) & \approx \frac{Me2^{2N-2}}{n^3_c} \underbrace{\sum_{z, n} \delta p^{(n)}_{\alpha t}(z)^2 p_0^{(n)}(z)}_{= \frac{2^N n_c}{2} \times \frac{1}{2^{2N-2}}\times \frac{1}{2^N}=\frac{n_c}{2^{2N-1}}} = \frac{M e}{2 n_c^2},
\end{align}
where we have used the fact that approximately $\frac{n_c}{2}$ of the $ \delta p^{(n)}_{\alpha t}(z)$ are nonzero, and that each entry of $\delta p^{(n)}_{\alpha t}(z)$ has magnitude $\frac{1}{2^{N-1}}$. We can therefore write
\begin{align}
    \nu_{\alpha t} = \frac{M \mathcal{C}_{\alpha t}}{n_c^2}, \,\, \alpha_{\alpha t}=\frac{\mathcal{A}_{\alpha t}}{n_c}, 
\end{align}
where both $\mathcal{C}_{\alpha t}$ and $\mathcal{A}_{\alpha t}$ are $O(1)$ numbers. Using the sub-exponential tail bound,
\begin{align}
     \text{Pr}\bigg( \bigg| \sum_{z, n} w_{\alpha t, z^{(n)}}\hat N_{z^{(n)}} -\sum_{z, n} w_{\alpha t, z^{(n)}}\frac{M}{n_c}p^{(n)}(z) \bigg| \geq \epsilon\bigg) &\leq 2 \exp\bigg( - \min\bigg\{\frac{n_c^2 \epsilon^2}{2 M \mathcal{C}_{\alpha t}},\frac{n_c \epsilon}{2 \mathcal{A}_{\alpha t}} \bigg\}\bigg)  \\
     \implies \text{Pr}(|v_{c, \alpha t} - \bar{v}_{c, \alpha t}| \geq \epsilon) = \text{Pr}\bigg( \bigg| \sum_{z, n} w_{\alpha t, z^{(n)}}\frac{\hat N_{z^{(n)}}}{M/n_c} -\sum_{z, n} w_{\alpha t, z^{(n)}}p^{(n)}(z) \bigg| \geq \epsilon\bigg) &\leq 2 \exp\bigg( - \min\bigg\{\frac{M \epsilon^2}{2  \mathcal{C}_{\alpha t}},\frac{M \epsilon}{2 \mathcal{A}_{\alpha t}} \bigg\}\bigg) \label{eq:expcoh}
\end{align}
The Gaussian-like bound occurs for small $\epsilon \leq \frac{\mathcal{C}_{\alpha t}}{\mathcal{A}_{\alpha t}}$.

Now, we further want to bound 
\begin{align}
   \big|\frac{v_{c, \alpha t}}{v_{ic, 0}} - \frac{\bar v_{c, \alpha t}}{\bar v_{ic, 0}}\big| = |\hat \theta_\alpha(t) - \theta_\alpha(t)|.
\end{align}
We work in a regime where $|v_{ic, 0} -  \bar v_{ic, 0}|\leq r$ and $|v_{c, \alpha t} - \bar  v_{c, \alpha t}| \leq r'$. By Eqs. \eqref{eq:expincoh2} and \eqref{eq:expcoh}, this can be achieved with probability $1-\delta_1 -\delta_2$ by choosing $M \geq \max\bigg\{\frac{2 \log(2/\delta_1)\mathcal{C}_0}{r^2},\frac{2 \log(2/\delta_1)\mathcal{A}_0}{r}, \frac{2 \log(2/\delta_2)\mathcal{C}_{\alpha t}}{r'^2},\frac{2 \log(2/\delta_2)\mathcal{A}_{\alpha t}}{r'} \bigg\}$. We choose $r$ and $r'$ appropriately such that $\frac{r}{\bar{v}_{ic, 0}} \leq 1$,$\frac{r'}{\bar{v}_{ic, 0}} \leq 1$, and we can Taylor expand 
\begin{align}\label{eq:texpcoh}
    \frac{v_{c, \alpha t}}{v_{ic, 0}} = \frac{\bar v_{c, \alpha t} +( v_{c, \alpha t}-\bar v_{c, \alpha t})}{\bar v_{ic, 0} + (v_{ic, 0} -  \bar v_{ic, 0})} = \frac{ \bar v_{c, \alpha t} +(v_{c, \alpha t}- \bar  v_{c, \alpha t})}{\bar v_{ic, 0}} \bigg(1 - \frac{ v_{ic, 0} - \bar 
 v_{ic, 0}}{\bar  v_{ic, 0}} +  c.\bigg),
\end{align}
where $c$ is the rest of Taylor expansion,
\begin{align}
    |c| &= \bigg|\bigg(\frac{\Delta v_{ic,0}}{\bar 
 v_{ic,0}}\bigg)^2 -\bigg(\frac{\Delta v_{ic,0}}{\bar 
 v_{ic,0}}\bigg)^3 \cdots\bigg| \leq\sum_{\alpha = 2} \bigg|\frac{\Delta v_{ic,0}}{\bar 
 v_{ic,0}}\bigg|^\alpha \\
    &= \bigg(\frac{\Delta v_{ic,0}}{\bar 
 v_{ic,0}}\bigg)^2\frac{1}{1-|\frac{\Delta v_{ic,0}}{\bar  v_{ic,0}}|} \leq \bigg(\frac{\Delta v_{ic,0}}{\bar 
 v_{ic,0}}\bigg)^2\frac{1}{1-\frac{r}{\bar 
 v_{ic,0}}} \equiv C\bigg(\frac{\Delta v_{ic,0}}{\bar  v_{ic,0}}\bigg)^2.
\end{align}
Here, we define $\Delta v_{ic, 0} \equiv v_{ic, 0} -  \bar v_{ic, 0} $ and $\Delta v_{c, \alpha t} \equiv v_{c, \alpha t} - \bar  v_{c, \alpha t} $, and we use the fact that  $\frac{\Delta  v_{ic, 0}}{ \bar  v_{ic, 0}} \leq \frac{r}{\bar 
 v_{ic,0}}\leq1$. So, we have $C\big(\frac{\Delta v_{ic, 0}}{\bar  v_{ic, 0}}\big)^2 \leq C\frac{\Delta v_{ic, 0}}{\bar  v_{ic, 0}}$ and $\frac{|\Delta v_{c, \alpha t}| |\Delta v_{ic, 0}|}{\bar  v_{ic, 0}^2} \leq \frac{|\Delta v_{c, \alpha t}|}{\bar v_{ic, 0}}$. Thus, using Eq. \eqref{eq:texpcoh},
\begin{align}
    \bigg|  \frac{v_{c, \alpha t}}{v_{ic, 0}}  - \frac{\bar v_{c, \alpha t}}{\bar v_{ic, 0}} \bigg| &\leq \frac{|\Delta v_{c, \alpha t}|}{\bar  v_{ic, 0}} + \frac{\bar v_{c, \alpha t}}{\bar v_{ic, 0}}\frac{ |\Delta v_{ic, 0}|}{\bar  v_{ic, 0}}+ \frac{|\Delta v_{c, \alpha t}| |\Delta v_{ic, 0}|}{\bar  v^2_{ic, 0}} + C\frac{|\Delta v_{c, \alpha t}| |\Delta v_{ic, 0}|^2}{\bar  v^3_{ic, 0}} + C\frac{\bar v_{c, \alpha t} |\Delta v_{ic, 0}|^2}{\bar  v^3_{ic, 0}} \\
    &\leq (C+2) \frac{|\Delta v_{c, \alpha t}|}{\bar  v_{ic, 0}} +(C+1) \frac{\bar v_{c, \alpha t}}{\bar v_{ic, 0}}\frac{ |\Delta v_{ic, 0}|}{\bar  v_{ic, 0}} \label{eq:boundcohcliff}.
\end{align}
Putting Eqs. \eqref{eq:expcoh}, \eqref{eq:expincoh2}, and \eqref{eq:boundcohcliff} together and using the union bound,
\begin{align}
    \text{Pr}\bigg( \bigg|  \frac{v_{c, \alpha t}}{v_{ic, 0}}  - \frac{\bar v_{c, \alpha t}}{\bar v_{ic, 0}} \bigg| \geq \epsilon \bigg) &\leq (C+2) \text{Pr}\bigg(\frac{|\Delta v_{c, \alpha t}|}{\bar  v_{ic, 0}} \leq \frac{\epsilon}{2C+3}\bigg) +(C+1) \text{Pr}\bigg(\frac{\bar v_{c, \alpha t}}{\bar v_{ic, 0}}\frac{ |\Delta v_{ic, 0}|}{\bar  v_{ic, 0}} \leq \frac{\epsilon}{2C+3}\big) \\
    &\leq 2(C+2) \exp\bigg( - \min\bigg\{\frac{M \epsilon^2 \bar  v_{ic, 0}^2}{2  \mathcal{C}_{\alpha t} (2C+3)^2},\frac{M \epsilon \bar  v_{ic, 0}}{2 \mathcal{A}_{\alpha t} (2C+3)} \bigg\}\bigg) \notag \\ 
    &+ 2(C+1) \exp\bigg( - \min\bigg\{\frac{M \epsilon^2 \bar  v_{ic, 0}^4}{2  \mathcal{C}_{0} (2C+3)^2 \bar v_{c, \alpha t}^2},\frac{M \epsilon \bar  v_{ic, 0}^2}{2 \mathcal{A}_{0} (2C+3)\bar v_{c, \alpha t}} \bigg\}\bigg), \\
    &\leq 2(2C + 3)\exp\bigg( - \min\bigg\{\frac{M \epsilon^2 \bar  v_{ic, 0}^2}{2  \mathcal{C}_{\alpha t} (2C+3)^2},\frac{M \epsilon \bar  v_{ic, 0}}{2 \mathcal{A}_{\alpha t} (2C+3)} , \notag \\
    & \hspace{3cm} \frac{M \epsilon^2 \bar  v_{ic, 0}^2}{2  \mathcal{C}_{0} (2C+3)^2 \theta_{\alpha}(t)^2},\frac{M \epsilon}{2 \mathcal{A}_{0} (2C+3)\theta_{\alpha}(t)} \bigg\}\bigg)
\end{align}
where we have used the fact that $\frac{\bar  v_{c, \alpha t}}{\bar  v_{ic, 0}} = \theta_\alpha(t)$.
Thus, to achieve $|\hat \theta_\alpha(t) - \theta_\alpha (t)| \leq \epsilon$ with probability $1-\delta_1-\delta_2-\delta_3$, we can take
\begin{align}
    M \geq \max&\bigg\{\frac{2 \log(2/\delta_1)\mathcal{C}_0}{r^2},\frac{2 \log(2/\delta_1)\mathcal{A}_0}{r}, \frac{2 \log(2/\delta_2)\mathcal{C}_{\alpha t}}{r'^2},\frac{2 \log(2/\delta_2)\mathcal{A}_{\alpha t}}{r'}, \frac{2 (2C+3)^2 \log(\frac{2(2C+3)}{\delta_3})\mathcal{C}_0}{\bar  v_{ic, 0}^2\epsilon^2}, \notag \\ &\frac{2 (2C+3) \log(\frac{2(2C+3)}{\delta_3})\mathcal{A}_0}{\bar  v_{ic, 0}\epsilon}, \frac{2 (2C+3)^2 \log(\frac{2(2C+3)}{\delta_3})\theta_{\alpha}(t)^2\mathcal{C}_{\alpha t}}{\bar  v_{ic, 0}^2\epsilon^2},\frac{2 (2C+3)\log(\frac{2(2C+3)}{\delta_3})\theta_{\alpha}(t)\mathcal{A}_{\alpha t}}{\bar  v_{ic, 0}\epsilon} \bigg\}.
\end{align} 
Setting $\delta_1=\delta_2=\delta_3 = \delta/3$, we see that we can achieve the desired precision with probability $1-\delta$ with $M= O(\max\{\frac{\log(\frac{1}{\delta})}{\epsilon^2}, \frac{\log(\frac{1}{\delta})}{\epsilon}\})$. To guarantee $\max_{\alpha, t}|\hat \theta_\alpha(t) - \theta_\alpha (t)| \leq \epsilon$ with probability $1-\delta$, we then need 
\begin{align}\label{eq:midcohsamp}
    M = O(\max\{\frac{\log(\frac{K_{c}}{\delta})}{\epsilon^2}, \frac{\log(\frac{K_c}{\delta})}{\epsilon}\}),
\end{align} where we take $\delta \to \frac{\delta}{K_{c}}$ and union bound. 

We now translate this analysis back to a deterministic number of samples per circuit rather than a probabilistic number $X\sim \text{Poisson}(M/n_c)$. By Eq. \eqref{eq:depoisincoh}, we see that with probability $1-\delta'-\delta ''$, the protocol succeeds and the true number of samples for each circuit lies within $[\frac{M}{2n_c}, \frac{3M}{2n_c}]$ if $M  \geq \max\{{2n_c\frac{\log(\frac{\delta''}{\sqrt{5}})}{\log(\frac{8e}{27})}}, n_c40\}=O(\log(\frac{K_{c}}{\delta'})\log(\frac{1}{\delta''}))$. Setting $\delta'=\delta''=\frac{\delta}{3}$ and combining with Eq. \eqref{eq:midcohsamp} with $\delta \rightarrow \frac{\delta}{3}$ using a union bound, we require 
\begin{align}
    M=O(\max\{\frac{\log(\frac{K_{c}}{\delta})}{\epsilon^2}, \frac{\log(\frac{K_{c}}{\delta})}{\epsilon}, \log(\frac{K_{c}}{\delta})\log(\frac{1}{\delta})\})
\end{align} samples to ensure $\max_{\alpha, t}|\hat \theta_\alpha(t) - \theta_\alpha (t)| \leq \epsilon$ with total probability $1-\delta$, where the the $\frac{1}{\epsilon^2}$ rate dominates for small $\epsilon$. Thus,  $M = O(\frac{\log(\frac{K_{c}}{\delta})}{\epsilon^2})$ for relevant scenarios with small $\epsilon$ $\qed$
 \subsubsection{Strong robustness of the incoherent signal protocol}

As discussed in the main text and in Property \ref{num:gcliff_rob}, classical error correction techniques can be used to error-correct our estimator $\hat \gamma_{\beta}(t)$ in the presence of readout error. Here, we show that our incoherent sensing protocol is correctable for large enough $N$ scaling with $\log K_{ic}^2$.

By Theorem \ref{thm:thm1}, each signal $\gamma_{\beta}(t)$ is mapped to a single bitstring. We thus denote the set of bitstrings which are in correspondence with the set of signals for the $n$th circuit as $\mathcal{Z}_n = \{z | k^{(n)}_\beta(z) = 1\}$. For each circuit $n$, readout error occurs when we measure a bitstring $z \notin \mathcal{Z}_n$.

Let $d_{min}(\mathcal{Z}_n)$ be the minimum pairwise Hamming distance between bitstrings in $\mathcal{Z}_n$. If the closest Hamming distance between the measured bitstring $z$ and and any bitstring in $\mathcal{Z}_n $ is $d_{H}(z,\mathcal{Z}_n)\leq \frac{d_{min}(\mathcal{Z}_n)}{2} $, then we can correct $z$ by simply replacing it with the closest bitstring in $\mathcal{Z}_n$ in Hamming distance. We call compatibility with active error correction as \textit{strongly robust} against readout error. We note that as our Clifford circuits are randomly selected, the random encoding of signals to bitstrings is inherent in our protocol. We can bound the probability that for a given random encoding, $\mathcal{Z}_n$ achieves $d_{min}(\mathcal{Z}_n) \geq d$, where $d$ is the desired minimum distance between codewords. The result is given in Theorem~\ref{thm:cliffrob}.

\begin{theorem}\label{thm:cliffrob}
    For randomly selected Clifford circuits $\{C_t\}_{t = 1}^T$, and $k_{\alpha, t}(z)$ defined in Eq. \eqref{eq:kalphacliff}, the probability 
    \begin{align}
        \text{Pr}(d_{min}(\mathcal{Z}) \geq d) \geq\frac{\prod_{m=0}^{K_{ic}-1}(2^{N}-mV(N, d-1))}{(2^N)^{K_{ic}}},
    \end{align} where $\mathcal{Z} = \{z | k_{\alpha, t}(z) = 1\}$, $d_{min}(\mathcal{Z})$ is the minimum pairwise Hamming distance between bitstrings in $\mathcal{Z}$, $K_{ic}$ is the total number of $k_{\alpha, t}(z)$, and $V(N, d-1) = \sum_{j = 0}^{d-1}\begin{pmatrix}
        N \\ j
    \end{pmatrix}$ is the volume of a ball with radius $d-1$.
\end{theorem}
\textit{Proof. } We can think of this problem as randomly choosing $K_{ic}$ bitstrings to form $\mathcal{Z}$, and determining the probability $\text{Pr}(d_{min}(\mathcal{Z}) \geq d)$. We calculate this by counting the number of valid bitstrings  in $\{0, 1\}^N$.

For the first selected bitstring $z^*$, there are no constraints, and the probability of choosing a valid bitstring is $\frac{2^N}{2^N} = 1$. For the second selected bitstring, we cannot choose any bitstrings $z$ s.t. $d_H(z, z^*) \leq d-1$. This excludes $V(N, d-1) = \sum_{j = 0}^{d-1}\begin{pmatrix}
        N \\ j
    \end{pmatrix}$ possible choices, leading to a probability of valid selection of $\frac{2^N-V(N, d-1)}{2^N}$. We repeat this $K_{ic}$ times, leading to $\text{Pr}(d_{min}(\mathcal{Z}) \geq d) \geq\frac{\prod_{m=0}^{K_{ic}-1}(2^{N}-mV(N, d-1))}{(2^N)^{K_{ic}}}$. $\qed$

Now, we note that $V(N, d-1) = \sum_{j = 0}^{d-1}\begin{pmatrix}
        N \\ j
    \end{pmatrix} \leq 2^{N H(\frac{d-1}{N})}$, where 
\begin{align}
    H(x) = - x \log_2(x) - (1-x)\log_2(1-x)
\end{align}
is the binary entropy. 

Thus, letting $q = 2^{-N \big(1-H\big(\frac{d-1}{N}\big)\big)}$, 
\begin{align}
    \prod_{m=0}^{K_{ic}-1}(2^{N}-mV(N, d-1)) &\geq \prod_{m=0}^{K_{ic}-1}2^N(1-mq) \\
    &\geq2^{K_{ic}N}\exp\bigg(-q\sum_{m=0}^{K_{ic}-1}m - q^2\sum_{m=0}^{K_{ic}-1}m^2\bigg) 
\end{align}
where we work in the regime of large $N$ and $\frac{d}{N} < 1$, so $q$ is small and $q K_{ic} \leq \frac{1}{2}$. In the second inequality, we have used the fact that $\ln(1-x^2) \geq x - x^2$ when $0 \leq x \leq \frac{1}{2}$.


Thus, neglecting terms higher-order in $qK_{ic}$, we have
\begin{align}
     \text{Pr}(d_{min}(\mathcal{Z}) \geq d)  \geq \exp\bigg(-2^{-N \big(1-H\big(\frac{d-1}{N}\big)\big)}\frac{K_{ic}(K_{ic}-1)}{2}\bigg).
\end{align}
This decreases exponentially in $K^2_{ic}$, but tends towards $1$ double-exponentially with increasing $N$. With large enough system size $N \sim \log \big(\frac{K_{ic}^2}{\delta}\big)/(1-H(\alpha))$, we thus expect to achieve $\frac{d_{min}}{N} > \alpha$ with probability $1-\delta$.

Although strong robustness only applies to our incoherent protocol, we note that Eq. \eqref{eq:estincoh1cliff} can also be used to sense coherent signals quadratically, up to $\theta_\alpha(t)^2$. Thus, in a situation where robustness is crucial, the incoherent protocol could be used to detect both $\hat \gamma_\alpha(t)$ and $\theta_\alpha(t)^2$. Once the forms of the nonzero signals are determined, a more targeted sensing procedure can be used to pinpoint the precise magnitude and sign of the coherent signals $\theta_\alpha(t)$. 

Moreover, even without using active error correction, we expect our estimators to be weakly robust against readout error---readout error will not affect the SQL scaling of our estimators. For our incoherent estimator Eq. \eqref{eq:estincoh1cliff}, this is true following the discussion in Section \ref{sec:qreadout} as our random Clifford circuits will generally map signals to high-weight bitstrings (c.f. Theorem \ref{thm:thm1}), such that they are minimally affected by readout error. For our coherent estimator Eq. \eqref{eq:estcohcliff}, weak robustness follows from an analogous discussion to Section \ref{sec:linreadout}.

\subsubsection{Thresholding}\label{sec:thres}
 When our signals are guaranteed to be sparse, we can enforce sparse solutions and improve the performance of our estimators (Eq. \eqref{eq:estincoh1cliff}, Eq. \eqref{eq:estcohcliff}) in practice by \textit{hard thresholding}, a type of statistical regularization \cite{Hastie2001}:
\begin{align}
\hat \theta_{\alpha, thres}(t) =
\begin{cases}
\hat \theta_\alpha(t) & \text{if } |\hat \theta_\alpha(t)|\geq \theta_{thres} \\
0 & \text{if }  |\hat \theta_\alpha(t)| < \theta_{thres}.
\end{cases}
\end{align}
$\hat \gamma_{\beta, thres}(t)$ is defined analogously. When we are given that all significant signals $\theta_\alpha(t) \geq \theta_{min}$, $\gamma_\beta(t) \geq \gamma_{min}$, we can choose $\theta_{thres}$ and $\gamma_{thres}$ based on $\theta_{min}$ and $\gamma_{min}$. For instance, using the results of Sec. \ref{sec:sccliff}, we can set $\theta_{thres} = \theta_{min} - 2\times \sqrt{\mathbb{E}_{\alpha}\Var(\hat \theta_\alpha)}$ and $\gamma_{thres} = \gamma_{min} - 2\times \sqrt{\mathbb{E}_{\beta}\Var(\hat \gamma_\beta)}$. In this manner, all estimated signals which have magnitude more than two standard deviations lower than the minimum are set to zero. Thresholding is applied for the numerical results shown in the main text. If $\theta_{min}$ and $\gamma_{min}$ are unknown but expected to be nonzero, than we can set $\theta_{thres}$ and $\gamma_{thres}$ to $2\times \sqrt{\mathbb{E}_{\alpha}\Var(\hat \theta_\alpha)}$ or $2\times \sqrt{\mathbb{E}_{\beta}\Var(\hat \gamma_\beta)}$, respectively, thus zeroing signals within two standard deviations of zero.

\subsubsection{Overlapping signal generators}\label{sec:ovlapcliff}

In our estimators Eqs. \eqref{eq:estincoh1cliff} and \eqref{eq:estcohcliff}, we assumed for simplicity that incoherent and coherent signals couple to the system via different sets of Pauli operators. However, our results can be easily modified to account for when the same set of generators $\{P_\alpha \}$ gives rise to both incoherent $\gamma_\alpha$ and coherent $\theta_\alpha$, as is the case in Fig. 3 of the main text. In this case, Eq. \eqref{eq:cliffpincoh} becomes
\begin{align}
    p(z|\vec{\theta}, \vec{\gamma}) \approx A\big(p_0(z) + \sum_{\alpha, t}\bigg(\frac{\gamma_\alpha}{1-\gamma_\alpha} + \theta^2_\alpha\bigg)k_{\alpha, t}(z)\big),
\end{align}
while Eq. \eqref{eq:cliffpcoh} stays the same. Our estimator $\hat \theta_\alpha$ (Eq. \eqref{eq:estcohcliff}) thus doesn't change, while $\hat \gamma_\alpha(t)$ becomes
\begin{align}
    v_{{ic},{\alpha t}}= \sum_{i, \tilde z}(V^TV)^{-1}_{\alpha t, i} V^{T}_{i, \tilde z}\frac{\hat N_{\tilde z}}{M/n_c}, \;\; \hat \gamma_\alpha(t) = \frac{v_{{ic},{\alpha t}} -  A\theta_\alpha(t)^2}{v_{{ic},{\alpha t}} -  A\theta_\alpha(t)^2+ v_{{ic},{0}}}, \;\;  \hat A = v_{ic, 0}.
\end{align}

We can take the values of $\theta_{\alpha}(t)$ to be the best estimates $\hat \theta_\alpha (t)$ we have from the coherent procedure.

\subsubsection{Bias scaling}\label{sec:biascliff}

\begin{figure}
    \centering
    \includegraphics[width=1\columnwidth]{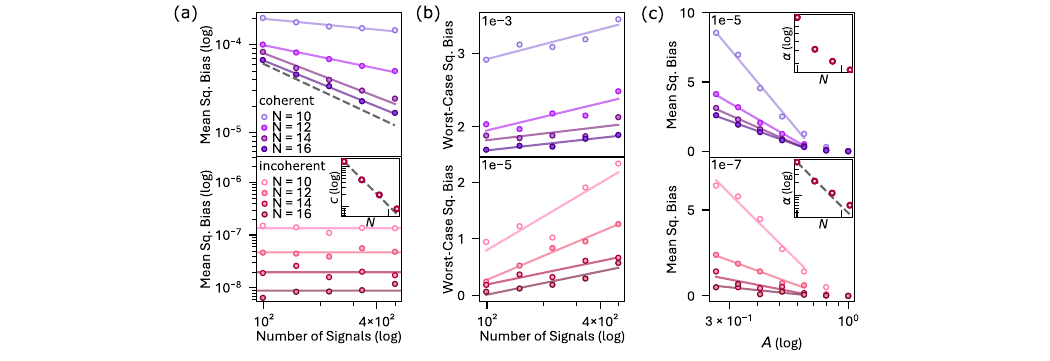}
    \caption{\textbf{(a)} Scaling of the mean squared bias (Eq.~\ref{eq:meansqb})
    for coherent signals (top) and incoherent signals (bottom) as a function of the total number of signals. Both coherent and incoherent signals are drawn randomly from single, two, three, and four-body Pauli $X$, $Y$, and $Z$ strings. A single time step $T = 1$ is considered. Coherent signals satisfy $A = 0.91$, while incoherent signals satisfy $A = 0.70$. $n_c = 18$ circuits are used to estimate coherent signals, and each data point is obtained by averaging over 60 random signal instances. $n_{c} = 5$ circuits are used to estimate incoherent signals, with each data point obtained from an average over 120 random signal instances. The mean squared bias for coherent signals decreases as $(K_c)^{-\gamma}$ with increasing $K_c$, with $\gamma$ approaching $1$ at large $N$. Solid lines indicate best fits to $aK_c^{-\gamma}$, while the dotted line demonstrations $K_c^{-1}$ scaling. The mean squared bias for incoherent signals is constant as $K_{ic}$ increases, with solid lines indicating the corresponding constant values $c$. The inset demonstrates an exponential decrease of $c$ with increasing $N$, with the dotted line indicating the best exponential fit. 
    \textbf{(b)} Scaling of the worst-case squared bias (Eq.~\ref{eq:wcasesqb}) for coherent signals (top) and incoherent signals (bottom) as function of the total number of signals, for the same settings as in (a). For both coherent and incoherent signals, the worst-case squared bias scales respectively as $\log(K_{c})$ and  $\log(K_{ic})$; solid lines show best fits to $\alpha \log(K_{\bullet}) +\beta$. \textbf{(c)} Scaling of the mean squared bias for coherent signals (top) and incoherent signals (bottom) as function of the signal fidelity $A$ for the same settings as in (a). For both coherent and incoherent signals, the bias decreases with $\log(A)$ as $A$ approaches 1; solid lines indicate best fits to $-\alpha \log(A) +\beta$, excluding points near $A = 1$ to avoid edge effects. As shown in the insets, $\alpha$ decreases with increasing $N$; for incoherent signals, the best exponential fit is indicated with a dotted line.}
    \label{fig:biascliff}
\end{figure}

Here, we analyze how the bias of the global Clifford sensing protocol scales with the number of coherent signals $K_c$, the number of incoherent signals $K_{ic}$, and the signal fidelity $A$.

The mechanism enabling the detection of incoherent signals is identical to that of the quadratic Ramsey protocol: each signal is mapped to a distinct bitstring, and the rate at which these bitstrings are detected provides an estimator for the signal magnitude. Accordingly, we expect the bias associated with incoherent signals to follow the same scaling behavior as in the quadratic Ramsey protocol, as discussed in Sec.~\ref{sec:biasqram}. These trends are confirmed numerically in Fig.~\ref{fig:biascliff}.

In contrast, the sensing of coherent signals relies on the same underlying principle as the tilted Ramsey protocol. Each coherent signal $\theta_\alpha(t)$ perturbs the measurement distribution to first order, distributing its effect evenly across all bitstrings $z$ with a random sign structure. By orthogonalizing these sign patterns across different circuits, we can reconstruct the magnitude of each coherent signal. We therefore expect the bias scaling for coherent signals to mirror that of the tilted Ramsey protocol, discussed in Sec.~\ref{sec:biastram}. This expectation is likewise confirmed numerically in Fig.~\ref{fig:biascliff}.

For both types of signal, the worst-case squared bias grows asymptotically as $\log(K_{\bullet})$ with the number of coherent and incoherent signals $K_{c}$ and $K_{ic}$---the same scaling behavior as the statistical error (Sec.~\ref{sec:cliffbounds})---as shown in Fig.~\ref{fig:biascliff}(b). Thus, for any $K_{c}$ and $K_{ic}$, there exists a broad regime of $M$ smaller than the inverse squared bias scale in which statistical error dominates---exactly the setting we focus on in this work. Moreover, bias can be further suppressed---exponentially, in the case of incoherent signals---by increasing the system size $N$.

\subsubsection{Details of numerical simulations}

Here, we give further detail on the simulation results shown Fig. 3 of the main text. 

Panel (b) is a reconstruction of signals $\hat \theta_\alpha(t)$ and $\hat \gamma_\alpha(t)$ estimated using global Clifford unitaries over $T=10$ time steps. The signal generators of both incoherent and coherent signals are Pauli $X_i$, $Y_i$, and $Z_i$ at each qubit $i$, as well as two-body $X_{i}X_{i+1}$ and $Z_{i}Z_{i+1}$ for nearest-neighbors $i, i+1$ on $N = 12$ qubits. There are a total of $K_{ic} = K_{c} =58\times10$ signals. $10$ circuits are used to determine $\hat \theta_\alpha(t)$ and $3$ circuits used for $\hat \gamma_\alpha(t)$. Estimators are plotted at $M = 10^4$, and their values are thresholded according to Section \ref{sec:thres}. We choose minimal signal strengths $\theta_{min} = 0.1$ and $\gamma_{min} = 0.07$ and have $A = 0.46$ (Eq. \eqref{eq:Acliff}). Specifically, nonzero signals are generated uniformly randomly, $|\theta| \in [0.1, 0.15]$ with random sign and $\gamma \in [0.7, 0.1]$. Following the discussion in Section \ref{sec:ovlapcliff}, 
the estimator for $\hat \gamma_\alpha(t)$ subtracts the coherent contribution from $\hat\theta_\alpha(t)$ at $M = \infty$. We note that using $\hat \theta_{\alpha}(t)$ at finite $M$ will result in the same scaling behavior; we choose $M=\infty$ to better compare our results with the theoretical prediction Eq.~\eqref{eq:varcliffincoh}.

Panel (c) (left) shows scaling of the RMS error for $\hat \theta_\alpha(t)$ and $\hat \gamma_\alpha(t)$ in (b), both without readout error and with readout error of $\gamma_r = 0.05$ per qubit. Theory curves plot the approximate variances calculated in Eqs. \eqref{eq:varcliffincoh} and \eqref{eq:varcliffcoh}. 

Panel (c) (right) shows the scaling of sample complexity with number of signals $K_{c}$ and $K_{ic}$ for both $\hat \theta_\alpha(t)$ and $\hat \gamma_\alpha(t)$. Here, sample complexity is defined as $\beta_c =MA^2 \mathbb{E}_\alpha [(\hat \theta_\alpha(t) -\theta_\alpha(t))^2]$ for coherent signals and $\beta_{ic} =MA\mathbb{E}_\alpha [(\hat \gamma_\alpha(t) -\gamma_\alpha(t))^2]$ for incoherent signals. By Eqs. \eqref{eq:varcliffincoh} and \eqref{eq:varcliffcoh}, we have theory predictions $\beta_c =0.5$ for $\hat \theta_\alpha(t)$ and $\beta_{ic}=\mathbb{E}_{\alpha}\gamma_\alpha(t)$ for $\hat \gamma_\alpha(t)$. The sample complexity for $\hat{\gamma}_\alpha(t)$ is averaged only over nonzero signals; averaging over all (including zero) signals in the sparse regime would cause $\beta_{ic}$ scale as $1/K_{ic}$. Signals included in the simulations are generated from randomly selected Pauli operators of the form $X_i, Y_i, Z_i$; $X_iX_j, Y_iY_j, Z_iZ_j$; and $X_iX_jX_k, Y_iY_jY_k, Z_iZ_jZ_k$ with no locality constraints. A single timestep $T=1$ is used so that larger signal sets correspond to a greater number of Pauli generators rather than to multiple timesteps. 15 circuits are used for coherent estimation and 3 for incoherent estimation. We note that the number of circuits used for coherent sensing is expected to be larger than that for incoherent sensing due to the difference between Theorem~\ref{thm:failurencincoh} and ~\ref{thm:failurecoh}.

\subsection{Applications}
In this section, we discuss applications of our multiparameter sensing protocol based on global Clifford unitaries to Hamiltonian learning. In particular, we first show how the protocol can be used as a subroutine for time-independent Hamiltonian learning with Heisenberg-limited scaling, as defined in Refs.~\cite{Huang2023, Hu2025}. We then formalize time-dependent Hamiltonian learning and introduce a simple class of time-dependent Hamiltonians for which our protocol achieves optimal performance.
\subsubsection{Time-independent Hamiltonian learning}

Broadly, time-independent Hamiltonian learning from dynamics addresses the following problem: given an unknown $H$ and the ability to evolve by $e^{-i H t }$ for any $t>0$, determine $H$. Algorithms for accomplishing this task generally use the total evolution time $T$ under the Hamiltonian $H$ as the figure of merit.  A variety of methods have been proposed to tackle this task, including estimating derivatives of observables \cite{Caro2024, Zubida2021}, reshaping the Hamiltonian to a simpler form \cite{Huang2023, Ma2024, Bakshi2024, Hu2025}, and leveraging potential sparsity of the Hamiltonian \cite{Yu2023, Ma2024, Hu2025}. \textit{Heisenberg-limited} Hamiltonian learning is achieved when $T$ scales with the error of estimation $\epsilon$ as $T \sim 1/\epsilon$ \cite{Huang2023, Bakshi2024, Hu2025}, in contrast to the standard scaling $1/\epsilon^2$. 

Our protocol can be used as a tool to reach the Heisenberg limit in Hamiltonian learning. Suppose the Hamiltonian $H = \sum{h_\alpha}P_\alpha$ encodes coherent signals, where $P_{\alpha}$ are arbitrary Pauli strings. We can use our coherent multiparameter sensing procedure to determine each $\theta_\alpha = h_\alpha \tau$ for a given evolution time $\tau$ over one run, by placing the Hamiltonian evolution between layers of Clifford circuits. Using Eq. \eqref{eq:varcliffcoh}, we see that for a certain error $ \sqrt{\mathbb{E}_\alpha[(\hat \theta_\alpha -\theta_\alpha)^2}] = \epsilon^*_\theta$, the total evolution time of our procedure $T$ is 
\begin{align}
    T = M \tau \sim \frac{\tau}{(\epsilon^*_\theta)^2}.
\end{align}
Now, for Hamiltonian learning, we wish to analyze the scaling of $T$ with respect to the error, $\epsilon_h$, of estimating $h_\alpha$. As $\theta_\alpha$ = $h_\alpha \tau$,   $ \epsilon^*_\theta  = \epsilon_h \tau$. So, 

\begin{align}
    T \sim \frac{1}{(\epsilon^*_\theta)^2}\frac{\epsilon^*_\theta}{\epsilon_h} \sim \frac{1}{\epsilon_h}
\end{align}
for fixed $\epsilon_\theta^*$. Thus, by fixing the target phase error $\epsilon^*_\theta$ and increasing the interrogation time $\tau$, we achieve $T\sim \frac{1}{\epsilon_h}$ scaling. However, for sufficiently long evolution times $\tau$, the accumulated phases $\theta_\alpha$ may become large, violating the perturbative regime assumed by our protocol.

This limitation can be addressed by embedding our scheme into a larger algorithm for learning a time-independent Hamiltonian $H$, which can achieve Heisenberg-limited scaling. Since our procedure is formulated for perturbatively small signals, such an algorithm may employ iterative or adaptive steps to manage Hamiltonians with larger strengths. 

One example is \emph{hierarchical learning}, introduced in Ref.~\cite{Hu2025}, which partitions Hamiltonian amplitudes into
$J = \lceil \log(1/\epsilon_h) \rceil$ levels. At each level $j$, all terms $h_\alpha$ with amplitudes
$2^{-(j+1)} < |h_\alpha| < 2^{-j}$ are estimated to precision $\epsilon_h$ using a subroutine such as the procedure described above. Concretely, the protocol estimates the accumulated phases $h_\alpha \tau$ to a target error $\epsilon_\theta^*$, which includes both bias and statistical error and is determined by the number of measurement samples $M$. The corresponding amplitude precision is then $\epsilon_h = \epsilon_\theta^*/\tau$. Moreover, at level $j$, estimates of all Hamiltonian terms with amplitudes larger than $2^{-j}$ are already available, defining an effective Hamiltonian $H_j$. These larger contributions can therefore be canceled, allowing the protocol to estimate the residual Hamiltonian $H - H_j$ at each step. As the remaining amplitudes decrease exponentially with $j$, this reduction is compensated by increasing the interrogation time to $2^j \tau$, keeping the accumulated phase at a fixed, perturbative scale. Using this approach, our protocol can be straightforwardly applied to estimate successively smaller terms,  though a different method may be required to estimate the largest Hamiltonian coefficients, yielding an overall Heisenberg-limited learning strategy.

\subsubsection{Time-dependent Hamiltonian learning}

In conventional time-independent Hamiltonian learning, one assumes access to unitary evolutions $e^{-iHt}$ for a tunable interrogation time $t>0$. This assumption does not straightforwardly extend to the learning of time-dependent Hamiltonians, where it becomes necessary to specify both the structure of the time dependence and the nature of the available access to the system’s evolution.

Broadly, we distinguish between two models of time-dependent Hamiltonian learning: \emph{absolute} and \emph{relative}. In the absolute setting, the Hamiltonian $H(t)$ is defined with respect to an absolute time reference, and each query provides access only to an evolution operator of the form
\begin{equation}
U(t_2,t_1)=\mathcal{T}\exp\!\left(-\int_{t_1}^{t_2} H(t')\,dt'\right),
\end{equation}
where $\mathcal{T}$ denotes time ordering. In contrast, in the relative setting the Hamiltonian can be reset, so that each query yields an evolution
\begin{equation}
U(t)=\mathcal{T}\exp\!\left(-\int_{0}^{t} H(t')\,dt'\right).
\end{equation}

Absolute time-dependent Hamiltonian learning is relevant for sensing applications such as dark-matter detection~\cite{Ebadi2022,Shi2023}, where one has no control over when a time-dependent signal is encountered. By contrast, relative time-dependent Hamiltonian learning is more appropriate for controlled settings, such as quantum process tomography~\cite{Chuang1997,Levy2024} or experiments probing the response of a system (e.g., a biomolecular mechanism~\cite{Aslam2023}) to a well-defined trigger. In both cases, constraints on the time dependence and the specific estimation task must be specified more precisely, since an unrestricted formulation would require estimating an arbitrary function $H(t)$, which entails infinitely many parameters.

In this section, we show that our multiparameter estimation protocol is optimal for learning a simple class of relatively time-dependent Hamiltonians, which we define below. Consider the task of learning a time-dependent Hamiltonian for time $ 0 \leq t \leq \tau$, where the Hamiltonian is piecewise-constant over $n$ intervals of $\frac{\tau}{n}$. The Hamiltonian takes the form
\begin{align}\label{eq:Htdep}
    H(t_i) = \sum_{\alpha}h_{\alpha, t_i} P_\alpha, \,\, i \in [0, 1, \cdots n-1]
\end{align}
where $\{P_\alpha\}$ are Pauli strings. Moreover, the Hamiltonian satisfies $\sum_{\alpha,i} (h_{\alpha, t_i}\frac{\tau}{n})^2 \ll 1 $, and the total size of the set $|\{P_\alpha\}| < 2^N/n$, where $N$ is the number of qubits. No other constraints on the weight or locality of the $P_\alpha$ are imposed. As in the time-independent Hamiltonian case, we use the total evolution time $T$ as the figure of merit for the learning algorithm. Moreover, we restrict our analysis to ancilla-free algorithms that utilize all available qubit sensors for sensing and perform a complete measurement of the system after each iteration of the protocol.

\begin{theorem}\label{thm:tdeplow}
    Consider the task of learning a Hamiltonian of the form Eq. \eqref{eq:Htdep} using an ancilla-free algorithm that employs all available sensor qubits and performs a complete measurement after each iteration. Assume arbitrarily fast control and no decoherence. 
    The lower bound to achieve error $\mathbb{E}_{\alpha, i} (\hat h_{\alpha, i} - h_{\alpha, i})^2 \leq \epsilon^2$  is given by $T = \Omega(\frac{n^2}{\tau \epsilon^2})$. 
\end{theorem}
\textit{Proof. } Suppose $H(t_n) = hZ^{\otimes N}$, and $H(t_{i\neq n}) = 0$. Then, the Hamiltonian is effectively time-independent and given by $H = hZ^{\otimes N}$ acting over the duration $\frac{n-1}{n} \tau\leq t \leq \tau$. A straightforward Quantum Fisher Information (QFI) calculation yields the bound 
\begin{align}
    \Var(\hat h) \geq \frac{n^2}{4 M \tau^2}.
\end{align}
Now, the total evolution time is $T = M \tau$, where we emphasize that the Hamiltonian must be evolved until $t=\tau$ for each iteration in order to sense the term in the last time interval. Thus, 
\begin{align}
    \epsilon^2 \geq \frac{n^2}{4 (T/\tau) \tau^2} \implies T \geq \frac{n^2}{4  \tau \epsilon^2}.
\end{align}
As we have shown the existence of a Hamiltonian which requires total evolution time $T \geq \frac{n^2}{4  \tau \epsilon^2}$ to learn, it follows that any algorithm for learning a generic $H$ of the form Eq \eqref{eq:Htdep} must take at least $T = \Omega(\frac{n^2}{\tau \epsilon^2})$ in the worst case $\qed$.

Now, our coherent multiparameter estimation procedure using Clifford circuits can be directly used to learn the time-dependent $H$.  By Eq. \eqref{eq:varcliffcoh}, we can learn $\theta_\alpha(t_i) \equiv h_{\alpha, t_i} \frac{\tau}{n}$ with error 
\begin{align}
    \Var(\hat \theta_\alpha(t_i)) \approx \frac{1}{2 M A^2},
\end{align}
where $A = \prod_{i,\alpha}\cos^2(\theta_{\alpha}(t_i))$. This scaling is demonstrated in Fig. 3(c) of the main text. Thus, the error of estimating the Hamiltonian terms is
\begin{align}
    \epsilon^2 \approx \frac{n^2}{2 A^2 \tau^2 M} =\frac{n^2}{2 A^2 \tau T},
\end{align}
where we have used $T = M \tau$. This achieves $T \sim O(\frac{n^2}{\tau \epsilon^2})$, which is optimal according to Theorem \ref{thm:tdeplow}. It is worthwhile to explore whether a better algorithm can be developed which further optimizes the constant from $\frac{1}{2}$ to $\frac{1}{4}$.

\subsection{Clifford circuits}\label{sec:loc_cliff}
\begin{figure}
    \centering
    \includegraphics[width=1\columnwidth]{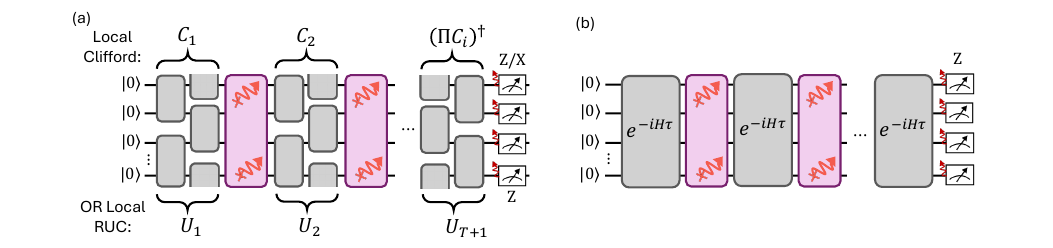}
    \caption{\textbf{(a)} Multiparameter sensing procedure using local Clifford unitaries or local random unitaries. Coherent rotations $e^{-i \sum_\alpha\theta_\alpha(t) P_\alpha}$ and incoherent dissipation $\rho \rightarrow (1-\gamma_\beta(t))\rho + \gamma_\beta(t) P_\beta \rho P_\beta$ are accumulated at each time step $t$, between layers of unitary evolution. \textbf{(b)} Multiparameter sensing procedure via evolution under an ergodic Hamiltonian $H$. $\tau$ is chosen to achieve appropriate scrambling between layers of signal. }
    \label{fig:localfig}
\end{figure}
Although our analysis in Section \ref{sec:glob_cliff} utilizes global Clifford unitaries, we can replace these global Clifford unitaries in practice with brickwork layers of nearest-neighbor two-qubit Clifford gates (Fig. \ref{fig:localfig}(a)).  Specifically, to ensure sufficient scrambling between signal accumulation, we apply two brickwork layers of randomly selected two-qubit Clifford gates between layers of signal. The final Clifford implements the inverse of all layers before, thus guaranteeing that in the absence of signal, the output distribution is $p(0)=1$ when measured in the $z-$basis. The procedure for incoherent and coherent signal estimation, as well as our estimators for $\hat \gamma_\beta(t)$ (Eq. \eqref{eq:estincoh1cliff}) and $\hat \theta_\alpha(t)$ (Eq. \eqref{eq:estcohcliff}), are the same as in the global Clifford case. The only difference is that local dynamics may result in less operator scrambling, meaning that the probabilistic bounds invoked in Theorems \ref{thm:failurencincoh}, \ref{thm:failurecoh},  and \ref{thm:cliffrob} may not be as accurate. In practice, this may result in more than one signal being mapped to the same bitstring, in the sense of Theorem \ref{thm:thm1}. To counter this effect, we may need to choose more sets of circuits $n_c$, and work with a larger system size $N$ to achieve the same target relative distance $\frac{d_{min}}{N} \geq \alpha$. Regarding the latter, although our error-correction analysis does not explicitly invoke bitstring collisions, it nevertheless relies on the assumption that signal bitstrings are encoded uniformly at random, which may not be strictly satisfied by local Clifford circuits.

\begin{figure}
    \centering
    \includegraphics[width=1\columnwidth]{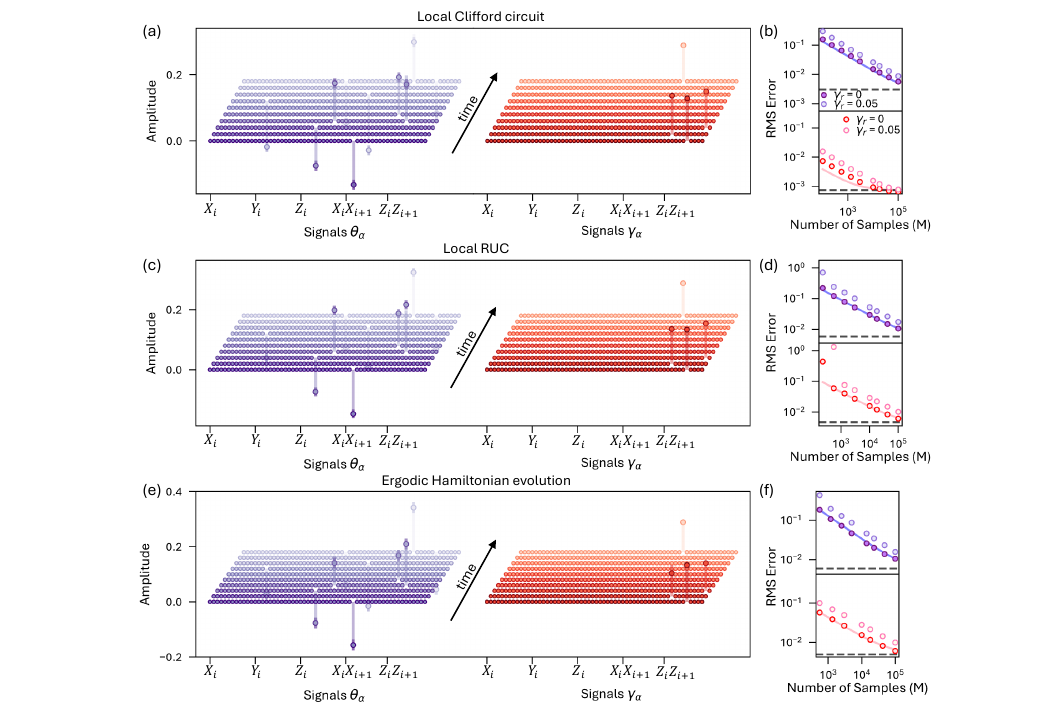}
    \caption{\textbf{(a)} Reconstruction of coherent $\hat \theta_\alpha(t)$ and incoherent $\hat \gamma_\alpha(t)$ using local Clifford circuits over $T=10$ time steps. The signal generators of both types of signal are Pauli $X_i$, $Y_i$, and $Z_i$ at each qubit $i$, as well as two-body $X_{i}X_{i+1}$ and $Z_{i}Z_{i+1}$ for nearest-neighbors $i, i+1$. The simulation is conducted on $N = 12$ qubits, for a total of $K_{ic} = K_{c} =58\times10$ signals. $15$ circuits are used to determine $\hat \theta_\alpha(t)$ and $5$ circuits used for $\hat \gamma_\alpha(t)$. The plot is shown at $M = 10^4$, where values are thresholded according to Sec. \ref{sec:thres}. Signals obey $\theta_{min} = \gamma_{min} = 0.1$, with a total strength $A = 0.53$ (Eq. \eqref{eq:Acliff}). Estimators $\hat \gamma_\alpha(t)$ are corrected by subtracting the coherent contribution from $\hat \theta_{\alpha}(t)$ at $M=\infty$ (see Sec. \ref{sec:ovlapcliff}).  \textbf{(b)}  Scaling of the RMS error for $\hat \theta_\alpha(t)$ and $\hat \gamma_\alpha(t)$ in (a), both without readout error and with readout error of $\gamma_r = 0.05$ per qubit. Theory curves plot the approximate variances for the global Clifford case, Eqs. \eqref{eq:varcliffincoh}, and \eqref{eq:varcliffcoh}. \textbf{(c)} Reconstruction of the same set of signals as in panel (a), but using the local RUC protocol. The plot is shown at $M = 4.2\times10^4$. Thresholding is conducted as in Sec. \ref{sec:thres}. \textbf{(d)} Scaling of the RMS error for $\hat \theta_\alpha(t)$ and $\hat \gamma_\alpha(t)$ in (c), both without readout error and with readout error of $\gamma_r = 0.05$ per qubit. Curves are fitted according to Eq. \eqref{eq:lrucvar}, yielding $\beta_1 = 2.3$ and $\beta_2 = 0.57$ (c.f. $\beta_1 = 0.5$ and $\beta_2 = 1.5\times10^{-3}$ for global Clifford unitaries). \textbf{(e)} Reconstruction of the same set of signals as (a), but using using evolution under Eq. \eqref{eq:kimhuse} for time $\tau = 5$ over $T=10$ time steps. The plot is shown at $M = 1.8\times10^4$, where values are thresholded according to Sec. \ref{sec:thres}. \textbf{(f)} Scaling of the RMS error for $\hat \theta_\alpha(t)$ and $\hat \gamma_\alpha(t)$ in (e), both without readout error and with readout error of $\gamma_r = 0.05$ per qubit. Curves are fitted according to Eq. \eqref{eq:lrucvar}, yielding $\beta_1 = 2.12$ and $\beta_2 = 0.53$ (c.f. $\beta_1 = 2.3$ and $\beta_2 = 0.57$ for local RUCs). }  
    \label{fig:localres}
\end{figure}

Numerical results of this procedure are shown in Fig. \ref{fig:localres}(a, b), corroborating our expectation that similar performance as the global Clifford procedure can be achieved using local gates. We note that a slightly higher number of circuits was used compared to the global Clifford case shown in the main text. Moreover, the RMS error of $\hat \gamma_\beta(t)$ is also slightly higher than that predicted using global Clifford unitaries---this is expected as the probabilistic assumptions made in deriving Eq. \eqref{eq:varcliffincoh} now no longer fully apply. Nevertheless, the estimators still exhibit SQL scaling, even in the presence of nonzero readout error.

\subsection{Random unitary circuits}\label{sec:lruc}
Rather than using random brickwork Clifford circuits, we can take the brickwork layers to be composed of random two-qubit unitaries drawn from the Haar measure (Fig \ref{fig:localfig}(a)). Let $\mathrm{U} = \{U_i\}_{i=1}^{T+1}$ indicate the layers of local unitaries applied between signal accumulation of $T$ timesteps. The use of RUCs to learn incoherent signals is studied in depth in Ref.~\cite{manole2025}. Here, we define
\begin{align}
    U_i = \big(U^i_{12}\otimes U^i_{34}\cdots \otimes U^i_{N-1,N} \big)\big(U^i_{23}\otimes U^i_{34}\cdots \otimes U^i_{N-1 ,1}\big),
\end{align}
where each $U^i_{j, j+1} \sim \text{Haar}(U(4))$, and $U(4)$ is the unitary group on $4\times4$ matrices.
At the end of the circuit, we measure in the $z-$basis, \textit{without} applying the circuit inverse as the last unitary layer. Unlike the Clifford case, we do not need to take multiple sets of circuits, nor do we need the last unitary to be the inverse of the previous layers. Although this results in less structure than the Clifford-circuit protocol, classical simulation here remains difficult regardless of the inverse constraint, so we omit it and analyze the general case. We note that as only two nearest-neighbor layers are used between signal layers to scramble signals, the requisite circuit depth is quite small. We justify why shallow circuits suffice in Sec.~\ref{sec:rucinvert}.

As before, we can consider how the presence of signals affects the output probability distribution $p(z|\vec{\theta}, \vec{\gamma})$. Now, we define
\begin{align}
p_0(z) &\equiv |\braket{z|U_{T+1}\cdots U_1|0}|^2, \label{eq:p0ruc}\\
    \delta p_{\alpha, t}(z) &\equiv\partial_{\theta_\alpha(t)}p(z|\vec{\theta}, \vec{\gamma}) = -i \braket{z|\cdots U_{t+1}P_\alpha U_t\cdots|0}\braket{0|U^\dagger_1\cdots U^\dagger_{T+1}|z} + h.c. , \label{eq:dpruc}\\
    k_{\alpha, t}(z) &\equiv \partial_{\gamma_\alpha(t)}p(z|\vec{\theta}, \vec{\gamma}) = |\braket{z|\cdots U_{t+1}P_\alpha U_t\cdots|0}|^2, \label{eq:kruc}
\end{align}
where $P_\alpha$ is a Pauli operator that generates the signal.

We can then approximate
\begin{align}\label{eq:problruc}
    p(z|\vec{\theta}, \vec{\gamma}) \approx A\bigg(p_0(z) + \sum_{\alpha, t}\theta_{\alpha}(t)\delta p_{\alpha, t}(z)+ \sum_{\beta, t}\frac{\gamma_\beta(t)}{1-\gamma_\beta(t)}k_{\beta, t}(z) \bigg),
\end{align}
where $A = \prod_{\alpha, t}\cos^2(\theta_\alpha(t))\prod_{\beta, t'}(1-\gamma_\beta(t'))$ is the probability that no signal occurs. Here, we also assume coherent and incoherent signals couple to the system via different operators $\{P_\alpha\}$, $\{P_\beta\}$ for simplicity, but our analysis can be easily extended to when they are present with the same Pauli generators (see Section \ref{sec:ovlapcliff}). Now, as before, our estimators are constructed by inverting Eq. \eqref{eq:problruc} using linear regression.

Specifically, we think of $p_0(z)$, $\delta p_{\alpha, t}(z) $, and $k_{\beta, t}(z)$ as vectors over $2^N$ bitstrings. We can concatenate them horizontally into a $2^N\times(K_c + K_{ic} + 1)$ dimensional matrix $V$, with columns $V_{z, 0} = p_0(z)$,  $V_{z, \alpha} = \delta p_{\alpha, t}(z)$, and $V_{z, \beta} = k_{\beta, t}(z)$. We can then rewrite Eq. \eqref{eq:problruc} as
\begin{align}
     p(z|\vec{\theta}, \vec{\gamma}) \approx A\bigg(V_{z,0} + \sum_{\alpha, t}\theta_{\alpha}(t)V_{z,\alpha}+ \sum_{\beta, t}\frac{\gamma_\beta(t)}{1-\gamma_\beta(t)}V_{z,\beta} \bigg).
\end{align}
Defining
\begin{align}
    v_l = \sum_{i, z}(V^TV)^{-1}_{l, i}V^T_{i, z}\frac{\hat N_z}{M}, 
\end{align}
where $\hat N_z$ is the number of times bitstring $z$ is observed and $M$ is the total number of samples, we then have
\begin{align}\label{eq:estruc}
    \hat \theta_{\alpha}(t) = \frac{v_\alpha}{v_0}, \;\; \hat \gamma_{\beta}(t) = \frac{v_\beta}{v_0 + v_\beta}.
\end{align}
These estimators are unbiased up to higher-order corrections in the signals. We expect these estimators to have sample complexity
\begin{align}\label{eq:lrucvar}
    \Var(\hat \theta_{\alpha}(t)) = \frac{\beta_1}{A^2M}, \;\; \Var(\hat \gamma_{\beta}(t)) = \frac{\beta_2}{A^2M},
\end{align}
where $\beta_1$ and $\beta_2$ are $O(1)$ constants. By the same logic as in Section \ref{sec:linreadout}, we also expect these estimators to be weakly robust to readout error.

Numerical results of this procedure are shown in Fig. \ref{fig:localres}(c),(d). Though our estimators Eq. \eqref{eq:estruc} do not perform as well as in the global Clifford or local Clifford case---namely, the scaling of RMS error with $M$ is worse by a constant factor---we still observe SQL scaling, even in the presence of nonzero readout error. Moreover, sparse signals are faithfully reconstructed.

For a more detailed treatment of incoherent signal estimation using RUCs, we refer the reader to Ref.~\cite{manole2025}, which considers this problem in the context of benchmarking and derives provably optimal estimators.

\subsubsection{Invertibility}\label{sec:rucinvert}


Here, we show through random matrix calculations why we expect $V^TV$ to be invertible when using local random unitary circuits.

To be clear, we first write the following expressions for the entries of $V^TV$:

\begin{align}
    (V^TV)_{0, 0} &= \sum_z p_0(z)^2 \\
    (V^TV)_{0, \alpha t} &= \sum_z  p_0(z) \delta p_{\alpha, t}(z) \\
    (V^TV)_{0, \beta t} &= \sum_z  p_0(z) k_{\beta, t}(z)  \\
    (V^TV)_{\alpha' t', \alpha t} &= \sum_z  \delta p_{\alpha, t}(z) \delta p_{\alpha', t'}(z) \\
    (V^TV)_{\alpha t, \beta t'} &= \sum_z  \delta p_{\alpha, t}(z) k_{\beta', t'}(z) \\
    (V^TV)_{\beta t, \beta' t'} &= \sum_z  k_{\beta, t}(z) k_{\beta', t'}(z).
\end{align}

Our goal will be to calculate the expected value of each of these entries over random unitary circuits.

\begin{figure*}
    \centering
    \includegraphics[width=\linewidth]{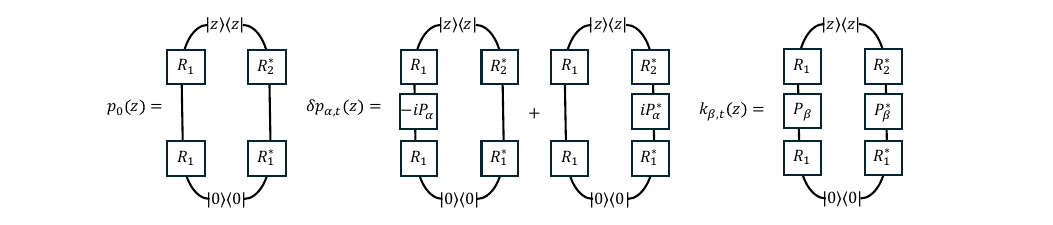}
    \caption{Tensor network diagrams depicting Eqs. \eqref{eq:lrucp01} to  \eqref{eq:lruck1}.}
    \label{fig:td1}
\end{figure*}

\begin{figure*}
    \centering
    \includegraphics[width=\linewidth]{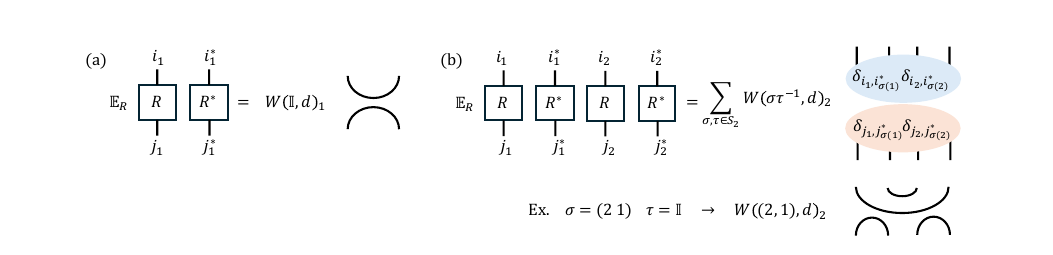}
    \caption{\textbf{(a)} Weingarten calculus calculation for a \text{1-design}, or when one copy of $U, U^*$ is averaged over the Haar measure. Here, $d = 2^N$ is the dimension of the $N$-qubit system. \textbf{(b)} Weingarten calculus calculation for a \text{2-design}. Higher moment calculations follow analogously. For more information, see Ref. \cite{Collins2006}.}
    \label{fig:td2}
\end{figure*}

\noindent\textbf{Diagonal Entries ---} To proceed, we carry out the calculation assuming signals which are sufficiently in the bulk of the RUC, such that that we can replace the brickwork circuits before and after the location of the signal with global Haar random unitaries. More specifically, for calculating the diagonal entries of $V^TV$, we approximate
\begin{align}
p_0(z) &= |\braket{z|R_2R_1|0}|^2 \label{eq:lrucp01}\\
    \delta p_{\alpha, t}(z) &= -i \braket{z|R_2 P_\alpha R_1|0}\braket{0|R^\dagger_1 R_2|z} + h.c. \label{eq:lrucdp1}\\
    k_{\alpha, t}(z) &= |\braket{z|R_2 P_\alpha R_1|0}|^2 \label{eq:lruck1}
\end{align}
where $R_1, R_2 \sim \text{Haar}(2^N)$ and average over $R_1$, $R_2$. We can compute this average by using tensor network diagrams, shown for the expressions above in Fig. \ref{fig:td1}, as well as Weingarten calculus \cite{Collins2006, Collins2022}, explained graphically in Fig. \ref{fig:td2}. Letting $d = 2^N$, this yields
\begin{align}
    \mathbb{E}_{R_1, R_2}  (V^TV)_{0, 0} &= \frac{2}{d} + O(\frac{1}{d^2}) \\
      \mathbb{E}_{R_1, R_2}  (V^TV)_{\alpha t, \alpha t} &= \frac{2}{d} + O(\frac{1}{d^2}), \label{eq:vdiaglrucp}\\
       \mathbb{E}_{R_1, R_2}  (V^TV)_{\beta t, \beta t} &= \frac{2}{d} + O(\frac{1}{d^2}),
\end{align}
where we have expanded the results in $\frac{1}{d}$. Fluctuations about these means can be shown to be suppressed by a factor of $\frac{1}{\sqrt{d}}$ though a similar calculation, and are thus exponentially smaller. Thus, even when not averaging over random circuits, we expect these expectations to hold for individual circuit instances.

\noindent\textbf{Off-diagonal Entries ---} Results for each of these entries are given in Eqs. \eqref{eq:p0pd}, \eqref{eq:p0k}, \eqref{eq:papb}, \eqref{eq:pakb}, and \eqref{eq:kbkbp}. For off-diagonal terms $(V^TV)_{0, \alpha t}$ and $(V^TV)_{0, \beta t}$, we continue to use the diagrams in Fig. \ref{fig:td1}. This yields
\begin{align}
     \mathbb{E}_{R_1, R_2}  (V^TV)_{0, \alpha t} &=0 \label{eq:p0pd}\\
     \mathbb{E}_{R_1, R_2}  (V^TV)_{0, \beta t} &= \frac{1}{d} + O(\frac{1}{d^3}) \label{eq:p0k}
\end{align}
Fluctuations about these means are suppressed by $\frac{1}{\sqrt{d}}$. As $\mathbb{E}_{R_1, R_2}  (V^TV)_{0, \alpha t} =0$, we further compute the variance exactly to quantify the magnitude of these entries,
\begin{align}
    \Var\bigg((V^TV)_{0, \alpha t}\bigg) = \mathbb{E}_{R_1, R_2} \bigg((V^TV)^2_{0, \alpha t}\bigg)= \frac{4}{d^3} + O(\frac{1}{d^4}).
\end{align}
So, the magnitude of $(V^TV)_{0, \alpha t}$ is $\frac{1}{\sqrt{d}}\frac{2}{d}$, or smaller by a factor of $\frac{1}{\sqrt{d}}$ with respect to entries on the diagonal.

\begin{figure}
    \centering
    \includegraphics[width=\linewidth]{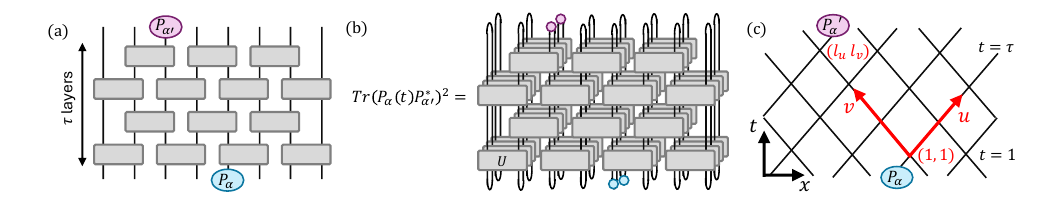}
    \caption{\textbf{(a)} Example of how we treat two space-time separated signals $P_\alpha$ and $P_{\alpha'}$. \textbf{(b)} Tensor network diagram depicting Eq. \eqref{eq:paruc}. Here, purple dots represent $P_{\alpha'}$ and blue dots $P_{\alpha}$. The unitaries within the brickwork circuit are ordered, from front to back, by $U$, $U^*$, $U$, and $U^*$. \textbf{(c)} Lightcone coordinates shown for our RUC system. A two-qubit Haar-random unitary lies at each point $(u, v)$.} 
    \label{fig:td3}
\end{figure}

We now calculate the expected value of $(V^TV)_{\alpha' t', \alpha t}$. Here, to account for the fact that the circuit separating signals $\theta_\alpha(t)$, $\theta_\alpha'(t')$ may not be deep enough to be considered a Haar random unitary, we consider the modified expressions
\begin{align}
    \delta p_{\alpha, t}(z) &= -i \braket{z|R_2 (RUC)P_\alpha R_1|0}\braket{0|R^\dagger_1 (RUC)^\dagger R^\dagger_2|z} + h.c. ,\\
    \delta p_{\alpha', t'}(z) &= -i \braket{z|R_2 P_{\alpha'} (RUC)R_1|0}\braket{0|R^\dagger_1 (RUC)^\dagger R^\dagger_2|z} + h.c.,
\end{align}
where $RUC$ is the brickwork random local unitary circuit separating $P_\alpha$ and $P_{\alpha'}$. For simplicity of analysis, we assume $P_\alpha$  and $P_{\alpha'}$ to be single-qubit operators.  This is illustrated in Fig. \ref{fig:td3}(a). Specifically, if signals $\theta_\alpha(t)$,  $\theta_\alpha'(t')$ are $\tau$ layers apart in the circuit (i.e. $t' = t + \tau$), then RUC is a brickwork circuit of depth $\tau$. We aim to average $(V^TV)_{\alpha' t', \alpha t}$ over both $R_1, R_2$ and the gates of $RUC$. From a similar calculation as before,
\begin{align}
     \mathbb{E}_{R_1, R_2}(V^TV)_{\alpha' t', \alpha t} &= \frac{1}{d^2} \bigg(\Tr(RUC^\dagger P_\alpha RUC P_{\alpha'}) + c.c. \bigg) \\
     & = \frac{1}{d^2} \bigg(\Tr( P_{\alpha}(\tau) P_{\alpha'}) + \Tr( P^*_{\alpha}(\tau) P^*_{\alpha'}) \bigg)
\end{align}
where in the second line, we have defined 
\begin{align}
    RUC^\dagger P_{\alpha} RUC \equiv P_{\alpha}(\tau).
\end{align}
Moreover, we have used that for Pauli operators, $P = P^\dagger$. Now, 
\begin{align}
    \mathbb{E}_{RUC} \frac{1}{d^2} \bigg(\Tr( P_{\alpha}(\tau) P_{\alpha'}) + \Tr( P^*_{\alpha}(\tau) P^*_{\alpha'}) \bigg)\sim \Tr(P_\alpha)\Tr(P_{\alpha'})  = 0,
\end{align}
where the average over the local gates of the RUC can be performed using a tensor network diagram and the rules of Fig. \ref{fig:td2}. Thus, we have 
\begin{align}
    E_{R_1, R_2, RUC} (V^TV)_{\alpha' t', \alpha t}= 0.
\end{align}
We now explicitly compute the variance of this entry to quantify its magnitude.
\begin{align}
\mathbb{E}_{R_1, R_2} ((V^TV)_{\alpha' t', \alpha t})^2 = \frac{d^{-2}}{d^2} \bigg(\Tr( P_{\alpha}(\tau) P_{\alpha'}) + \Tr( P^*_{\alpha}(\tau) P^*_{\alpha'}) \bigg)^2 + O(\frac{1}{d^3}),
\end{align}
where we have explicitly included $d^{-2}$ to normalize the trace, which we expect to be $O(d^2)$. Importantly, if we were to assume a Haar random unitary between $P_\alpha$ and $P_{\alpha'}$, or $P_{\alpha'}$ not within the lightcone of $P_\alpha$, then the variance would be $O(\frac{1}{d^3})$. Thus, the $O(\frac{1}{d^2})$ component quantifies how a local brickwork RUC affects correlation between two different signals which are causally related.

We can expand $\bigg(\Tr( P_{\alpha}(\tau) P_{\alpha'}) + \Tr( P^*_{\alpha}(\tau) P^*_{\alpha'}) \bigg)^2$ = $\Tr( P_{\alpha}(\tau) P_{\alpha'})^2 + 2\Tr( P_{\alpha}(\tau) P_{\alpha'}) \Tr( P^*_{\alpha}(\tau) P^*_{\alpha'}) + \Tr( P^*_{\alpha}(\tau) P^*_{\alpha'})^2$. We can simplify this expansion by noting that
\begin{align}
    \Tr( P^*_{\alpha}(\tau) P^*_{\alpha'})  &= \Tr(P^\dagger_{\alpha'} P^\dagger_{\alpha}(\tau)) \\
    &= \Tr(P_{\alpha}(\tau) P_{\alpha'} ),
\end{align}
where we have used the cyclic property of the trace, as as well the hermiticity of Pauli operators.

So, we have the leading order term 
\begin{align}\label{eq:paruc}
    \mathbb{E}_{R_1, R_2, RUC} ((V^TV)_{\alpha' t', \alpha t})^2 =4 \mathbb{E}_{RUC} \frac{d^{-2} \Tr( P_{\alpha}(\tau) P_{\alpha'})^2}{d^2}.
\end{align}
A tensor network diagram for this term is depicted in Fig. \ref{fig:td3}.

Using techniques pioneered in Refs. \cite{Nahum2018} and \cite{Bao2020}, we can evaluate the average of this quantity over local gates in the brickwork RUC by mapping to a spin model. We find it convenient to state the final answer in terms of lightcone coordinates,
\begin{align}
    u \coloneqq \frac{(t + x + 1)}{2}, \,\, v \coloneqq \frac{(t - x + 1)}{2}, \,\, t = u + v - 1, \,\, x= u - v,
\end{align}
as shown in Fig. \ref{fig:td3}(c), as well as for arbitary local dimension $q$. Here, $x$ is the space coordinate, while $t$ is the vertical, or time coordinate. We assume $t = 1$ at the start of the circuit and $t = \tau$ at the end.

Then, similar to the result in Ref.~\cite{Nahum2018},
\begin{align}\label{eq:dxt}
    \mathbb{E}_{RUC} \bigg(d^{-2} \Tr( P_{\alpha}(\tau) P_{\alpha'}) \bigg) &=  \frac{q^2}{q^4 - 1}\bigg(\frac{q}{q^2 + 1}\bigg)^{2 \tau - 2} \sum_{u = 0}^{l_u - 1} \sum_{v = 0}^{l_v - 1} \bigg(\frac{1}{q}\bigg)^{2\tau - 2 u - 2 v}\bigg[\begin{pmatrix}
        \tau - 1 \\ v
    \end{pmatrix}\begin{pmatrix}
        \tau - 1 \\ u
    \end{pmatrix} - \begin{pmatrix}
        \tau - 1 \\ v - 1
    \end{pmatrix}\begin{pmatrix}
        \tau - 1 \\ u - 1
    \end{pmatrix} \bigg], \\
    &\equiv D(l_ u, l_v ) \notag,
\end{align}
where $(l_u, l_v)$ is the separation between $P_\alpha$ and $P_\alpha'$ in spacetime coordinates. The second term in brackets vanishes when $u-1$ or $v-1$ is less than $0$. $D(l_ u, l_v )$ thus quantifies the leading-order variance of the correlation between two spacetime-separated operators in a local brickwork RUC. As we work with qubits, $q = 2$. So, we finally have
\begin{align}\label{eq:papb}
    \mathbb{E}_{R_1, R_2, RUC} ((V^TV)_{\alpha' t', \alpha t})^2 = \frac{4D(l_ u, l_v )}{d^2}  + O(\frac{1}{d^3}),
\end{align}
and $(V^TV)_{\alpha' t', \alpha t}$ has fluctuations with magnitude $\frac{2\sqrt{D(l_ u, l_v )}}{d}$. Upon first glance, this is the same order in $\frac{1}{d}$ as the diagonal terms Eq. \eqref{eq:vdiaglrucp}. However, examining Eq. \eqref{eq:dxt}, we can see that it is only nonzero within the lightcone of $P_\alpha$, and it is furthermore exponentially suppressed when moving further away from $P_\alpha$ in time and space. In practice, as our signals are separated by at least two brickwork layers, we find that off-diagonal terms are small compared to the diagonal.

We follow a similar procedure for $(V^TV)_{\beta' t', \alpha t}$, using

\begin{align}
    \delta p_{\alpha, t}(z) &= -i \braket{z|R_2 (RUC)P_\alpha R_1|0}\braket{0|R^\dagger_1 (RUC)^\dagger R^\dagger_2|z} + h.c. ,\\
    k_{\beta', t'}(z) &= |\braket{z|R_2 P_{\beta'} (RUC)R_1|0}|^2.
\end{align}
This yields
\begin{align}\label{eq:pakb}
    \mathbb{E}_{R_1, R_2, RUC}(V^TV)_{\beta' t', \alpha t} = 0.
\end{align}
Furthermore, the magnitude of fluctuations of $(V^TV)_{\beta' t', \alpha t}$ is suppressed by a factor of $\frac{1}{\sqrt{d}}$ with respect to diagonal entries of $(V^TV)$. 

We finally compute $(V^TV)_{\beta' t', \beta t}$, using the same ideas as before. We have
\begin{align}
    \mathbb{E}_{R_1, R_2}(V^TV)_{\beta' t', \beta t} =  \frac{1 + d^{-2}|\Tr( P_\beta(t) P_{\beta'})|^2 }{d} + O\bigg(\frac{1}{d^2}\bigg).
\end{align}
Now, if $RUC$ were global Haar-random, or if $P_\beta$ and $P_{\beta'}$ are not within each others' light cones, then we would have $\mathbb{E}_{R_1, R_2}(V^TV)_{\beta' t', \beta t} = \frac{1}{d} + O(\frac{1}{d^3})$. Thus, the difference that the local RUC imparts lies in the $|\Tr( P_\beta(t) P_{\beta'})|^2$ term. From our previous calculations, we can write
\begin{align}\label{eq:kbkbp}
    \mathbb{E}_{R_1, R_2, RUC}(V^TV)_{\beta' t', \beta t} =  \frac{1 + D(l_u, l_v)}{d} + O\bigg(\frac{1}{d^2}\bigg),
\end{align}
where $(l_u, l_v)$ is the separation between $P_\beta$ and $P_\beta'$ in lightcone coordinates (see Fig. \ref{fig:td3}). $D(l_u, l_v)$ is defined in Eq. \eqref{eq:dxt}. $D(l_u, l_v)$ is exponentially suppressed the further  $P_{\beta'}$ and  $P_{\beta}$ are from each other. So, in practice, with two local brickwork layers between signals, this correction is quite small and $(V^TV)_{\beta' t', \beta t} \approx \frac{1}{d}$. Moreover, fluctuations about this mean are suppressed by a factor of $\frac{1}{\sqrt{d}}$.

\noindent\textbf{Invertibility analysis---} Gathering our results from above, for enough layers between signal operators, we expect $V^TV$ to have the following entries
 \begin{align}\label{eq:EVTVlruc}
     \mathbb{E}(V^TV) \approx
\bordermatrix{
          & 0 & \beta t_1 & \beta t_2 & \beta t_3 & \cdots & \alpha t_1 & \alpha t_2 & \alpha t_3 & \cdots \cr
  0       & \tfrac{2}{d} & \tfrac{1}{d} & \tfrac{1}{d} & \tfrac{1}{d} & \cdots & 0 & 0 & 0 & \cdots \cr
  \beta t_1 & \tfrac{1}{d} & \tfrac{2}{d} & \tfrac{1}{d} & \tfrac{1}{d} & \cdots & 0 & 0 & 0 & \cdots \cr
  \beta t_2 & \tfrac{1}{d} & \tfrac{1}{d} & \tfrac{2}{d} & \tfrac{1}{d} & \cdots & 0 & 0 & 0 & \cdots \cr
  \beta t_3 & \tfrac{1}{d} & \tfrac{1}{d} & \tfrac{1}{d} & \tfrac{2}{d} & \cdots & 0 & 0 & 0 & \cdots \cr
  \vdots   & \cdots       & \cdots       & \cdots       & \cdots       & \ddots & \cdots & \cdots & \cdots & \ddots \cr
  \alpha t_1 & 0 & 0 & 0 & 0 & \cdots & \tfrac{2}{d} & 0 & 0 & \cdots \cr
  \alpha t_2 & 0 & 0 & 0 & 0 & \cdots & 0 & \tfrac{2}{d} & 0 & \cdots \cr
  \alpha t_3 & 0 & 0 & 0 & 0 & \cdots & 0 & 0 & \tfrac{2}{d} & \cdots \cr
  \vdots   & \cdots & \cdots & \cdots & \cdots & \ddots & \cdots & \cdots & \cdots & \ddots \cr
}
 \end{align}
Matrix elements between $\beta' t'$ and $\beta t$ and  have corrections that are exponentially small in the spacetime distance of $\gamma_{\beta'}(t')$, $\gamma_{\beta}(t)$. Moreover, fluctuations about these means are generally of order $\frac{1}{d^{3/2}}$, with the exception of $(V^TV)_{\alpha' t', \alpha t}$--- in this case, $(V^TV)_{\alpha' t', \alpha t}$ has fluctuations of magnitude $O(\frac{1}{d})$ that also decrease exponentially with the spacetime distance of the signals $\theta_{\alpha'}(t')$, $\theta_{\alpha}(t)$. We omit these exponentially small corrections in Eq. \eqref{eq:EVTVlruc}. Exponentially small fluctuations indicate that even when using a single random unitary circuit instance, $V^TV$ should behave like its mean. 

Now, Eq. \eqref{eq:EVTVlruc} is invertible: its upper left block is a rank-one update of the identity, while its lower right block is diagonal. For a single circuit instance, the realized $V^TV$ will only have small--- in general, exponentially small-- perturbations to Eq. \eqref{eq:EVTVlruc}. This, we expect it to be invertible~\cite{Cook2018}.

\subsection{Ergodic Hamiltonian evolution} \label{sec:ham_ev}
Rather than using local RUC layers, our sensing protocol can also use layers of ergodic Hamiltonian evolution~\cite{Mark2023}, as illustrated in Fig. \ref{fig:localfig}. We can still run the same estimation procedure for coherent and incoherent signals described in Section \ref{sec:lruc}. Intuitively, this is because we expect local Hamiltonian dynamics to be roughly equivalent to shallow layers of a nearest neighbor brickwork RUC.

To be specific, a Hamiltonian $H$ is chosen for the entire protocol and ``turned on" in between signal accumulation layers. We choose a reasonable evolution time $\tau$, and the unitary $e^{-i H\tau}$ is implemented between signals. This can be achieved, for example, by selectively echoing out the Hamiltonian while probing desired signals~\cite{Zhou2020, Choi2020}.


At the end of the procedure, we perform measurements in the computational basis as usual. Our output probability distribution is still defined by Eq. \eqref{eq:problruc}, with definitions in Eqs. \eqref{eq:p0ruc}, \eqref{eq:dpruc}, and \eqref{eq:kruc}. In this case, we just replace unitary layers by evolution under a single Hamiltonian $H$
\begin{align}\label{eq:kimhuse}
    U_i \rightarrow \exp(-iH\tau),
\end{align}
where $\tau$ is a chosen period of evolution to ensure sufficient scrambling. The data analysis protocol is identical to Section \ref{sec:lruc}: we compute Eqs. \eqref{eq:p0ruc}, \eqref{eq:dpruc}, and \eqref{eq:kruc} using classical simulation of the Hamiltonian dynamics, and construct the estimator Eq. \eqref{eq:estruc} for coherent and incoherent signals.

To test this protocol, we choose to evolve under the mixed-field Ising model with Kim-Huse parameters \cite{Kim2013} and periodic boundary conditions,
\begin{align}
    H = -\sum_{i = 1}^{N}Z_{i}Z_{i+1} - h_x\sum_{i = 1}^{N}X_{i} -h_z\sum_{i = 1}^{N}Z_{i}, \;\; (h_x, h_z)=\bigg(\frac{\sqrt{5}+5}{8}, \frac{\sqrt{5}+1}{4}\bigg).
\end{align}
We begin with an initial product state of $\ket{+}_y^{\otimes N}$, where $\ket{+}_y$ is the $+1$ eigenstate of Pauli $Y$. This ensures that the initial state is at infinite temperature, and hence the dynamics explores a large effective Hilbert space. The results are shown in Fig. \ref{fig:localres}(e,f), for an evolution time $\tau = 5$. The performance is comparable to the local RUC protocol, with estimators faithfully reconstructing sparse signals and maintaining SQL scaling in the presence of nonzero readout error. We leave a more extensive optimization of these estimators in the case of Hamiltonian evolution to future work.


\bibliography{refs_sup}